\documentclass{aastex63}
\usepackage{gensymb} 
\usepackage{booktabs} 
\graphicspath{{./}}
\usepackage[running,right]{lineno}

\begin{document}


\title{The Robotic Multi-Object Focal Plane System of the Dark Energy Spectroscopic Instrument (DESI)}

\newcommand{\comment}[1]{\textcolor{red}{[#1]}}
\newcommand{\michael}[1]{\textcolor{blue}{\small [MS: #1]}}

\newcommand{\figref}[1]{(see fig.~\ref{#1})}

\newcommand{\BU}{\affiliation{Boston University (BU)}}
\newcommand{\CIEMAT}{\affiliation{Centro de Investigaciones Energéticas, Medioambientales y Tecnológicas (CIEMAT)}}
\newcommand{\EPFL}{\affiliation{École polytechnique fédérale de Lausanne (EPFL)}}
\newcommand{\DurhamCAI}{\affiliation{Centre for Advanced Instrumentation, Department of Physics, Durham University}}
\newcommand{\DurhamICC}{\affiliation{Institute for Computational Cosmology, Department of Physics, Durham University}}
\newcommand{\FNAL}{\affiliation{Fermi National Accelerator Laboratory (FNAL)}}
\newcommand{\HarvardCAF}{\affiliation{Center for Astrophysics $|$ Harvard \& Smithsonian}}
\newcommand{\IAA}{\affiliation{Instituto de Astrofísica de Andalucía (IAA)}}
\newcommand{\IAC}{\affiliation{Instituto de Astrofísica de Canarias (IAC)}}
\newcommand{\ICECSIC}{\affiliation{Institut de C\`{i}encies de l'Espai (ICE), IEEC-CSIC}}
\newcommand{\ICREA}{\affiliation{Instituci\'o Catalana de Recerca i Estudis Avan\c{c}ats (ICREA)}}
\newcommand{\IFAE}{\affiliation{Institut de F\'{i}sica d’Altes Energies (IFAE), The Barcelona Institute of Science and Technology (BIST)}}
\newcommand{\IRFUSaclay}{\affiliation{IRFU, CEA, Universit\'{e} Paris-Saclay}}
\newcommand{\Kavli}{\affiliation{Kavli Institute for Particle Astrophysics and Cosmology and SLAC National Accelerator Laboratory}}
\newcommand{\LBNL}{\affiliation{Lawrence Berkeley National Laboratory (LBNL)}}
\newcommand{\LLNL}{\affiliation{Lawrence Livermore National Laboratory (LLNL)}}
\newcommand{\OSUphysics}{\affiliation{Department of Physics, The Ohio State University (OSU)}}
\newcommand{\OSUcosmo}{\affiliation{Center for Cosmology and AstroParticle Physics, The Ohio State University (OSU)}}
\newcommand{\OSUastro}{\affiliation{Department of Astronomy, The Ohio State University (OSU)}}
\newcommand{\NAOCAS}{\affiliation{National Astronomical Observatories, Chinese Academy of Sciences}}
\newcommand{\NOIR}{\affiliation{NSF's National Optical-Infrared Astronomy Research Laboratory (NOIRLab)}}
\newcommand{\Sienna}{\affiliation{Department of Physics and Astronomy, Siena College}}
\newcommand{\Sorbonne}{\affiliation{Sorbonne Universit\'{e}, Laboratoire de Physique Nucl\'{e}aire et de Hautes Energies (LPNHE)}}
\newcommand{\SMU}{\affiliation{Southern Methodist University (SMU)}}
\newcommand{\SSL}{\affiliation{Space Sciences Laboratory (SSL), UC Berkeley}}
\newcommand{\Stanford}{\affiliation{Stanford University}}
\newcommand{\UAMHCTLab}{\affiliation{Grupo de Investigación HCTLab, Escuela Politécnica Superior, Universidad Autónoma de Madrid}}
\newcommand{\UAMCSIC}{\affiliation{Instituto de F\'{\i}sica Te\'{o}rica (IFT) UAM/CSIC, Universidad Aut\'{o}noma de Madrid}}
\newcommand{\UCI}{\affiliation{University of California, Irvine (UCI)}}
\newcommand{\UCL}{\affiliation{University College London (UCL)}}
\newcommand{\UMich}{\affiliation{University of Michigan (UM)}}
\newcommand{\Waterloo}{\affiliation{Department of Physics and Astronomy, University of Waterloo}\affiliation{Perimeter Institute for Theoretical Physics}}
\newcommand{\Yale}{\affiliation{Yale University}}
\newcommand{\AffilTBD}{\affiliation{(affiliation)}}
 
\correspondingauthor{Joseph H. Silber}
\email{jhsilber@lbl.gov}
\author{Joseph Harry Silber}\LBNL
\correspondingauthor{Parker A. Fagrelius}
\email{parker.fagrelius@noirlab.edu}
\author{Parker Fagrelius}\NOIR

\author{Kevin Fanning}\OSUphysics\OSUcosmo
\author{Michael Schubnell}\UMich
\author{Jessica Nicole Aguilar}\LBNL
\author{Steven Ahlen}\BU
\author{Jon Ameel}\UMich
\author{Otger Ballester}\IFAE
\author{Charles Baltay}\Yale
\author{Chris Bebek}\LBNL
% ...

\collaboration{1000}{(DESI Collaboration)}  

\begin{abstract}
    A system of 5,020 robotic fiber positioners was installed in 2019 on the Mayall Telescope, at Kitt Peak National Observatory. The robots automatically re-target their optical fibers every 10\,--\,20~minutes, each to a precision of several microns, with a reconfiguration time less than 2~minutes. Over the next five years, they will enable the newly-constructed Dark Energy Spectroscopic Instrument (DESI) to measure the spectra of 35~million galaxies and quasars. DESI will produce the largest 3D map of the universe to date and measure the expansion history of the cosmos. In addition to the 5,020 robotic positioners and optical fibers, DESI's Focal Plane System includes 6 guide cameras, 4 wavefront cameras, 123 fiducial point sources, and a metrology camera mounted at the primary mirror. The system also includes associated structural, thermal, and electrical systems. In all, it contains over 675,000 individual parts. We discuss the design, construction, quality control, and integration of all these components. We include a summary of the key requirements, the review and acceptance process, on-sky validations of requirements, and lessons learned for future multi-object, fiber-fed spectrographs.
\end{abstract}

\section{Introduction}
    The Dark Energy Spectroscopic Instrument (DESI) is a new multi-object, fiber-fed spectrograph, operating in the wavelength range of $0.36-0.98\,\micron$. It was constructed by a collaboration of hundreds of researchers around the world, and installed on the 4m Mayall Telescope at Kitt Peak National Observatory (KPNO). The DESI Survey will be completed over 5 years, during which the instrument will measure the precise spectra and redshifts of tens of millions of galaxies and quasars, up to a redshift of $\sim3.5$. This data set will enable the DESI collaboration to build the largest 3D map of the universe to date, providing insight into the expansion history of the universe and ultimately furthering our understanding of dark energy \citep{desi16a}. The DESI instrument was developed by a worldwide collaboration, led by Lawrence Berkeley National Laboratory (LBNL), and supported by the U.S. Department of Energy, Office of Science.
    
    The primary scientific goal of the DESI survey is to constrain possible models of dark energy. We are measuring a record-breaking catalog of spectroscopic redshifts, with which we will determine the Baryon Acoustic Oscillation (BAO) scale with sub-percent precision. We will also measure anisotropies in galaxy clustering, or Redshift Space Distortions, which facilitate the study of the growth of structure in the universe. Additionally, DESI will contribute to the theory of inflation, to measurements of the mass of neutrinos, and will conduct a Milky Way Survey that will inform our understanding of the distribution of dark matter and the assembly of our galaxy. 
    
    This paper describes one of the major DESI subsystems: the Focal Plane System (FPS)\footnote{In fact DESI's focal surface is an asphere, but we frequently refer to this surface as a ``plane''.}. The DESI Focal Plane System contains 5,020 robotic positioners, six guide cameras, four wavefront cameras, 123 fiducial sources, and a metrology camera. The system also includes associated structural, thermal, electrical systems, and software. Parallel papers describe the overall DESI instrument (DESI Collaboration et al. 2022 in prep.), the optical corrector (Miller et al. 2022 in prep.) to which the FPS is mounted and which establishes the aspherical focal surface, and the fiber system (Poppett et al. 2022 in prep.) which carries the light from the FPS to the spectrographs (Jelinsky et al. 2022 in prep.). 
    
\begin{figure} 
    	\includegraphics[width=\textwidth]{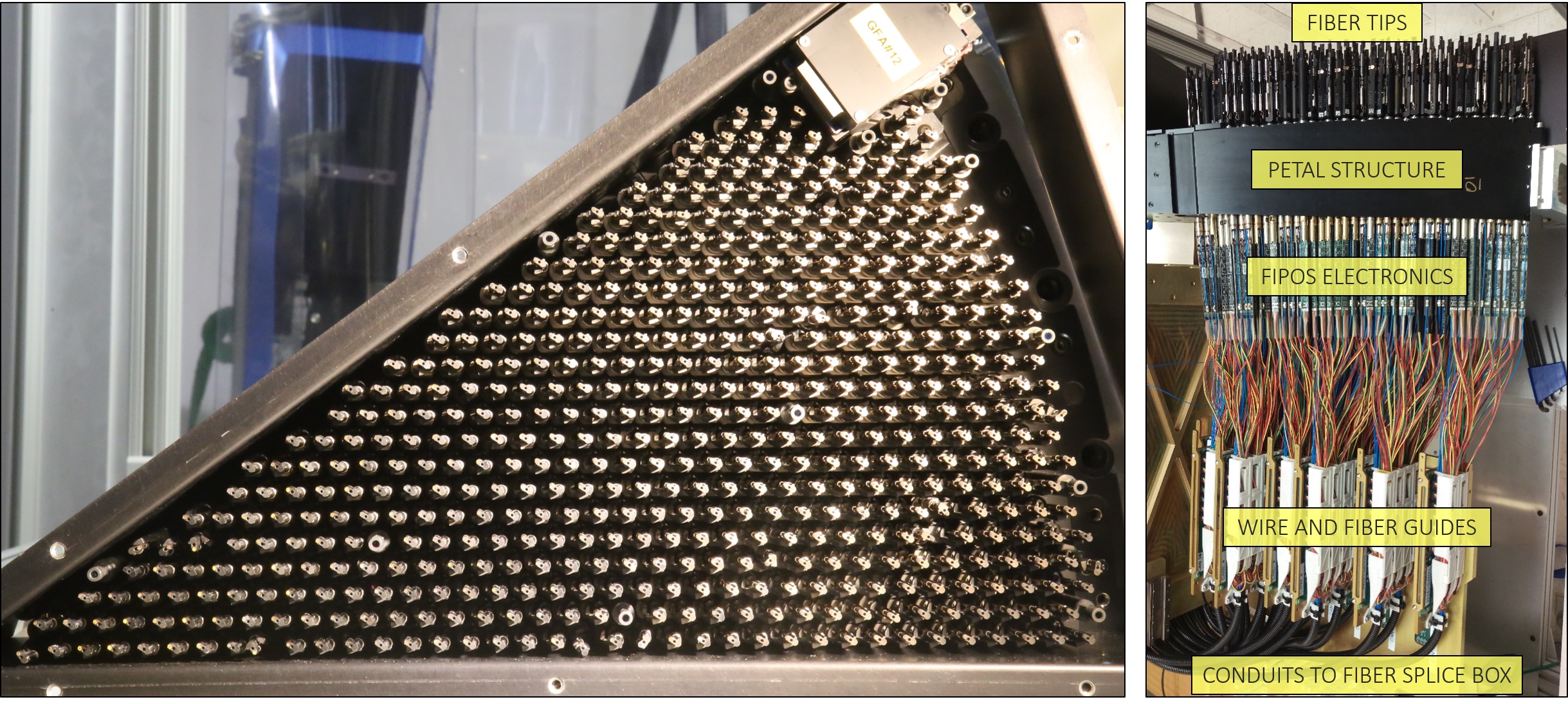}
    	\caption{\label{fig:petal1} Each of the 10 petals on DESI is a complete instrument with 502 robots, 12\,--\,14 fiducial point sources, a guide or wavefront camera, a protective enclosure for fiber splices, power supplies, and control electronics. At left, a fully-assembled petal is viewed from the front, looking toward the robotically-controlled fiber tips. Fiducial sources are mounted in cylindrical housings, interspersed throughout the robot array. A guide camera (covered by a protective plate in this image) is mounted at the upper right corner, and the whole assembly here is surrounded by a protective metal enclosure (which is removed just prior to installation in the telescope). At right, a side view shows the layout of fibers, electronics, and support structure.}
    \end{figure}

    The key feature of the Focal Plane System is a close-packed array of fiber positioning robots, each of which manipulates one optical fiber. For logistical flexibility and repetition of design elements, we subdivided the system into 10 sub-units, each with 502 robots. Mechanically, the sub-units are 10 identical 36\degree~wedge-shaped instruments, which together fill the circular focal plane. We refer to these 10 units as ``petals''. The fibers from each petal are transmitted in a cable to a single DESI spectrograph. Figure \ref{fig:petal1} shows a petal in the lab.

    Early conceptual designs for the instrument began in 2011, and by 2015 budgeting and planning was complete. Construction of the Focal Plane System took place from 2016 to 2018, and it was installed in 2019 on the Mayall Telescope \citep{besuner20, shourt20}, located atop Iolkam Du’ag (Kitt Peak) in the Tohono O’odham Nation (near Tucson, Arizona).  On-sky commissioning spanned October 2019 to March 2020, during which we demonstrated that all subsystems met or exceeded design requirements \citep{fagrelius20,poppett20,meisner20}. Operations were interrupted between March and November 2020 due to the global COVID-19 pandemic. We began a five-month Survey Validation phase starting on December 14, 2020 and began the DESI Main Survey on May 14, 2021. 
    
    Large astronomical spectroscopic surveys have long been made possible by increasingly complex fiber-fed spectrographs. This started with the Las Campanas Redshift Survey (LCRS) \citep{lascampanas} which used 100 optical fibers to measure the redshifts of 26,000 galaxies. This capability was expanded with the 640 fiber Sloan Digital Sky Survey (SDSS) focal plate system \citep{smee2013} and the 400 fiber two degree field system (2dF) with the AAOmega spectrograph on the Anglo-Australian Telescope (AAT) \citep{aaomega}. The number of fibers on the SDSS focal plane was increased to 1,000 and was used to measured more than 2 million galaxies and close to 1 million quasars during the BOSS \citep{dawson2013} and eBOSS surveys \citep{ahumada20}.
     
    For early systems, fibers were plugged by hand into custom-drilled plates, with a uniquely manufactured plate being used for each field of sky targets. To reduce reconfiguration time and labor, there has been a shift toward robotically-positioned fibers. Robotic systems have been demonstrated on both the Large Sky Area Multi-Object Fiber Spectroscopic Telescope (LAMOST) \citep{lamost} and the Fiber Multi-Object Spectrograph (FMOS) \citep{fmos} on the Subaru telescope. With its high speed ($<$~2~minute average reconfiguration time, \S \ref{sec:reconfig_performance}), large fiber count (5,020 fibers, see \S \ref{sec:layout}), and tight positioning accuracy ($<$~10\,\micron\,RMS, see \S \ref{sec:positioner_accuracy}), the DESI Focal Plane System described in this paper represents a significant technological development in the state-of-the-art for such multi-object fiber-fed designs. By January 2022, during its first year of survey operations, DESI had measured 7.5 million unique galaxy and quasar redshifts, exceeding the 3.8 million galaxy and quasar redshifts gathered during the first twenty years of operation of the Sloan Digital Sky Survey \citep{ahumada20}.
    
    This paper discusses the design, construction, quality control, and integration of the Focal Plane System. We intend for this paper to be of some utility as succeeding collaborations develop their own new technologies in the field. We describe the hardware elements in \S\ref{HARDWARE} and the software in \S\ref{SOFTWARE}. We then discuss the assembly and test of the instrument (\S\ref{ASSEMBLY_AND_TEST}), installation into the Mayall telescope (\S\ref{INSTALLATION}), and system performance (\S\ref{PERFORMANCE}). Finally, in \S\ref{LESSONS_LEARNED} we discuss key lessons-learned through the development process.
    
\section{Hardware}\label{HARDWARE}

    The DESI Focal Plane is considered a ``zonal'' system, in that each robot has a limited patrol region. This is in contrast to the 640 (upgraded to 1,000) fibers on the Sloan Digital Sky Survey focal plane \citep{smee2013}, which are hand-plugged into arbitrary holes on numerous custom-machined aluminum plates, or the 300 fibers on the MMT Hectospec \citep{hectospec}, which are positioned serially with two robots.  Despite the limitation in the reach of individual fibers, zonal systems typically offer the advantages of larger fiber count and faster reconfiguration. A zonal system was chosen for DESI, as it has been for other similar planned instruments, such as the Subaru Prime Focus Spectrograph (PFS), which contains 2,400 robotically-positioned fibers \citep{pfs2020}, the SDSS-V Focal Plane System, which will replace the manual plug-plates with 500 zonal fiber positioners \citep{sdssVfps}, and the 2,400 fiber 4MOST instrument on the VISTA telescope \citep{4most2012}.
    
    We installed the DESI Focal Plate Assembly (FPA) on the Mayall 4-m telescope, mounting it to the rear flange of DESI's prime focus corrector (Miller et al. 2022 in prep). The FPA includes the array of robotic positioners (\S \ref{sec:robots_hardware}), six guide cameras and four wavefront cameras (\S \ref{sec:GFA_hardware}), 123 fiducial sources (\S \ref{sec:fiducials}), fiber splice boxes (Poppett et al. 2022 in prep.), and the structural (\S \ref{sec:fp_structure}) and electrical systems (\S \ref{sec:electrical_system}) to support them.
    
\begin{figure} 
    	\includegraphics[width=\textwidth]{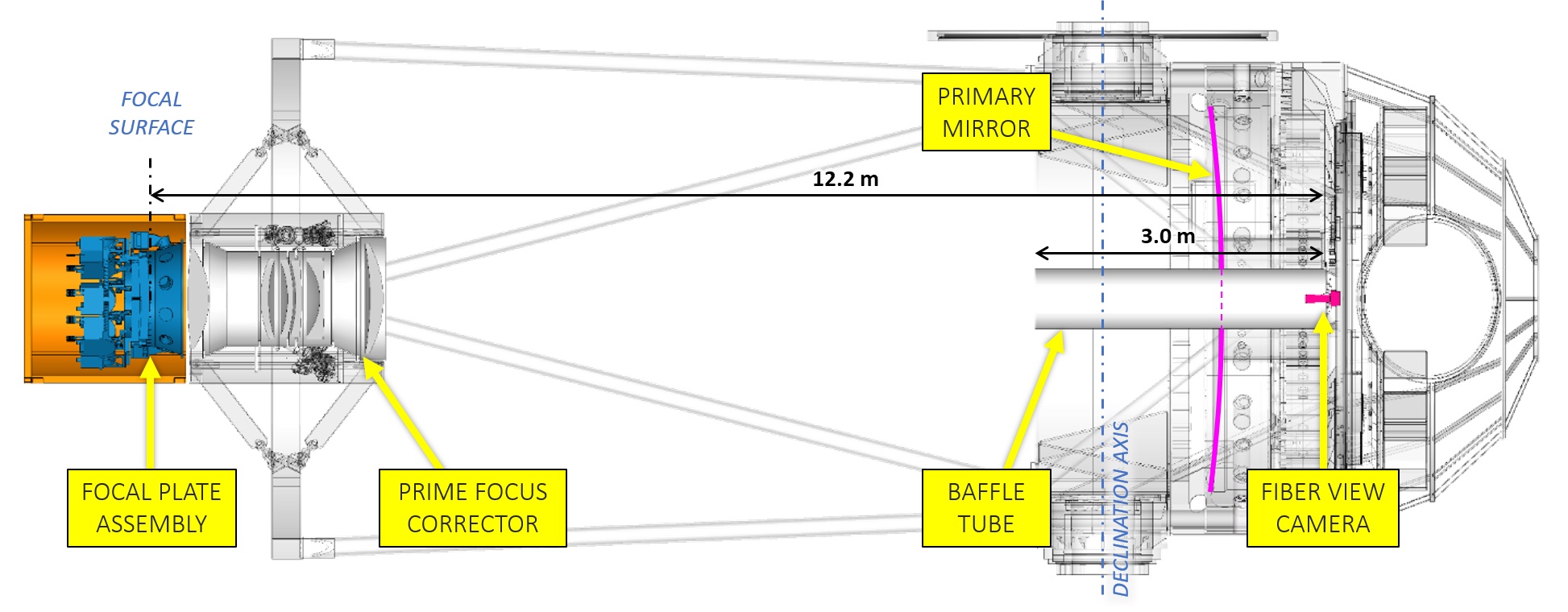}
    	\caption{\label{fig:fvc_diagram} This diagram shows the location of the focal plane system components as installed on the Mayall telescope. The focal plate assembly, which contains all robotic fiber positioners, guide and focus cameras, and fiducial sources, is mounted to the prime focus corrector. The Fiber View Camera (FVC) is mounted behind the primary mirror and measures the positions of the backlit fibers, as well as the fixed fiducials interspersed throughout the array.}
    \end{figure}

    The Focal Plane System includes the FPE, an environmental enclosure around the FPA (\S \ref{sec:thermal_mgmt}), a liquid chiller system for cooling, and a metrology camera (\S \ref{sec:fvc}), used to close the control loop on fiber positioning at the end of each reconfiguration. We mounted this Fiber View Camera (FVC) behind the primary mirror, $\sim$\,12\,m from the focal surface \figref{fig:fvc_diagram}. 

    Our survey needs required $\ge$~3,000 fibers on our instrument, each positionable to within 10\,\micron\,RMS relative to the guiding sensors. During planning, we weighed the risks of instrument downtime and the failure probabilities of individual robots against the unit costs of robots, cables, and spectrographs. From this analysis, we settled on 5,000 fibers\footnote{Twenty additional robotic fibers were added at the periphery of the instrument (hence the 5,020 total robots frequently referenced in this paper), which feed not into science spectrographs, but rather into a separate camera for monitoring sky background levels.} as a fairly optimal target for the hardware design. Additional key requirements on the Focal Plane hardware design were to operate in the ambient temperature, humidity, and dust levels of the Kitt Peak environment, and to keep the total mass of all components attached to the optical corrector $\le$~870\,kg. 

\subsection{Fiber positioner robots}
\label{sec:robots_hardware}
\begin{figure} 
    	\includegraphics[width=\textwidth]{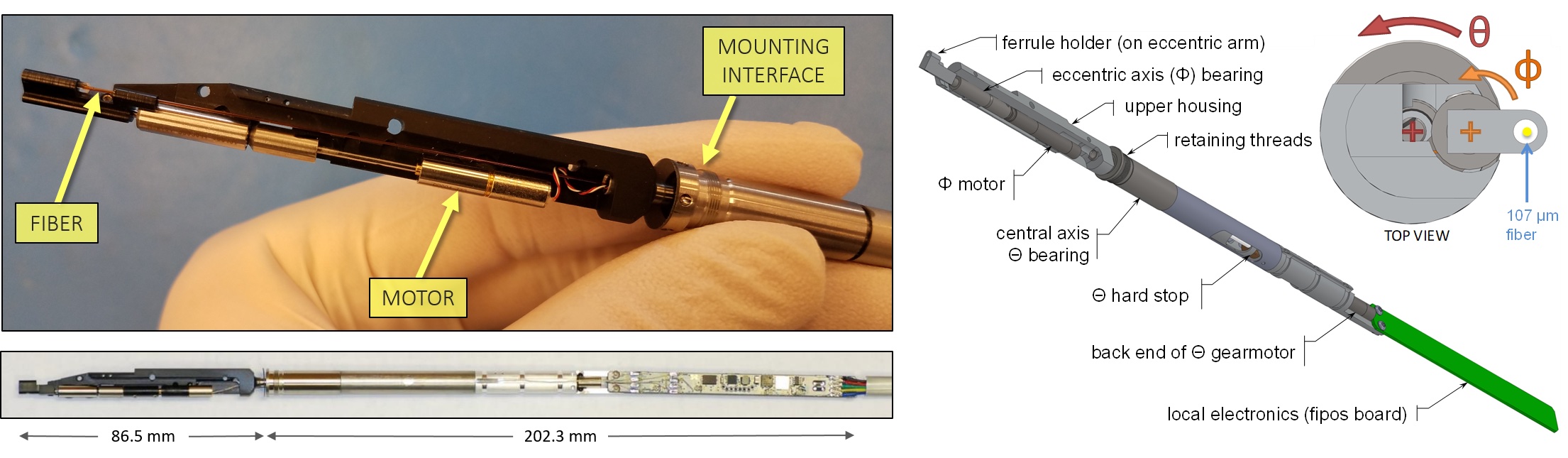}
    	\caption{\label{fig:positioner_design} The DESI fiber positioner robot is designed for a minimum 10.4\,mm pitch between neighboring units. It has 2 rotational axes, driven by independent \o\,4\,mm 337:1 gear motors. Drive electronics are integrated in a board mounted to the aft end. The assembly consists of 22 parts and 10 fasteners. The design was developed at LBNL and SSL, and mass-produced by UM and EPFL.}
    \end{figure}

\begin{figure} 
    	\includegraphics[width=\textwidth]{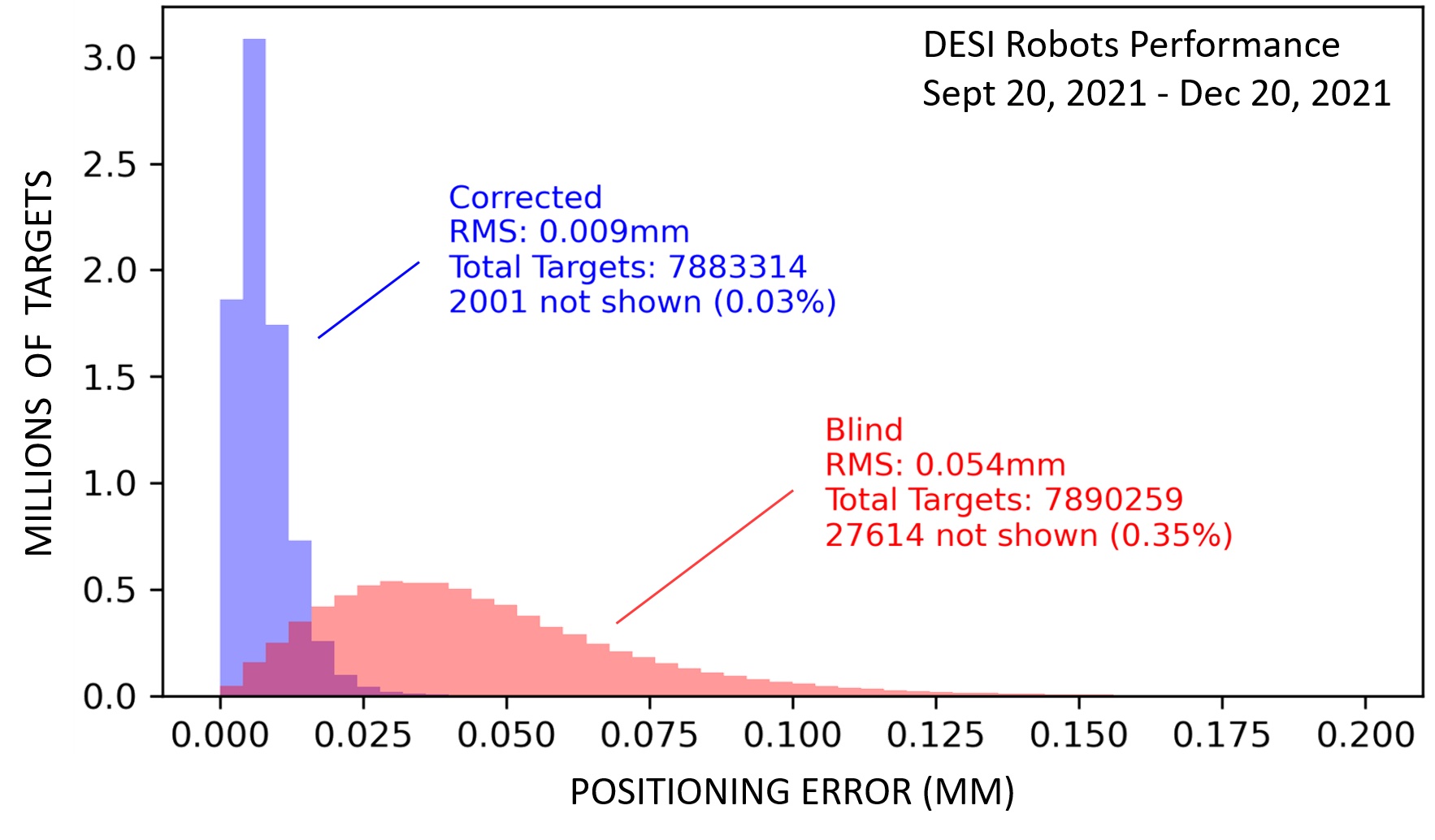}
    	\caption{\label{fig:positioner_accuracy_hist} Targeting accuracy for DESI fiber robots over a three-month period of survey operations in  autumn 2021, during which the Focal Plane System acquired more than 7.8 million targets. The shallower red histogram shows `blind' move accuracy, i.e. the first open-loop move made toward each target. The sharper blue histogram shows positioning error after a single correction move, based on feedback from the Fiber View Camera (FVC, \S \ref{sec:fvc}). Blind move RMS error is 54\,\micron, with 0.35\% of attempted targets having error $>$~200\,\micron. After the correction move, the RMS error is 9\,\micron, with 0.03\% outliers $>$~200\,\micron. Errors here are with respect to the commanded target positions, and include significant contributions of measurement noise (\S \ref{sec:fvc_performance}) due to centroiding precision limits ($\sim$\,3\,\micron) and air turbulence in the dome ($\sim$\,3\,--\,8\,\micron). Some turbulence subtraction code introduced in late 2021 further reduced the error to 6\,\micron\,RMS. \figref{fig:positioner_accuracy_with_turbcorr}}
    \end{figure}

    Each fiber positioner has two degrees of freedom, driven by two independent \o\,4\,mm DC brushless gearmotors \figref{fig:positioner_design}, and patrols a nominal \o\,12\,mm region with its fiber. The exact patrol diameter varies slightly from unit to unit, depending upon the as-built dimensions. Our gearmotors have an output ratio of $(46/14+1)^4 \approx$ 337:1 and no encoders. The robots are mounted in a nearly hexagonal close-packed pattern (\S \ref{sec:layout}). Minimum center-to-center pitch is 10.4\,mm. Their patrol regions overlap, preventing gaps in coverage. For each retargeting, anticollision software (\S\ref{sec:move_scheduling}) plans out a timed sequence of moves that each robot must perform to reach its target while avoiding hitting its neighbors.\footnote{No collision has yet been observed to harm the hardware, but such interference might prevent a fiber (and its neighbor) from getting to the planned destination.} Fiber targeting accuracy has been excellent, as shown in figures \ref{fig:positioner_accuracy_hist} and \ref{fig:positioner_accuracy_with_turbcorr}, and discussed in \S \ref{sec:positioner_accuracy}.
    
\begin{figure} 
    	\includegraphics[width=\textwidth]{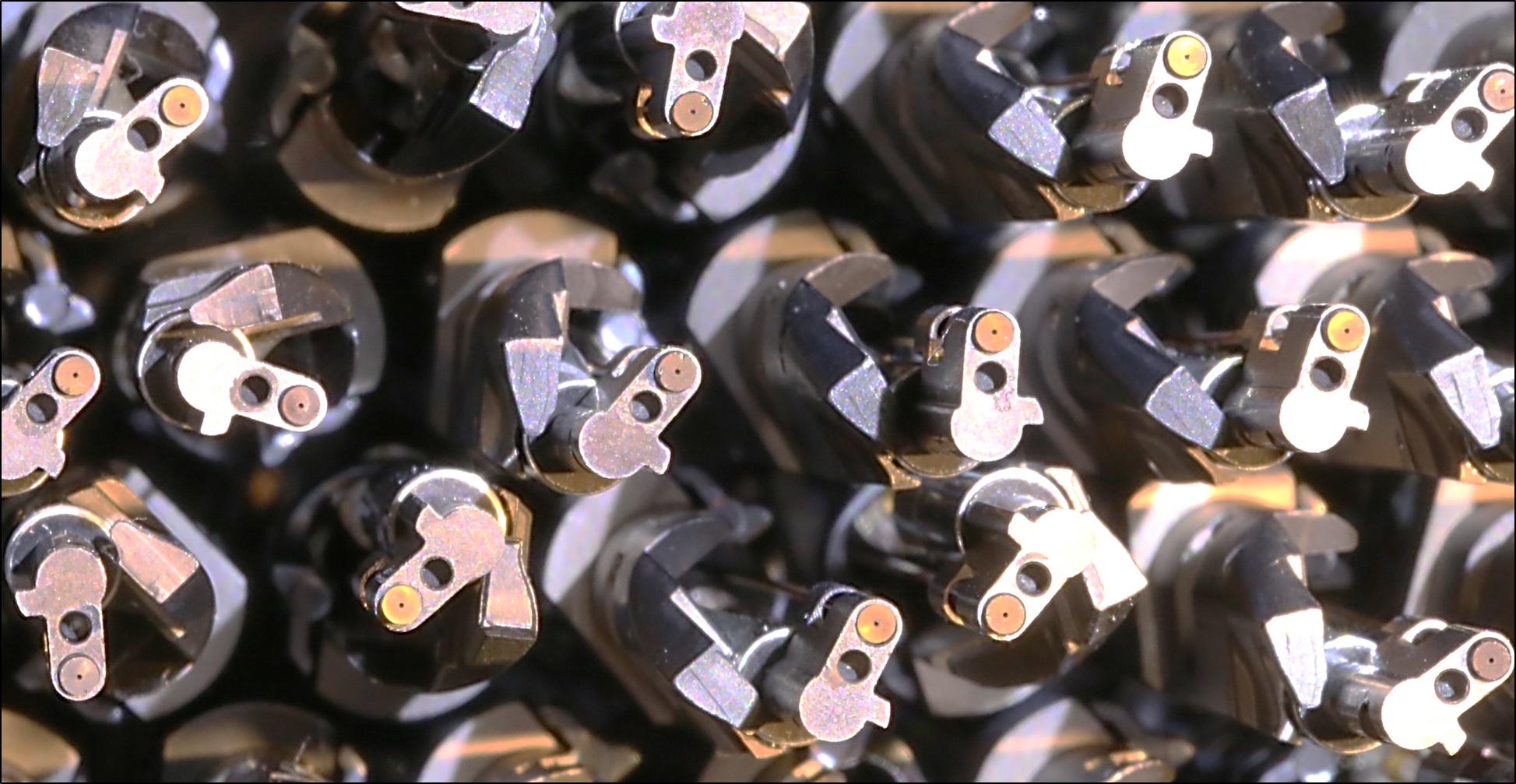}
    	\caption{\label{fig:robots_together} Fiber positioner robots are shown packed together on the focal plate. In addition to the fiber ferrule, each $\phi$ arm has a light-trap hole with a dark dye. The original concept for these light-traps was to be able to selectively extinguish bright stars in the field; in practice they have been most useful for identifying $\phi$ angles during hardware debugging. A projecting tab feature on each ferrule holder provides a simple but effective hard limit against clockwise over-extension.}
    \end{figure}

    Key to the alignment of the positioner in the focal plane is the central axis ($\theta$) bearing cartridge. The outer, cylindrical surface of the cartridge is machined to \o\,8.286\,--\,8.295\,mm, and contains two ball bearings at either end. The bearings maintain alignment of the hollow, rotating, inner shaft to within 0.005\,mm total run-out. When mounted in the matching precision holes of the focal plate (\o\,8.308\,--\,8.318\,mm), we are ensured of both maintaining alignment throughout the rotation range, and that the robot will not excessively intrude into the envelopes of its neighbors \figref{fig:robots_together}. The $\theta$ bearing cartridge has a precise front flange, with M8.7 x 0.35 threads just behind it. Screwing the positioner into the focal plate with these threads cinches the flange against shallow, machined spotfaces in the focal plate. This sets the focus alignment of the robot, such that the fiber tips will land on the focal surface, 86.5\,mm further towards the primary mirror.
    
    Glued to the shaft of the $\theta$ cartridge is an upper housing. The eccentric ($\phi$) motor and bearing cartridge are glued to this housing; thus they ride on the central axis rotation. The fiber and $\phi$ motor wires route together through the hollow shaft of the $\theta$ cartridge and hardstop mechanism, past the $\theta$ motor, and out the back end of the positioner. There, motor wires terminate in solder joints at the forward area of the electronics board, while fibers continue to a clip at the back. Throughout the positioner, the fiber is protected by a \o\,0.4\,mm polyimide sleeve. At the clip, we transition to a short length of \o\,0.9\,mm furcation tubing, made from a thermoplastic polyester elastomer material (often referred to by the trade name ``Hytrel''), spanning the gap between the robot and larger conduits to the rear.\footnote{We found the furcation tubing to have quite high friction against our polyimide-coated fibers. We initially had furcation tubing sleeves covering each fiber all the way from the robot to the splice box, a length of 1.7\,--\,2.5\,m, depending on location in the petal. A potentially catastrophic latent issue was averted during production, when we discovered (after having already built up two complete petal assemblies), that this length was sufficient to drive fibers forward, bowing them out of one positioner and into the adjacent envelope of its neighbor's rotating hardware. This effect could be driven either by relaxation of tension induced in the Hytrel during assembly, or by thermal contraction of the Hytrel in the presence of moderately cold temperatures. The high friction prevented the fiber from sliding back to its nominal state, leaving it permanently bowed outward. The issue was resolved by reducing the Hytrel length to 0.2\,m, and for good measure we added a fiber-retaining guard piece at the front end of each positioner. It was necessary to take out all the positioners from the first two petals to do this rework.}
    
    Both the $\theta$ and $\phi$ axes have hard mechanical limits in both directions of rotation. On the $\phi$ axis, the limits (nominally $-5\degree < \phi < +185\degree$) are imposed with machined features which contact the upper housing. On the $\theta$ axis, we designed a sliding idler mechanism, to achieve a range of $-195\degree < \theta < +195\degree$ (i.e. complete coverage of the circle). To transmit rotation from the solid $\theta$ motor shaft to its hollow bearing cartridge, while still allowing the fiber and $\phi$ wires to pass by, we designed a `lollipop'-shaped mechanical connection, with an arc slot covering 300\degree. Thus at the 195\degree~extremes of $\theta$ rotation, the fiber and wires are deflected no more than 45\degree~by the rotating hardware.
    
    We took care to ensure a gentle path for the fiber through the internal mechanisms of the robot. From early on in our prototyping process, we tested both near and far-field fiber performance, operating the robots over their full range of motion \citep{poppett14}. Direct prediction of fiber performance based on mechanical strain is a complex subject, and so for design purposes we set ourselves a minimum allowable fiber bend radius of 50\,mm. We carefully modeled 3D fiber paths through the complete robot mechanism, at the extrema of its travel range, and then measured the minimum radii on these 3D curves in CAD, to ensure margin. Wherever possible, we took care to chamfer or radius all sharp mechanical corners which might contact and unnecessarily constrain the fiber's motion. Ultimately this method (and similar attention to fiber path details throughout the system) paid off: all petals as-delivered met their throughput requirements, with $\ge$~99\% of fibers intact and $\ge$~90\% throughput from the prime focus corrector to the spectrograph \citep{poppett20}.
    
    To reduce stray light reflections off the positioners' leading surfaces, we specified a two-step black inorganic anodize, per ECSS-Q-70-03A, as the surface treatment for the eccentric arm and upper housing, which are made from 6061-T6 aluminum. We also included a small black acetal washer, riding on the $\phi$ bearing shaft, just below the eccentric arm. This washer reduces glints off the stainless steel bearing housing below it.
    
    The gearmotors naturally have $\sim 2\degree$ of internal backlash in the gears. We remove the backlash by always finishing each fast rotation with a slow rotation of 3\degree, always in the same direction. To maintain the stability of the fiber, each axis has a tab applying friction sideways to its shaft. The tabs have two layers: rubbing against the shaft is a 0.002" thick leaf of polyimide, which is backed up by a 0.002" leaf of beryllium copper.

\begin{figure} 
    	\includegraphics[width=\linewidth]{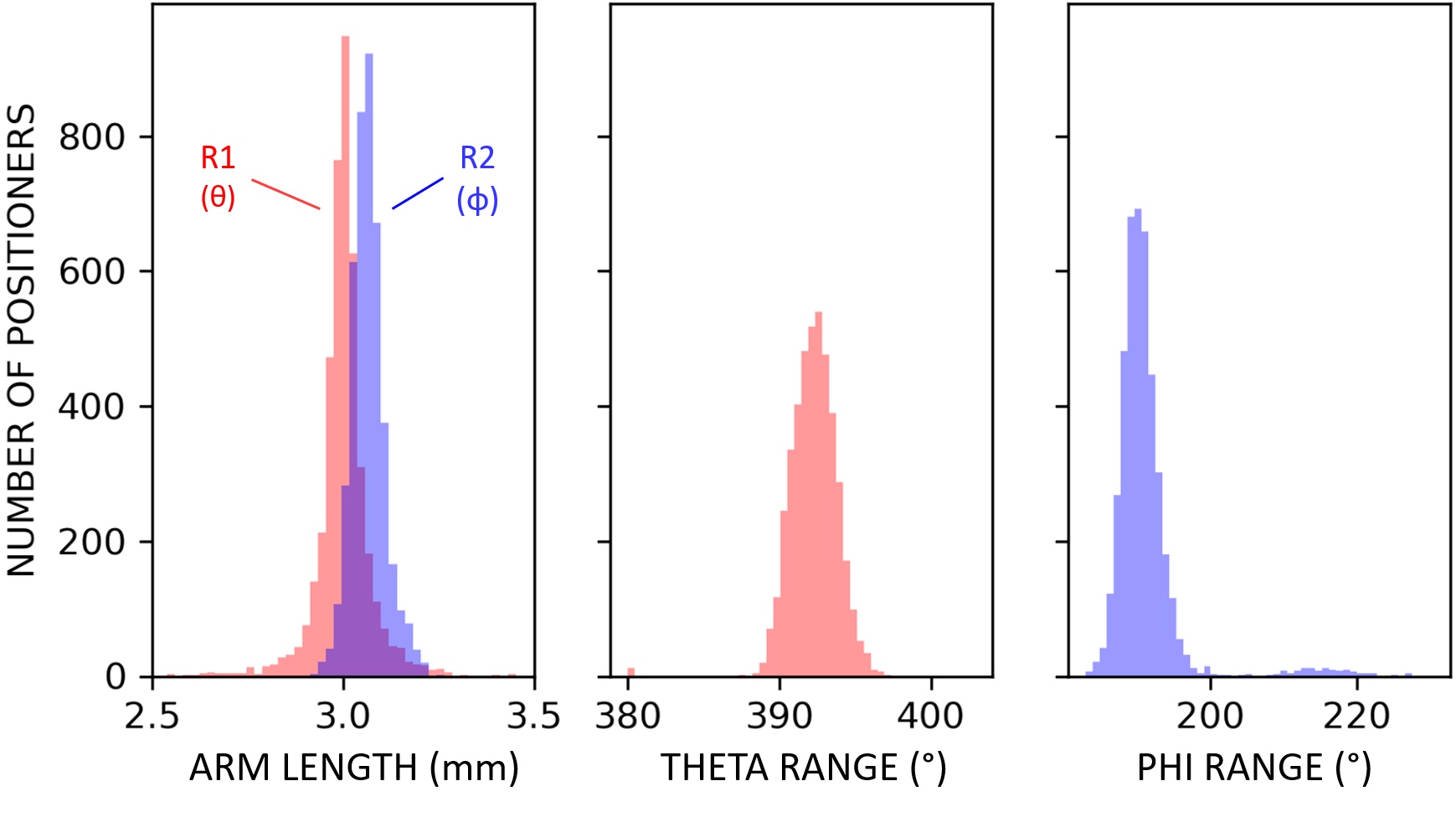}
    	\caption{\label{fig:positioner_calibrations_hist} Distributions for several key calibration parameters for the fiber positioner robots. At left are the kinematic arm lengths of the two robot axes. `R1' is the distance from the $\theta$ rotation axis to the $\phi$ rotation axis. `R2' is the distance from the $\phi$ axis to the fiber. The maximum radius to which a given robot unit can move its fiber is its particular value for $(R1 + R2)$. At the center of each unit there is an inaccessible zone of radius $|R1 - R2|$. The middle and right plots show the maximum travel ranges from hardstop-to-hardstop for the $\theta$ and $\phi$ axes, respectively. The $\phi$ range distribution is bimodal due to a period of several months during robot production where $\phi$ bearings were being glued at a spacing $\sim$\,0.5\,mm further than nominal off their housings, thus increasing the effective travel range. Other calibration parameters (not shown here) include center position of the $\theta$~rotation axis, angular zero points of $\theta$ and $\phi$, and effective gear output ratio for each axis (equal to 1.0 for nominally functioning robots).}
    \end{figure}

    Calibration of the complete array of robots requires more than 40,000 individual parameters.\footnote{Eight calibration parameters specify the geometric kinematics of an individual robot unit. We have two effective arm lengths, for the central and eccentric axes. A Cartesian coordinate pair specifies the location of the robot's central axis within the reference frame of its petal. Each of the two axes then has a calibrated angular zero point, and a total angular travel range from hardstop-to-hardstop. Beyond these eight geometric parameters, we also have numerous lower-level settings relating to the behavior and tuning of each gearmotor (e.g. speed, acceleration, duty cycle, as-soldered wiring polarity, true output ratio, etc), most of which are kept uniform across the robot array, but which we can (and do) adjust in special cases.} These can be measured on all robots in parallel, as installed in the focal plate, using the Fiber View Camera (\S \ref{sec:fvc}). The key calibration parameters are those used to transform between internally-tracked $(\theta, \phi)$ shaft angles and externally-measurable Cartesian $(x,y)$ coordinates. We determine these by moving the robots in parallel to a series of nominal commanded positions, using nominal transformation parameters, and measuring the resulting fiber positions at each point. We then calculate best-fit values for these parameters, minimizing error of fiber positions as predicted by the transform with respect to the measured actual values. Any set of points reasonably covering the patrol disk of the robot will work for this calibration method. In practice, the most efficient pattern is to measure independent circular arcs of points on the $\theta$ and $\phi$ axes. To ensure good arc fits, in particular on the range-limited $\phi$ axes, we typically measure $\sim$\,16 calibration points per arc, though in most cases half that number is sufficient. We usually also follow up the arcs with a rectilinear grid of 24 more calibration points, to ensure we are spatially covering the full patrol disk (not just two arc paths). Distributions of key positioner calibration parameters are shown in figure \ref{fig:positioner_calibrations_hist}.

\begin{figure} 
    	\includegraphics[width=\textwidth]{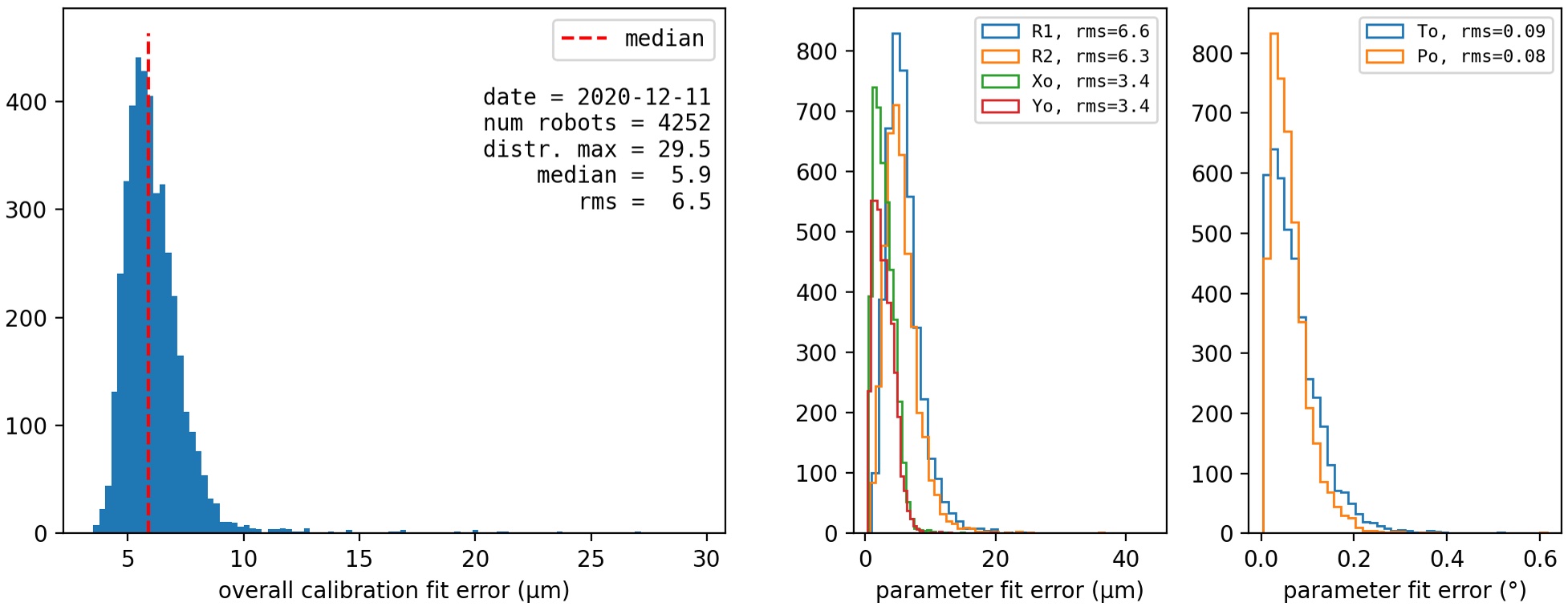}
    	\caption{\label{fig:calib_err_hist} Best-fit errors for a calibration measurement of 4,252 positioner robots, performed in December 2020. For each positioner, transformation parameters between $(\theta,\phi)$ and $(x,y)$ coordinates were calculated so as to minimize the error between measured and predicted fiber locations. The minimization in this case was done on a data set of 62 points per robot. The left plot shows the distribution of the transformation's overall fit error for each unit (6.5\,\micron\,RMS). The middle and right plots show errors on the six key kinematic parameters independently. `R1' and `R2' are respectively the kinematic arm lengths between central ($\theta$) and eccentric ($\phi$) axes, and between eccentric axis and fiber. `Xo' and `Yo' are the location of the central axis on the focal plane. `To' and `Po' are angular zero points for the $\theta$ and $\phi$ axes, respectively. The Xo and Yo fit errors are effectively a measure of FVC measurement precision.}
    \end{figure}

    Typical calibration error distributions are given in figure \ref{fig:calib_err_hist}. The overall fit error of the calibration (6.5\,\micron\,RMS) is a combined measure of the mechanical circularity of robot motion paths and of our measurement repeatability of the fiber tips. Comparing this both to the blind move accuracy ($\sim$\,50\,\micron\,RMS, see figures \ref{fig:positioner_accuracy_hist} and \ref{fig:positioner_accuracy_with_turbcorr}), and to the correction move accuracy (6\,--\,9\,\micron\,RMS)  one can see that the majority of initial blind move error comes from other factors, which may include: imperfect prediction of the plate scale as we slew, gravitational deflections at varying declination angles (estimated to be $\sim$\,5\,\micron\,RMS, based on measurements of a set of non-moving robots), and subtle non-linearities (at a length scale greater than the typical correction move, i.e. longer than $\sim$\,50\,\micron~$\approx$~0.5\,--\,1\degree) in the gears. For tests in the laboratory, with robots oriented in a fixed horizontal position and no corrector, robots typically had blind move errors of 20\,--\,30\,\micron\,RMS. Our only practical constraint on blind move accuracy is to be sufficiently predictable for anticollision path planning (see \S \ref{sec:move_scheduling}).

    Each positioner robot has its own microcontroller and motor drivers (see \S\ref{sec:electrical_system}). It is attached by two wires to a power bus, two wires to a Controller Area Network (CAN) communications bus \citep{CAN}, and by one wire to an additional common logic line (`SYNC'), independent from the CAN bus. In typical operation, the pre-calculated move sequences are uploaded via CAN to the microcontrollers, one unique move table going to each robot. Upon receiving a synchronized start signal, all the robots execute their particular sequences of motor rotations and intermediate pauses, each according to its own internal clock. The synchronized start signal can be either a level shift on the SYNC line, or alternately a broadcast CAN message sent simultaneously to all buses on the petal.
    
    From 2009\,--\,2016 we produced and tested 8 generations of 34 prototype units. In our early units (2009\,--\,2012), we used \o\,6\,mm DC brushless motors and ``R-$\theta$'' kinematics (i.e. a radial stage mounted upon a central rotation axis). These designs had 12\,mm center-to-center pitch and several variants of geared or flexured radial stage mechanisms \citep{silber12}. Such designs perform well and can naturally simplify anti-collision schemes while providing inherent anti-backlash preloads, however they tended to have high part counts (57\,--\,65 parts per unit) and assembly complexity. By 2013 several motor vendors were producing smaller \o\,4\,mm gearmotors, which enabled direct mounting of an eccentric $\phi$ motor and bearing to the rotating upper housing. At this point we switched to a simpler $\theta$-$\phi$ design, which despite its gear backlash and inconvenient paths of motion (both of which we recognized could be mitigated in software), cut the part count per unit in half, to 32. We were also able to reduce the minimum center-to-center pitch to 10.4\,mm, which relaxed some constraints on the optical corrector by allowing a small diameter focal surface. During the period 2013\,--\,2016 we built and tested 19 $\theta$-$\phi$ prototypes, progressively refining the design \citep{schubnell16}. Following these prototypes, from December 2016\,--\,March 2017 we fabricated a pre-production run of 440 units. Full production commenced in June 2017, and concluded by October 2018, with 7,148 robots produced (\S \ref{sec:pos_assy_test}).
    
    The design was developed at Lawrence Berkeley National Laboratory (LBNL) and Space Sciences Laboratory (SSL), and mass-produced by the University of Michigan (UM) and École Polytechnique Fédéral Lausanne (EPFL). To ensure the overall success of the project, alternative designs were developed in parallel and successfully tested by a Spanish-Swiss consortium of member institutions within the collaboration \citep{fahim2015}.

\subsection{Focal plate layout}
\label{sec:layout}
    The layout of the DESI focal plate is related to the pixel size and count of the spectrograph CCD sensors. Inside each spectrograph, fibers are arranged in a linear ``slithead'', and their individual spectra are dispersed in shallow, parallel arcs across the CCD area, mostly orthogonal to the line of fibers. We selected our CCD size early in project planning, and determined that 500 fibers would optimally cover the sensor. Considering construction cost, survey time, focal surface area, and feasible miniaturization of robots, we decided upon a total of 10 spectrographs, thus 5,000 total fibers.

    To match the architecture of the 10 DESI spectrographs, we segmented the focal plate into 10 wedges called `petals', each petal being 36\degree~wide. The fiber positioning robots are mounted to the petals such that their fiber tips patrol \o\,12\,mm disks tangent to the focal surface.
    
    The focal surface is aspheric rather than flat or spherical. This geometry relaxed our design constraints on DESI's large, six-element optical corrector \citep{doel14}. It came at the cost of increased complexity in designing, machining, and inspecting the physical focal plate (\S \ref{sec:fp_structure}, \S \ref{sec:final_petal_metrology}). With modern CAD, CNC, and metrology tools, we determined this to be an acceptable trade.
    
\begin{figure} 
    	\includegraphics[width=\textwidth]{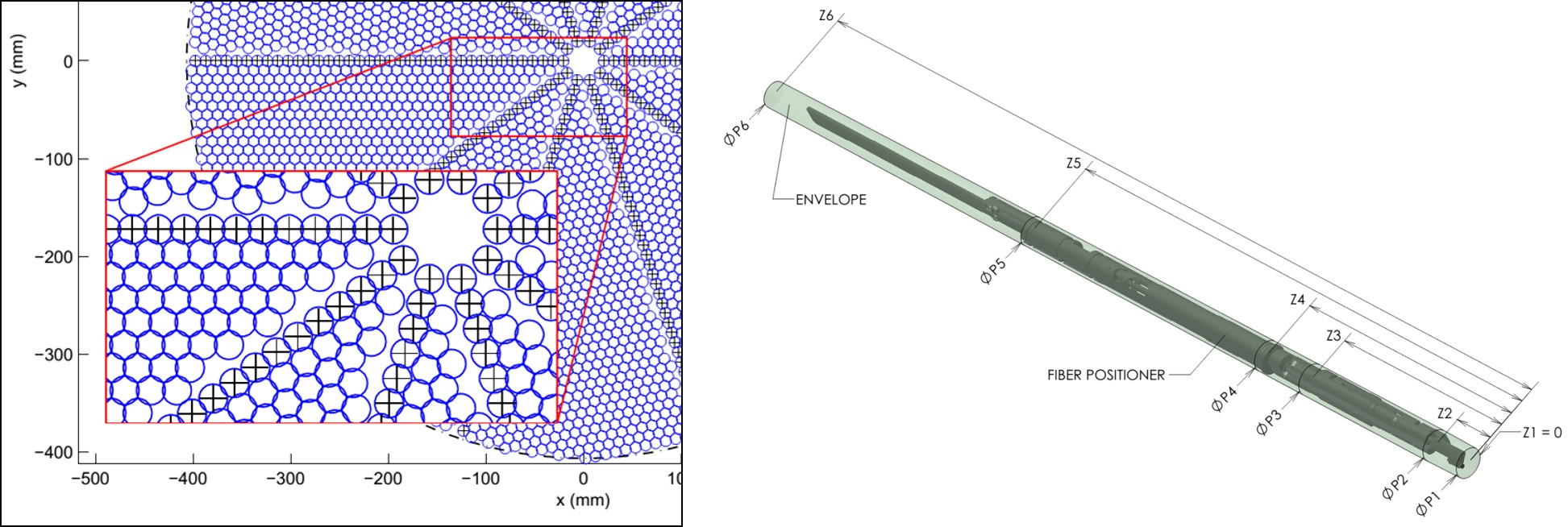}
    	\caption{\label{fig:close_packing} Layout of the focal plate. Left: The envelopes were iteratively squeezed together, with 3 key constraints: fiber tips stay on the asphere, central axis matches the chief ray, and the pattern along one edge of the 36\degree~petal is a straight line (see \S\ref{sec:layout}). Right: An axisymmetric tolerance envelope was defined by 5 conical sections.}
    \end{figure}

    Perfect hexagonal close-packing is not possible on such a curved surface. To generate the robot layout pattern, we developed an iterative code that maximizes the number of positioners on our \o\,812\,mm diameter aspheric focal surface. A key input to the code is an axisymmetric tolerance envelope, enclosing the maximum expected geometric deviations for the complete manufacturing run of robots \figref{fig:close_packing}. We produced this envelope by performing a detailed Monte Carlo analysis of component manufacturing tolerances, shaft and mounting hole alignments, and the spatial volumes swept out during repositioning.
    
    The layout code enforced three key constraints: fiber tips must remain on the asphere, the central axis of each positioner must align with the local chief ray, and there must be a straight line of positioners along one radial edge of each petal. The code first seeds the robot pattern on a wide-spaced, flat, hexagonal grid, and projects this pattern onto the asphere with a simple shift in Z coordinate. Then, respecting the constraints described above and the tolerance envelope for the positioner, the code iteratively nudges robot positions by small amounts, toward each other and toward the center of the focal surface. Iteration halts when the nudging process converges.
    
\begin{figure} 
    	\includegraphics[width=\textwidth]{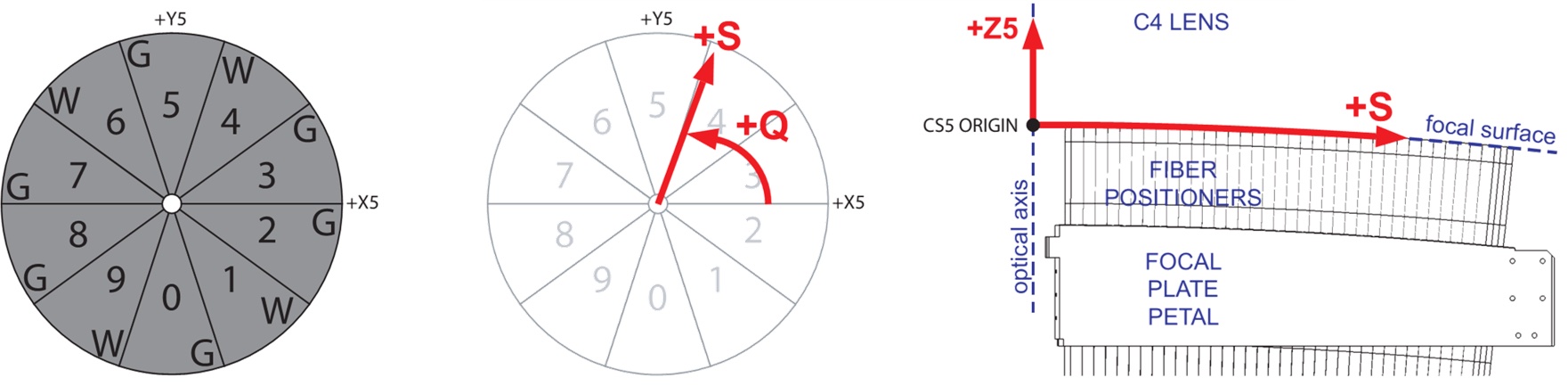}
    	\caption{\label{fig:gwcoord} Left: We selected the pattern of guider versus wavefront cameras (indicated by `G' and `W') to maximize spacing of the wavefront measurements, since they must control the tip/tilt degrees of freedom of the hexapod. Middle and right: At the interfaces between software systems---such as move scheduling (\S\ref{sec:move_scheduling}) and PlateMaker (\S\ref{sec:centroiding_spot_matching})---we use a modified polar coordinate system (Q, S), where `S' is distance along the aspheric focal surface within a plane that intersects the optical axis and is at angle `Q' with respect to the cartesian coordinates of the focal plane (X5, Y5, Z5). By matching the natural geometry of the focal surface, the coordinate system removed several possible sources of ambiguity between developers.}
    \end{figure}

    A further input to the code was an envelope for the GFA camera. The code optimizes the location of the camera to minimize loss of robots, while keeping the CCD within the \o\,812\,mm boundary. We ultimately placed a single guider or wavefront camera at a common, accessible location at the outer corner of each petal. Their distribution among the petals is shown in figure \ref{fig:gwcoord}.
    
    To produce a manufacturable 3D CAD model, the petal was first modeled with no robot mounting holes. We made a separate model of the positive knockout geometry for a single robot's mount hole. Then, driving the modeling program FreeCAD with a custom Python script, we patterned this knockout geometry according to a table of hole positions and angles from the layout code. The resulting positive hole pattern was boolean subtracted from the petal geometry.
    
    Finally, we visually inspected the pattern in CAD, and several more robot positions were manually removed, which were deemed by engineering judgment to have too little support material for practical mounting. These positions occur along one edge of the petal, where its 36\degree~wide envelope cannot perfectly align to natural $\sim$\,60\degree~pattern lines. We found the vacancies resulting from this mismatch to be useful locations to put accessory screw holes for handling fixtures and temperature sensors.
    
    On DESI, we ultimately packed 5,020 positioners at mean pitch of 10.525\,mm into the \o\,812\,mm aspheric focal surface. As a comparison guideline for future projects, one can view this as an overall 84\% packing efficiency, i.e. the ultimate number of science positioners $\sim 0.84 (d/p)^2$, where $d$ is the focal plate diameter and $p$ is the positioner-to-positioner pitch. This value includes all sources of spatial inefficiencies on the DESI focal plate: GFA camera envelopes, fiducial point sources, gaps between petals, and the central cap rings, which structurally connect the noses of all ten petals.

\begin{figure} 
    	\includegraphics[width=\textwidth]{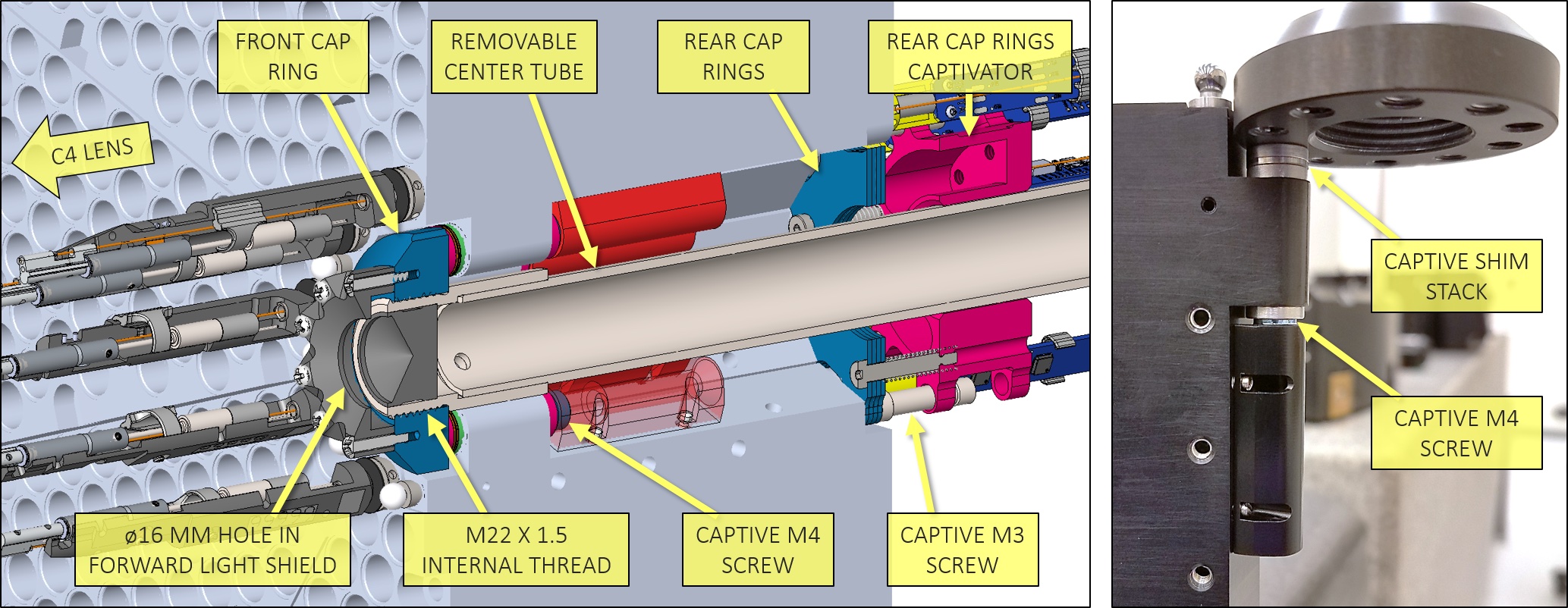}
    	\caption{\label{fig:nose_assembly} The ten petals are structurally connected at their noses by Front and Rear Cap Rings. Captive shims at the Front Cap Ring are selected to equalize the as-installed heights of the 10 petals. During assembly, long custom screw drivers are guided in from the rear of the focal plate along a removable central tube. Installation and removal of the Rear Cap Ring stack is facilitated by a special captivator that rides on this central tube. There is a \o\,16\,mm center hole, behind which lies an M22 x 1.5 mounting thread.}
    \end{figure}

    We provided the inner diameter of the central Front Cap Ring with an M22 x 1.5\,mm thread. The thread is used during assembly to attach a guide tube. At present, the hole is not instrumented, but could in principle hold a compact camera or integral field unit (IFU) at the center of the focal plane, so long as the unit fits through a \o\,16\,mm hole \figref{fig:nose_assembly}.

\subsection{Focal plate structure}
    \label{sec:fp_structure}
\begin{figure} 
    	\includegraphics[width=\textwidth]{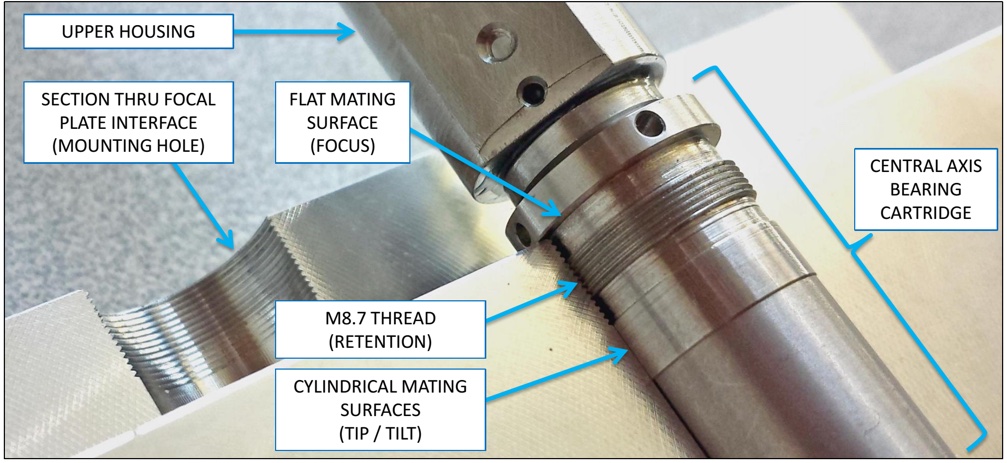}
    	\caption{\label{fig:sparkplug} The robots and the fiducials screw into the petal like a spark plug. This photo shows a prototype robot, placed in a cross-sectioned aluminum test block. The test block was fabricated by milling two faces flat, clamping them together, and then drilling, reaming, and tapping the custom threads.}
    \end{figure}

    The key structural component of each petal assembly is the petal itself. This is a 36\degree~aluminum wedge, 82\,mm thick at the outer radius and 103.8\,mm thick at the nose. Each petal weighs 8.6\,kg and supports a much larger mass than its self-weight in positioner robots and support electronics. The alloy is 7075, with a -T651 temper to strengthen the material while minimizing internal stress. The petal has 514 CNC-machined holes, each with a precision spotface and bore \figref{fig:sparkplug}. Ahead of the bore is a fine female thread. Positioners and fiducials screw into the bores like spark plugs, eliminating the need for additional small parts such as screws to hold each positioner in place. Contact of the device's flange against the spotface sets the focus position of the optical fiber installed in the robotic positioner. Fit of the bore to the device's outer diameter sets the position and angle of alignment.  The rotational alignment of each positioner is not controlled with any precision, but is measured precisely after assembly. The petals were machined by Boston University (BU). 
    
\begin{figure} 
    	\includegraphics[width=\textwidth]{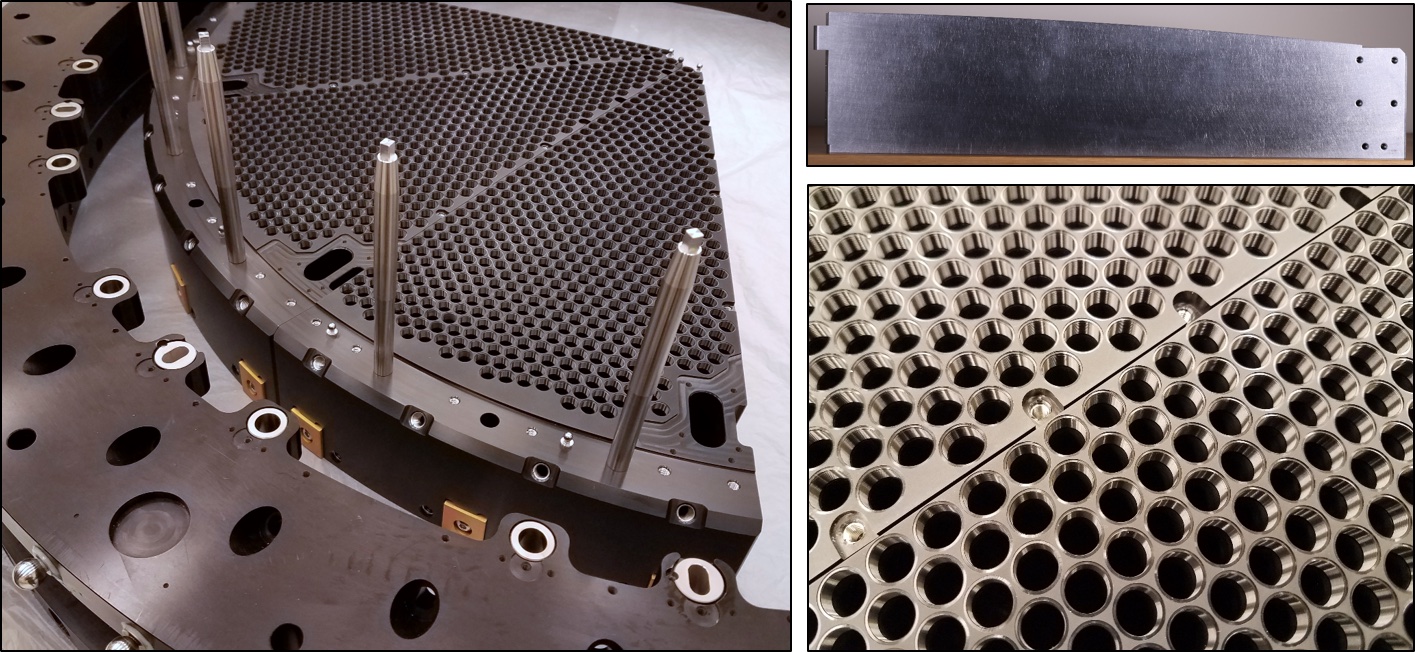}
    	\caption{\label{fig:structure1} The ten petals mount into a large integration ring (FPR), about $1/4$ of which can be seen in the photo at left. At upper right, a petal is shown from the side, illustrating the curved front surface. We made the back surface flat to facilitate machining, metrology, and mounting of hardware. At lower right, two petals are placed adjacent to each other at their separation distance (0.6\,mm) when installed in the ring. Individual robot/fiducial mounting holes can be seen, each with a precision spotface and M8.7 x 0.35 mounting thread. Also visible (in the upper of these two petals) are accessory mounting holes, which we located in the natural gaps which occur along the irregular edge of each 36\degree~petal.}
    \end{figure}

\begin{figure} 
    	\includegraphics[width=\textwidth]{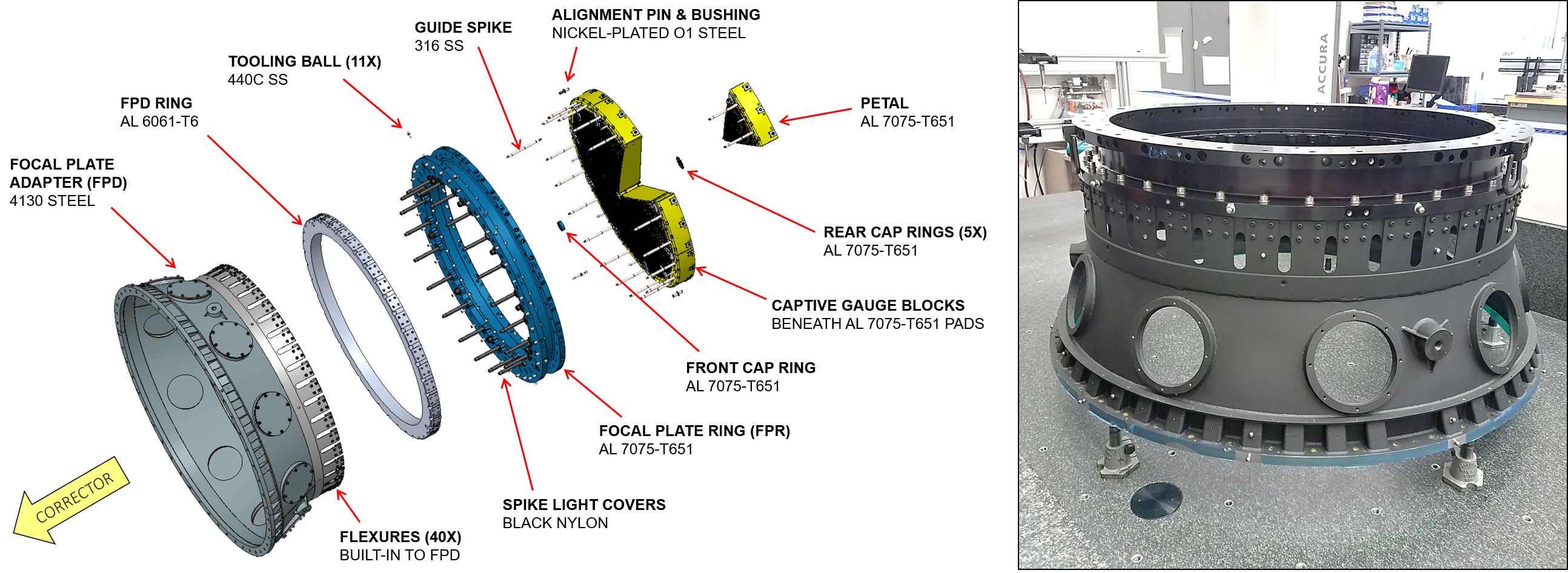}
    	\caption{\label{fig:fpa_assembly} Left: Key components of the focal plate structure include the steel FPD and aluminum FPD Ring, FPR, and petals. The FPD Ring is permanently bolted to the FPD via 40 flexural tabs, which accommodate the differential thermal expansion between the dissimilar metals of the Corrector Barrel and the Focal Plate. Right: FPD and FPR during alignment on a coordinate measuring machine.}
    \end{figure}

    The petals are retained in a large aluminum integration ring \figref{fig:structure1}, which ultimately connects to the steel DESI corrector barrel, via a match-drilled and flexured steel cone called the `Focal Plate Adapter' (FPD) \figref{fig:fpa_assembly}. To accommodate thermal expansion mismatch, the FPD has 40 flexural features around the diameter, 2\,mm thick x 58\,mm long, across the unsupported gap. Each flexure clamps against the outer diameter of the Focal Plane Ring (FPR) via 4 bolts. Fabrication and alignment of the FPD structure were done by Fermi National Accelerator Lab (FNAL) at the same time as they produced the Corrector Barrel \citep{gutierrez18}.
    
\begin{figure} 
    	\includegraphics[width=\textwidth]{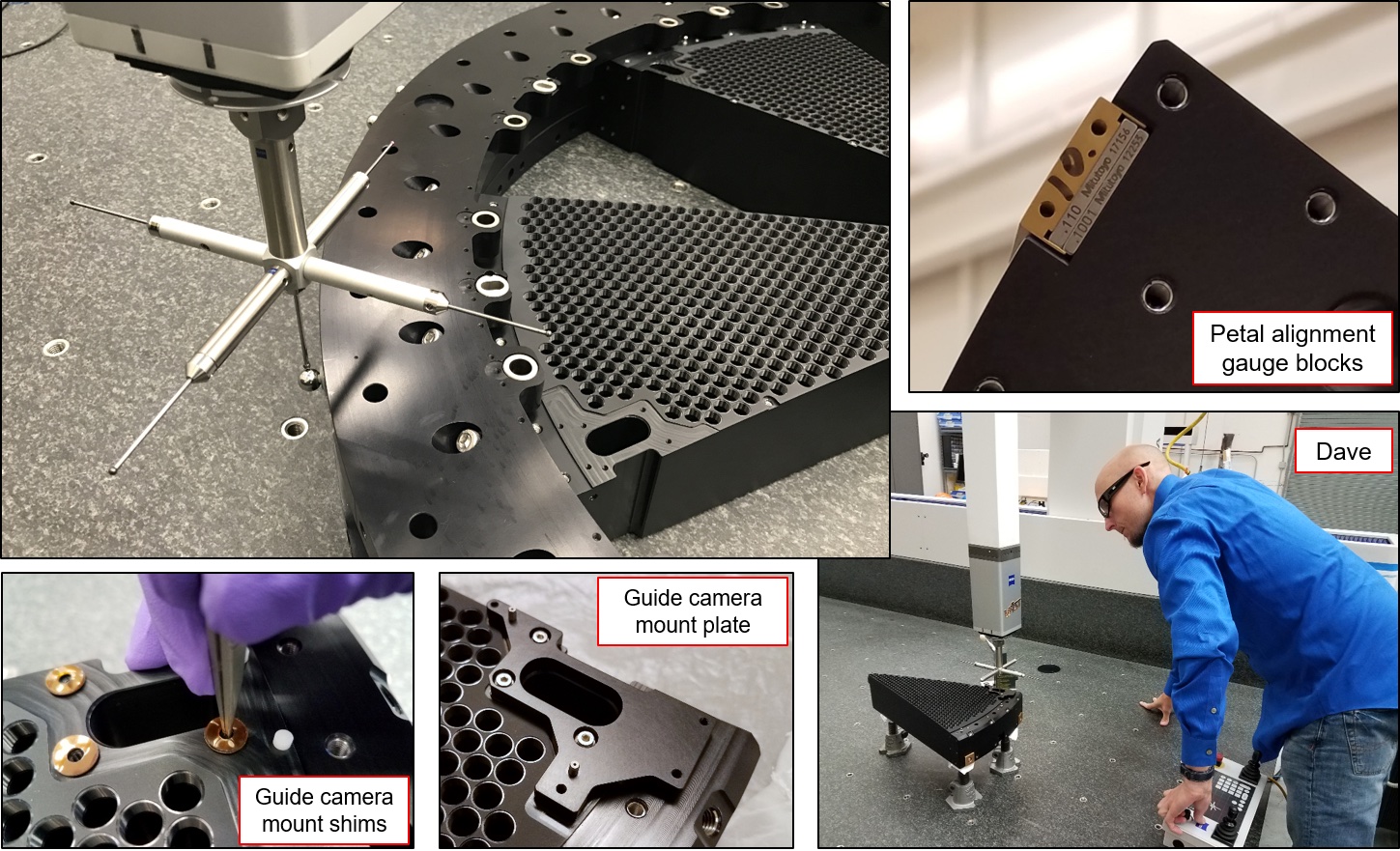}
    	\caption{\label{fig:structure2} Each as-built petal was robotically surveyed and aligned such that fiber tips, fiducials, and cameras, when installed, would land in focus on DESI's nominal aspheric focal surface. The upper left and lower right photos show the FPR and petals being surveyed in a coordinate measuring machine (CMM) at LBNL. The two photos at lower left show the GFA camera mount plate on a petal, which is adjusted by shimming at its three mount screws. At upper right, a stack of two gauge blocks and an aluminum retention/contact pad are seen at the corner of a petal. Each petal has two of these, which contact the FPR, and give us both stiff mounting and fine tilt adjustment. Not shown in these photos are two additional adjustment features of the petals: these are captive shim stacks at (1) the outer radius, to set focal position, and (2) at the inner radius \figref{fig:nose_assembly}, to equalize the petals with respect to each other at their tips. Over the set of all 12 petals (10 installed plus 2 spares), we achieved 12\,\micron\,RMS, 33\,\micron~max error in our alignment.}
    \end{figure}

    Subsequently, the FPD and FPR were shipped to LBNL. A system of permanent gauge blocks and shims was used for a dry-fit alignment of each as-built petal, installed in the ring \figref{fig:structure2}. This allowed us to achieve consistent focus positions of the 10 positioner arrays and the Guide/Focus/Alignment cameras with respect to one another.
    
\subsection{Guide, focus and alignment cameras}
\label{sec:GFA_hardware}
\begin{figure} 
        \frame{\includegraphics[width=0.999\textwidth]{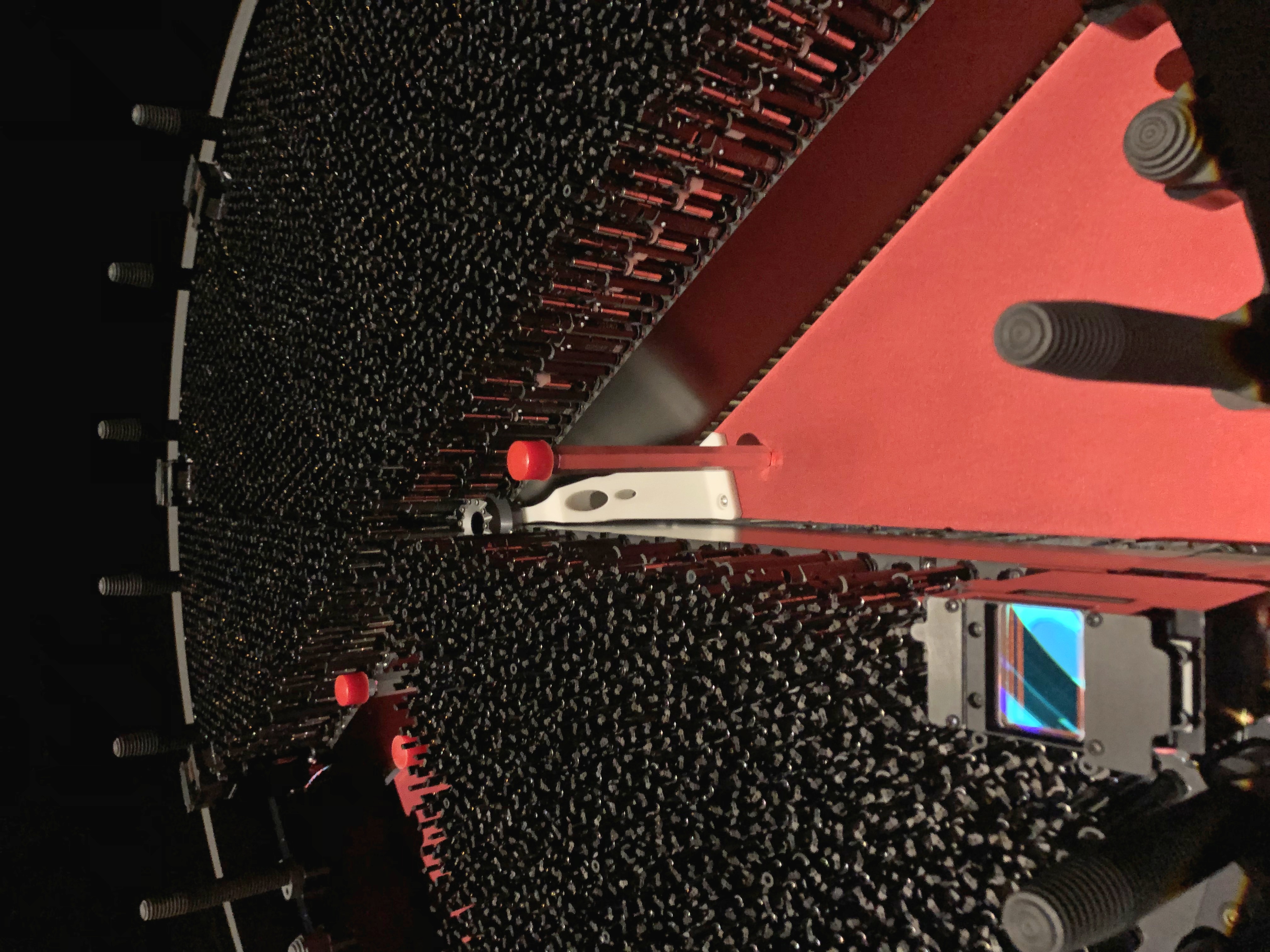}}
    	\caption{\label{fig:seven_petals} View of focal plane looking through a side porthole. Seven of ten petals had been installed at the time of this photograph. One guider GFA camera is visible in the foreground, while three more cameras can be seen in the far background. The textured posts are light blockers covering stainlesss steel guide pins, used during petal insertion. The front central cap ring is also visible, structurally connecting and co-aligning the 10 petals at their tips.}
    \end{figure}

    Ten custom Guide, Focus \& Alignment (GFA) cameras are mounted in the focal plane, one at the outer edge of each petal \figref{fig:seven_petals}. Six of the cameras are dedicated to the task of guiding the telescope by centroiding on astrometric calibration stars. They also provide real-time tracking of atmospheric transparency and seeing. The other 4 cameras are used as wavefront sensors, for focus and alignment of the corrector and focal plane on a 6 DOF hexapod (\cite{gutierrez18}; Miller et al. (2022) in prep.). The GFA cameras were built by a consortium of four Spanish institutions: Institut de Física d'Altes Energies (IFAE), Institute of Space Sciences (ICE-CSIC, IEEC), Centro de Investigaciones Energéticas, Medioambientales y Tecnológicas (CIEMAT), and Instituto de Física Teórica (IFT-UAM/CSIC).
    
\begin{figure} 
    	\includegraphics[width=\textwidth]{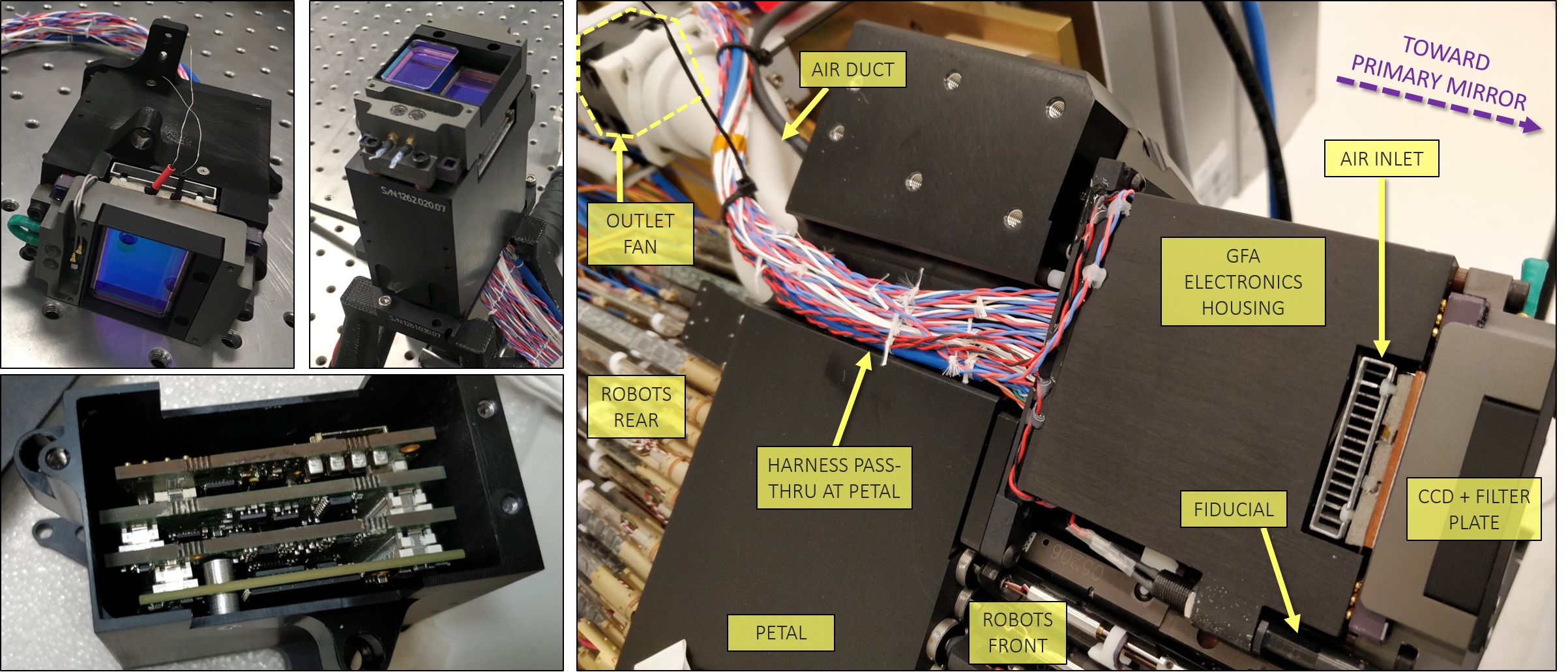}
    	\caption{\label{fig:gfa_front_and_rear} Guider and wavefront cameras (upper left) are identical other than their flat versus stepped filters. A compact electronics design (lower left) minimizes the footprint of the camera so that it consumes little more than the size of the CCD package itself in the focal plane. Cooling air and electrical harness pass through the petal (right).}
    \end{figure}

    The guider and wavefront cameras are identical in all respects except for their optical filters. The guiders have a flat filter, 5\,mm thick, while the wavefront cameras have a stepped filter instead, with one half of the active area covered by a 1.625\,mm thick filter, and the other half at 8.375\,mm \figref{fig:gfa_front_and_rear}. The filters have an index of refraction of 1.8, and the CCD is positioned to be in focus when the 5\,mm glass is bonded to the cameras.
    
    We glued the filters and CCDs into Ti 6Al-4V plates. Titanium was chosen for the good match of its coefficient of thermal expansion (CTE $\sim 8.6$\,ppm/K) to both the dense flint filter glass (N-SF6, $\sim$\,9.0\,ppm/K) and alumina ($\sim$\,6\,ppm/K) substrate on which the CCD is mounted. Below the filter plate, the GFA housing is made from 7075 aluminum, so as to match the CTE of the robot upper housings, thus keeping the sensor in the same surface of focus as the fiber tips despite any ambient temperature fluctuations. The $\sim$\,15\,ppm/K mismatch between titanium and aluminum causes an elastic strain, predominantly in the relatively thin-walled (1.6\,mm) aluminum housing. Because of the importance to targeting of knowing the relative positions of the GFA fiducials (GIFs; see \S\ref{sec:fiducials}) with respect to the CCD pixels, we took some care in the design of the bolted joint between the filter plate and housing, to ensure that the GIFs (mounted to the aluminum) would never permanently shift, even minutely, due to joint slippage under thermal extrema. To reduce stray light reflections, we anodized the titanium filter plates per AMS 2488 Type II.
    
    In the wavefront camera, the thinner and thicker filters shift the focus positions of stars to surfaces 1.5\,mm above and below the CCD, respectively. The resulting defocused `donut' shapes captured by the CCD are analyzed to determine focus and tip/tilt corrections for the hexapod.
    
    GFA optical filters transmit light between 570\,--\,717\,nm (designed to be 567\,--\,716\,nm). This red band was selected for being both in the middle of DESI's spectral range, while also excluding any light from the 470\,nm fiber backlight and fiducial point sources (\S\ref{sec:fiducials}). This latter feature allows operation of the GFAs simultaneously while the robot positions are being measured. These custom SDSS r\textquotesingle-band filters were produced by Asahi Spectra Co.
    
    An early study of guide star density indicated the need for at least 5 guider cameras, to ensure a sufficient number of guide stars per field. Under most weather conditions, at least 10 high-quality astrometric standards from the Gaia Data Release 2 \citep{gaia18} could be observed if we assume a magnitude range of $14 < R < 16$. In practice, we use guide stars in the range of $12 < R < 18$. We additionally need a minimum of 3 wavefront cameras, well-separated on the focal plate, in order to control the 3 DOF of defocus plus two directions of tilt. Given the 10 petal architecture, we therefore found it natural to put a GFA on each petal, and enjoy the redundancy of having a spare each of the guiders and wavefront cameras on the instrument.
    
    The GFA CCD sensors are back-illuminated devices from e2v (CCD230-42-1-143). There are 2048 x 2064 pixels, $15\,\micron$ square, covering an image area 30.7\,mm x 30.7\,mm. A significant selection criteria was the small overall package size of 42\,mm x 61\,mm. A key constraint on the mechanical design of the camera was to minimize any excursions beyond the package, thus minimizing loss of area in the overall focal layout of the focal plane. Further selection factors included dark current and noise versus readout speed.\footnote{During unit testing of CCDs, we found that an error in the data sheet indicated a much lower dark current than the devices actually can achieve at our operating temperatures. It was still insignificant for meeting our guiding requirements.} The requirement for readout time for the GFAs was established at 0.2 Hz, with a frame transfer time of 100\,ms. The read noise was expected to remain below 20\,$e^{-}$/pixel, with a goal of 20\,$e^{-}$/pixel when operated at the GFA’s nominal readout rate.
    
    The cameras have no physical shutters. The central half of the CCD is used for image capture, imaging a region of $3.34' \times 7.27'$ or 24.27 sq. arcmin when accounting for the meridional and saggital rays. The outer zones are masked for frame transfer and the CCDs are read out from four quadrants. Readout electronics are located physically close to the sensor, in the volume below the CCD package. The power supplies are in a separate box, called the GFA Controller Box (GXB), mounted behind each petal. Built into the body of each camera are mounting points for two illuminated fiducials. Positions of the fiducials were carefully measured with respect to the CCD sensor on each camera (see \S\ref{sec:GFA_integration}).
    
    The GFA electronic design is based on digital correlated double sampling. The electronics consist of 4 stacked boards, each with a specific task. At the top, an Enclustra Mercury ZX5 board contains the Xilinx Zynq-7000 SoC (``system on-a-chip'') that implements all logic and communication. The Zynq-7000 SoC contains 2 ARM Cortex A9 cores, which run a Linux image and a server to receive commands and send out the acquired images through TCP sockets and a 1\,Gb Ethernet link. It also has a programmable logic area in which we implemented the logic to control all the elements of the electronics as well as safety measures related to power to the CCD. Three custom boards were developed to complete the stack: one for the voltages and clocks, another for power, and one for the front end electronics that contains the 100\,Msps analog to digital converter (ADC) (TI ADS5263) and preamplifiers. 
    
    The GFA is cooled by air, flowing through the camera in the direction from the front to the rear of the focal plane. The pressure differential driving the flow has two sources: (1) positive pressure between the C4 lens and the FPA, produced by the single, large FPD fan (see \S\ref{sec:thermal_mgmt}) and (2) suction produced by 10 smaller GFA fans (one per camera) mounted via 3D printed ducts, 225\,mm to the rear. The suction fans are two-stage, counter-rotating models (SanAce 9CRA0312P4K03) to ensure sufficient pressure to push the air through the densely-packed readout electronics. The ducts (as well as many other lightweight airflow adapters, brackets, and tie points in the FPS) were printed by Selective Laser Sintering (SLS) in unfilled, white Nylon 12. We mounted a thermoelectric cooler to the underside of each CCD package, with cooling fins in the air stream, but in practice have not found it necessary to turn the coolers on.

\subsection{Fiber View Camera and Illuminated fiducials}
\label{sec:fvc}
\label{sec:fiducials}
    The Fiber View Camera (FVC) closes the control loop when repositioning the robots. The robot motors themselves are driven open loop. For each retargeting, the positioners are initially moved open-loop to their rough target positions, with typical accuracy of $\sim$\,50\,\micron\,RMS \figref{fig:positioner_accuracy_hist}. LEDs at the spectrographs\footnote{These are LED strips behind a blue filter and diffuser. The strips are mounted on the forward face of the spectrograph shutter, so that they shine directly into the linear array of fibers at the slithead.} are illuminated, backlighting the fibers. Additionally, an array of 123 fiducial devices with well-determined fixed positions on the focal surface are illuminated. In the FVC image, pixel centroids are measured for fibers and fiducials. A set of small, open-loop correction moves is then sent to the positioners, based on the FVC measurement. Because the FVC guides the robots to their final target positions, and because in practical terms the motors have infinitesimal resolution, the limiting factors on positioning accuracy are measurement precision of the FVC, turbulence in the dome between FVC and prime focus, and the accuracy of our optical model of the corrector and FVC lens when converting between camera pixels and focal surface locations.
    
    The FVC is mounted behind the primary mirror in the Cassegrain cage, in the central obscuration. The camera is fit with its own 25.4\,mm diameter lens with 600\,mm focal length, and looks through the optical corrector toward the focal plane 12.2\,m away \figref{fig:fvc_diagram}. The first 3\,m of this air gap are baffled by a \o\,0.7\,m tube with \o\,0.3\,m aperture at its end. This reduces the length over which air turbulence in the dome can affect the measurement. The FVC is mounted with a fixture that allows us to rotate the camera, thus allowing us to map out distortions of the lens.
    
    The Fiber View Camera is a FLI Microline 50100 using a Kodak KAP-50100 CCD camera with 6132 x 8176 pixels at 6\,\micron~pitch. Initially, a 600\,mm f/4 Canon telephoto Lens was used, as was tested during ProtoDESI \citep{fagrelius18a}. During commissioning of the focal plane, we found that the non-axisymmetric distortions in this lens were sufficiently complex to be difficult to take out in image processing. Therefore, we decided to replace the telephoto lens with a simpler singlet design. The newer lens system was built using Thorlabs lens tubes and a single Newport Plano-Convex BK7 600\,mm focal length lens.
    
\begin{figure} 
    	\includegraphics[width=\textwidth]{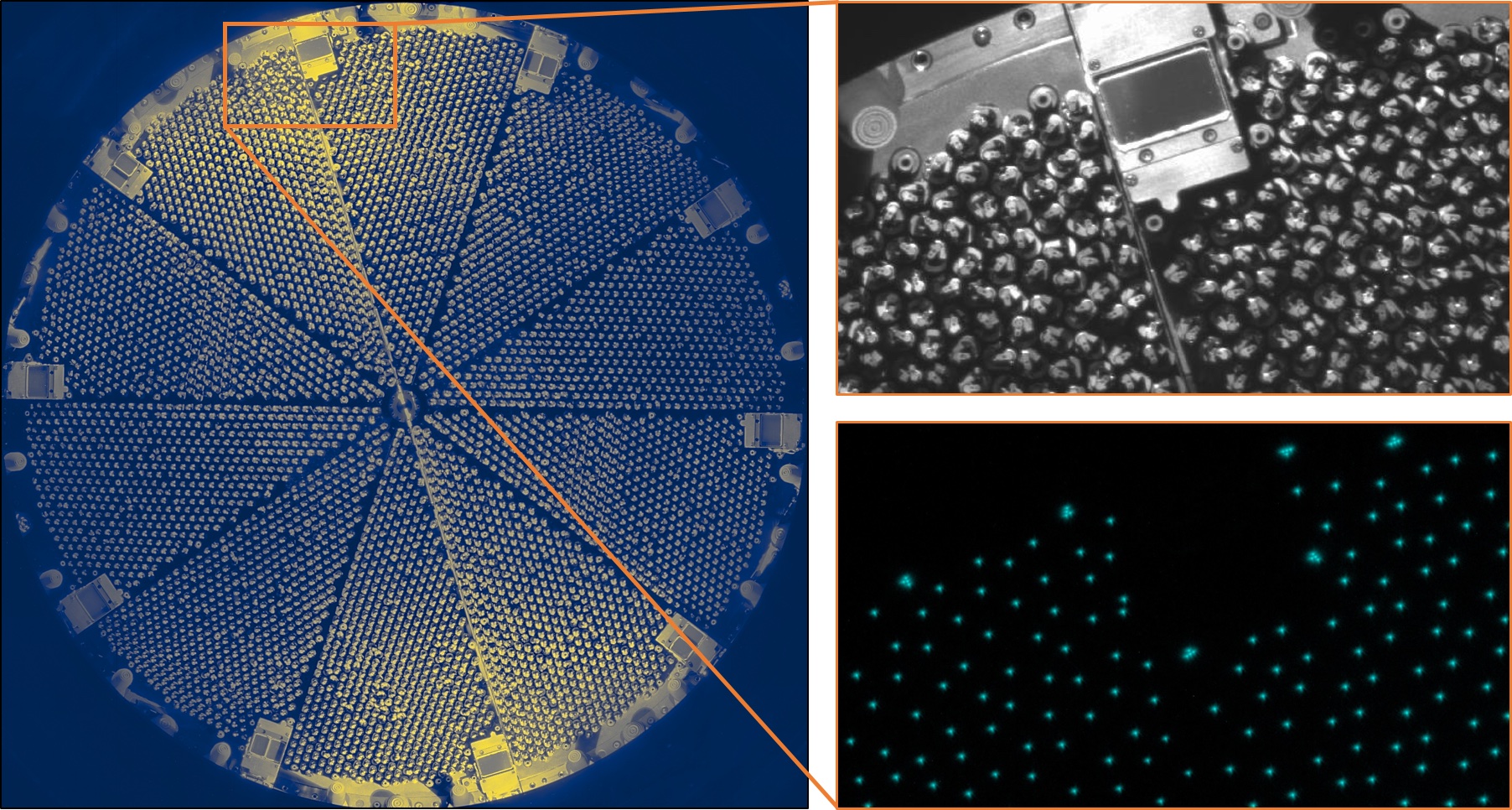}
    	\caption{\label{fig:fvc_brightdark} Images taken by the Fiber View Camera (FVC). At left and upper right are bright images of the focal plane, taken with side-lighting LEDs turned on. The lower right image shows the same region as in the upper right panel, except with the fiber tips and fiducials illuminated for centroid measurements of their positions. Note that there are four illuminated pinholes per fiducial.}
    \end{figure}

    We intentionally image the sources with a small aperture so that projected sizes of the sources are smaller than the Airy disk. This yields a uniform PSF across the FVC field, minimizing centroiding biases. Figure \ref{fig:fvc_brightdark} gives several views of the focal plane as taken by the FVC. The camera does have a shutter, which we have had to replace several times due to mechanical failures.
    
\begin{figure} 
    	\includegraphics[width=\textwidth]{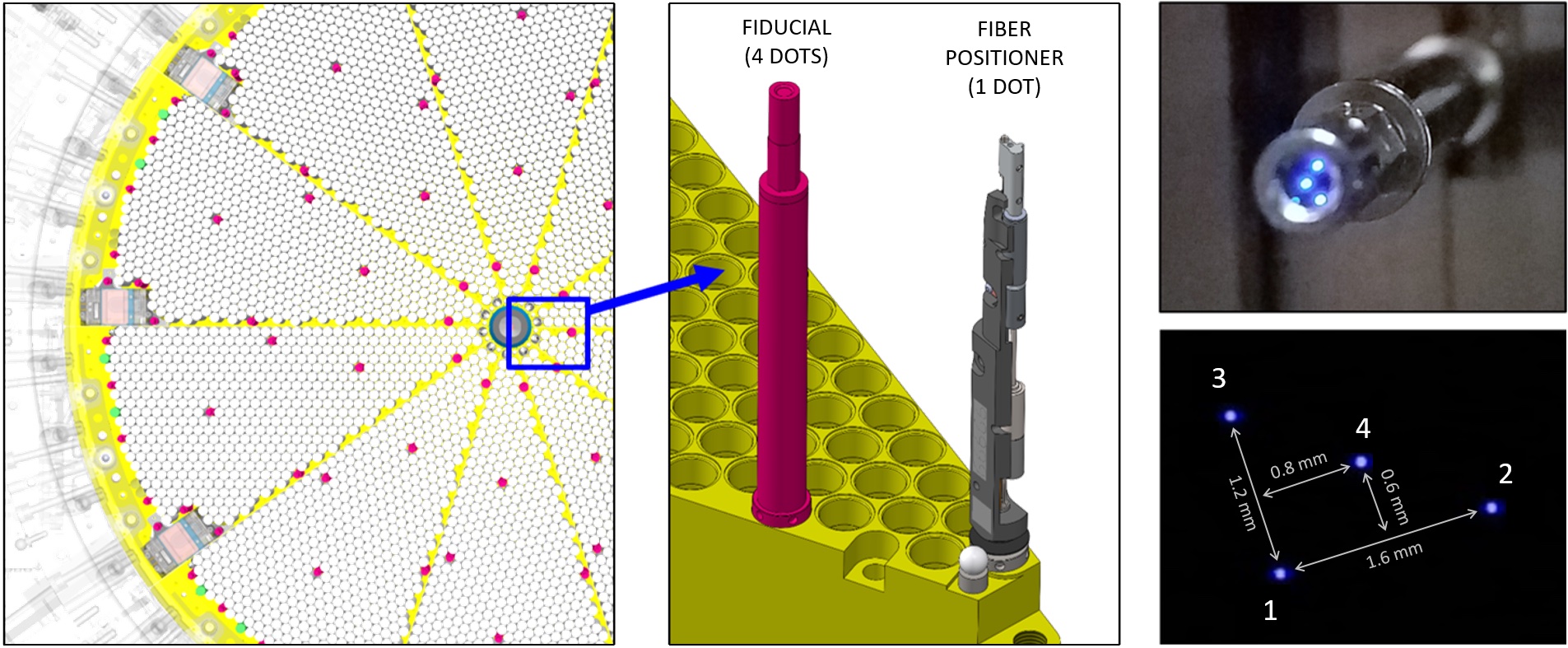}
    	\caption{\label{fig:fiducials} Fiducial point sources, embedded throughout the array, constrain optical plate scale and distortion polynomials. Each fiducial has four etched pinholes and is backlit by a 470\,nm LED. Field Illuminated Fiducials (FIFs) are mounted directly to the petal like the positioners. Two fiducials per petal are mounted to the GFA housing (GIFs; not pictured).}
    \end{figure}

    There are a total of 123 fiducials \figref{fig:fiducials} on the focal plane: 10 for each petal, except for one petal with 11 fiducials and another with 12. The extra fiducials break symmetry in the image seen by the FVC, to remove any ambiguities as to orientation. Each fiducial contains a glass disk, masked with chrome on one side and frosted on the other. On the forward-facing, masked side, four \o\,10\,\micron~pinholes are etched. The rear of the disk is illuminated by a 470\,nm LED, the same wavelength as the backlit fibers, mounted inside the fiducial housing. The housing mimics the length and material of the fiber positioners, such that the face of the glass lands in the same focal surface as the fiber tips. The fiducials mount to the petal via the same spark plug design as the robotic positioners. The locations of the pinholes were measured relative to the outer diameter of the fiducials with 1.5\,\micron~precision using a CMM (Coordinate Measuring Machine) prior to shipment to LBL for installation. 
    
    Two fiducials on each petal were reserved for integration with the GFA cameras and are referred to as GFA Illuminated Fiducials (GIFs), while the remaining are referred to as Field Illuminated Fiducials (FIFs). The two types of fiducials are identical at the front end, containing the LED and masked glass disk. Toward the rear, however, the FIFs have a mechanical body that mimics the fiber positioner's interface to petal and FIPOS boards, while the GIFs screw directly into the GFA housing. FIPOS boards are mounted at the rear of the petal to drive the GIF backlights. The FVC and fiducials were all delivered to the project by Yale University \citep{baltay19}.
    
\subsection{Thermal management}
    \label{sec:thermal_mgmt}

\begin{figure} 
    	\includegraphics[width=\textwidth]{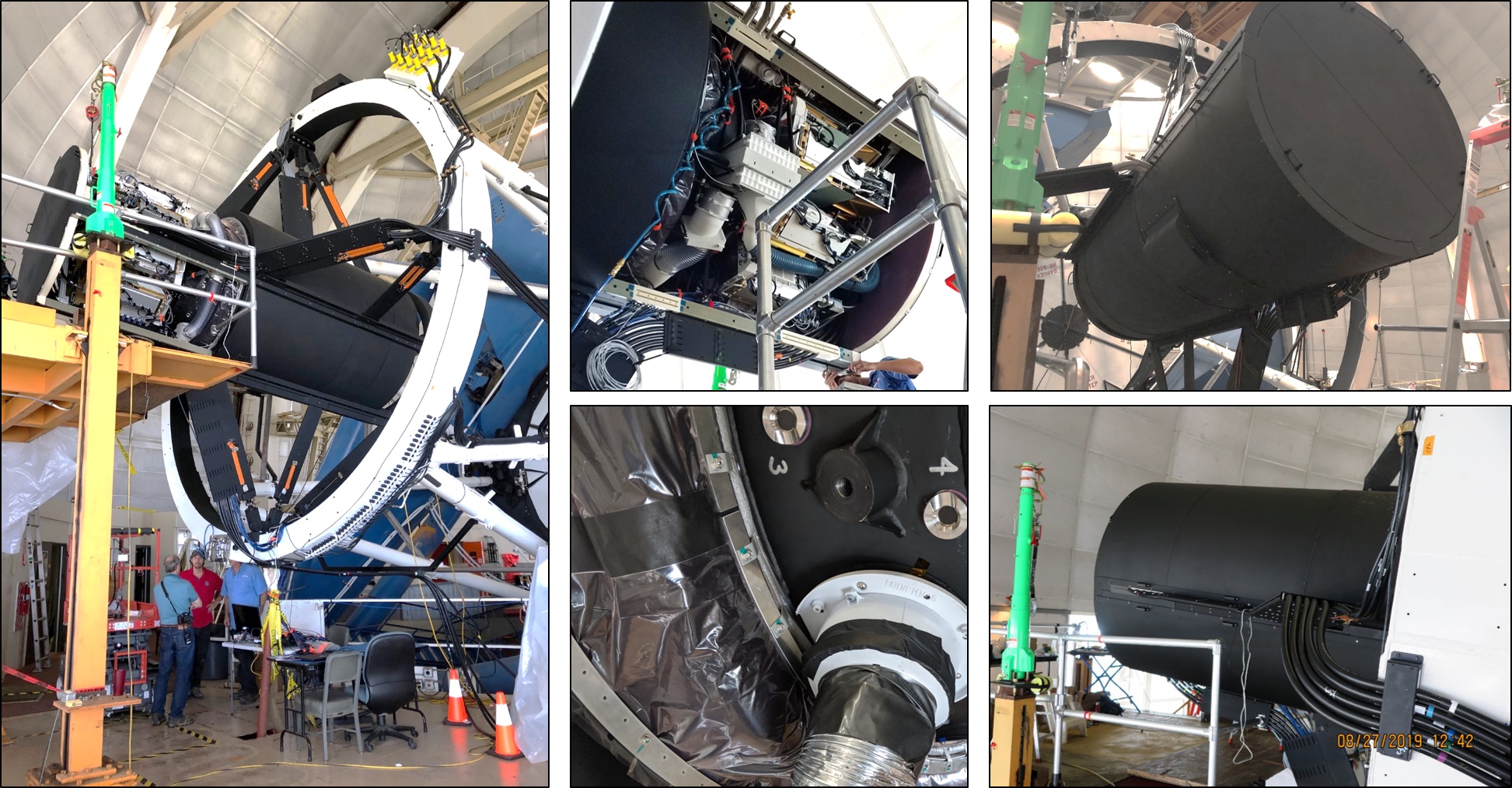}
    	\caption{\label{fig:thermal_system_photos} The thermal enclosure, shown with side panels off (left and center) and on (right). The heat exchanger can be seen in the upper center photo, surrounded by white MERV filter holders. The duct just below the heat exchanger wraps around the focal plane, injecting diffuse air flows at several ports into the volume between the C4 lens and the robots. At lower center, the flexible air skirt is visible, adjacent to an air duct feeding into the FPD (the support structure that surrounds the volume between the C4 lens and the robot array).}
    \end{figure}

    The focal plate assembly is surrounded by an insulating enclosure, made of 3\,inch thick, rigid polyurethane foam, with fiberglass facings \figref{fig:thermal_system_photos}. The enclosure serves two purposes:
    
    \begin{enumerate}
        \item Thermally isolate the focal plane from ambient conditions, to prevent air turbulence in the dome (which can spoil astronomical seeing).
        \item Provide a controlled temperature and humidity environment for focal plane components.
    \end{enumerate}

    When spinning thousands of motors simultaneously, the instantaneous power dissipated by the FPA can be 5\,--\,10\,kW. This occurs over a brief period of time, generally less than $\sim$\,10 seconds. For a typical observation duty cycle of 1000 seconds, the average dissipated power is $\sim$\,1.2\,kW.

    Early on during planning of the system, we recognized dome turbulence as a significant risk to science performance. We set ourselves a requirement to keep exterior surfaces of the enclosure to within $\pm$ 1\degree C of ambient. This value came primarily from operational experience at the Mayall Telescope.\footnote{We additionally noted that for a \o\,1.8\,m cylinder in air at $\sim$\,300\,K and Kitt Peak altitude, a surface temperature 1\degree C above ambient indicates a Rayleigh number $Ra \sim 10^8$. Typically $Ra \sim 10^9$ is considered to be onset of turbulence.} We selected the thickness and material of our enclosure wall to insulate the internal loads at this level.
    
    The corrector and focal plate assembly move on a hexapod with respect to the enclosure, up to $\sim 10$\,mm in any direction. To accommodate this motion, the annular gap between the focal plate and the thermal enclosure is closed out by a flexible air skirt, rather than a rigid ring. This skirt is made of 4 layers of 0.26\,inch thick fabric-covered neoprene, with a final vapor barrier layer on top, of 0.002\,inch thick Airtech Stretchlon 800 Bagging Material.

\begin{figure} 
    	\includegraphics[width=\textwidth]{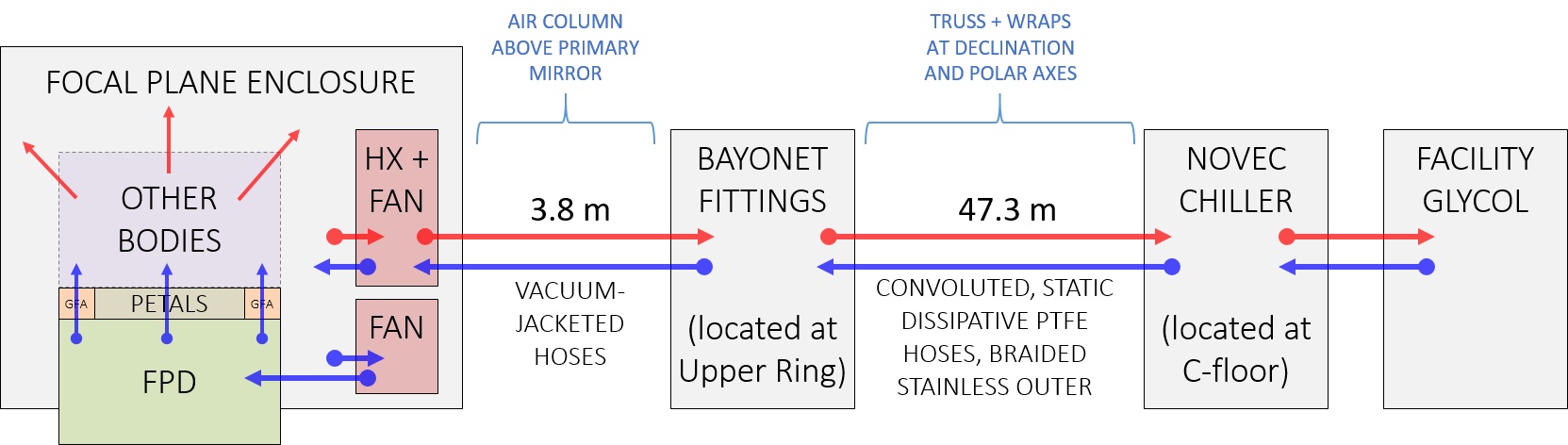}
    	\caption{\label{fig:thermal_system_diagram} Cooling system for the focal plane. Blue and red arrows indicate flows of colder and warmer fluids, respectively. Within the Focal Plane Enclosure (FPE) there are two fans. One blows air through a heat exchanger (HX), thus transferring energy from the FPE's interior air to the liquid colant. This fan is oriented to blow air in an annular route around the outside of FPD and petals. A second fan blows air via four ducts into the FPD, where it travels axially through the GFAs and positioners, exhausting past the electronics, fibers, and routing structures (located in the `OTHER BODIES' zone of the diagram). A fraction of the cooled output air from the heat exchanger is snorkeled into the upper zone of the FPE to ensure good thermal mixing.}
    \end{figure}

    A chiller unit (Thermonics LC-5) located on the telescope control floor (``C-floor'') delivers 3M Novec 7100 coolant to the focal plane via 51\,m of hose \figref{fig:thermal_system_diagram}. The chiller is rated to deliver coolant within a range of -40\degree C to +50\degree C, at a flow rate of up to 4 gpm at 50 psi, and can remove $> 3000$\,W at -15 \degree C. The unit is itself cooled by a Mayall facility loop of 50/50 ethylene glycol/water (EGW). DESI purchased two chiller units, both of which are kept operable on the C-floor. We can readily plumb in the spare as-needed, and have done so several times for routine maintenance and software upgrades. The chiller is powered by a 480\,VAC uninterruptible power supply (UPS).
    
    Novec 7100, like other hydrofluoroether (HFE) fluids, is a relatively expensive coolant. It has a very low viscosity, increasing its ability to leak through poor fittings, and may affect soft, rubbery seal materials. However, we found these to be relatively minor issues in comparison to several key benefits. It is non-flammable, non-conductive, low toxicity, has a low Global Warming Potential, and zero ozone depletion potential. It is commonly used as a cleaning agent and has a low flash point. Should any spill occur, there is no risk of either staining the optics or shorting the electronics, and no clean-up is required. We do, however, have to regularly top off our chillers with more Novec ($\sim 1.5$ liters every two weeks), due to its ease of evaporation. One potential issue we considered is that as a low-viscosity solvent, it has the ability to remove oil from the gearboxes of motors.\footnote{We tested this by submerging a positioner in the fluid. It operated well for some time until binding up. When we submerged a positioner in 50/50 EGW, its electronics instantly died.} However, our robots are well-separated from the coolant system by barriers of metal, and so we saw little risk of this occurring in practice.

    Once inside the enclosure, the coolant passes through a tube-fin heat exchanger (Lytron 4310G10). Air is forced through the heat exchanger (HXA) by a 172\,mm wide 48\,VDC counter-rotating fan (San-Ace 9CR5748P9G001). The fan is rated for flow rates up to 636\,CFM and static pressure of 5.62\,in\,H$_2$O. To eliminate dust in the system, we drive the air through two MERV-13 filters, positioned immediately ahead of and behind the heat exchanger. At the inlet we additionally have a charcoal filter to pick up volatile organic compounds. For ease of replacement, we made special holding cartridges for the filters, and located them near a small access panel in the enclosure. In practice, at duty cycles of 30 to 50\%, the HXA fan drives 306 to 365\,CFM through the heat exchanger and filters, drawing 135 to 251\,W of electrical power. Testing with and without the filters in place, we found they reduce flow rate by only 2 to 5\%.
    
    A separate fan of the same make, referred to as the FPD fan, takes air from the perimeter zone inside the enclosure and drives it through 4 large portholes into the volume between the C4 lens and the robot array. A MERV-13 filter is mounted immediately after the FPD fan. After passing through the filter, the air routes around the outer perimeter of the FPD structure via a \o\,150\,mm aluminum duct. The portholes are also \o\,150\,mm and each have a filter and diffuser screen at their outlets. Any time the enclosure is opened for maintenance, after closing it we run the HXA fan for $\sim$\,1/2 hour prior to operating the FPD fan. This cleans up any dust in the air prior to driving it into the volume where the optical surfaces reside.
    
    From the C4-FPD volume, the air takes three paths. Approximately 1/3 of the flow goes through the GFA cameras, boosted by the suction fans mounted to each. The remainder of the flow goes either through the robots (which have hollow central shafts) and past their electronic boards behind the petal, or else through the 0.6\,mm interstitial gaps between petals.

\begin{figure} 
    	\includegraphics[width=\textwidth]{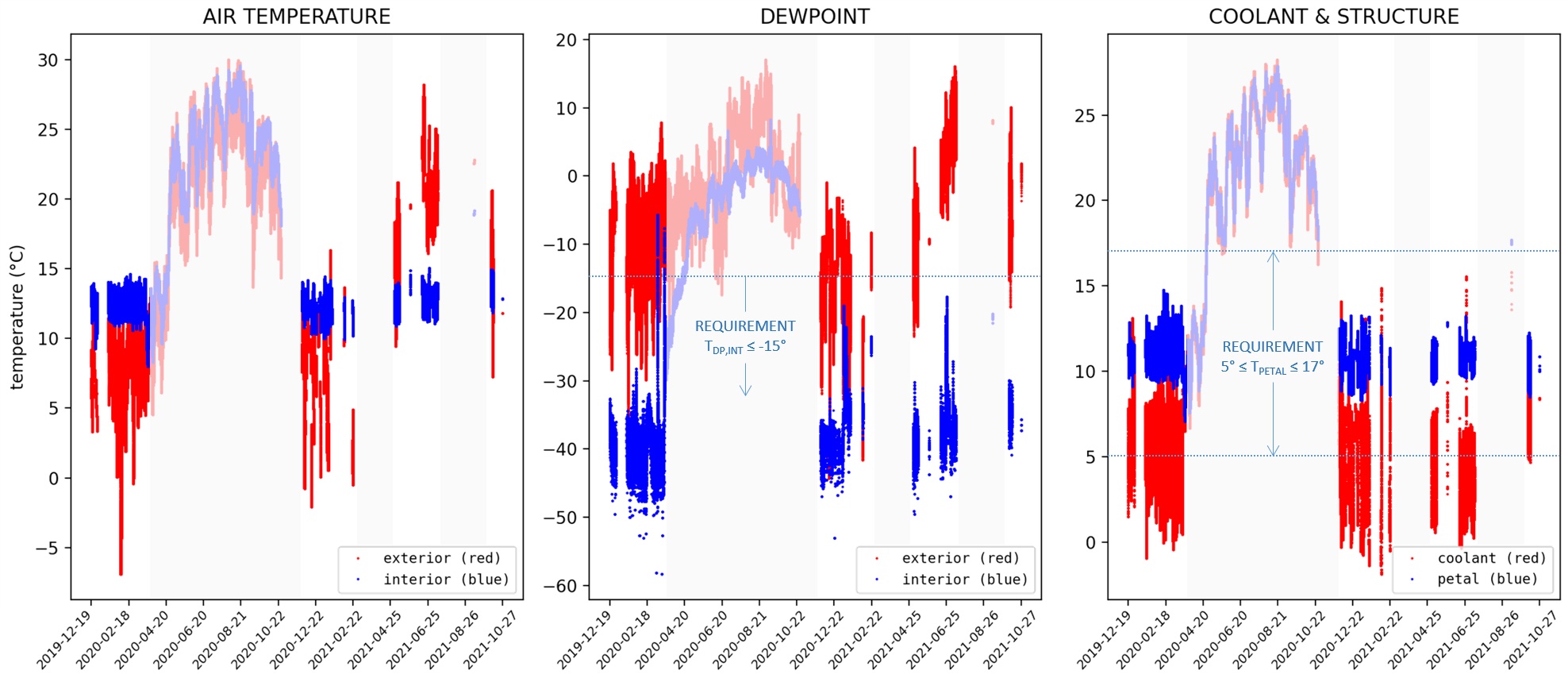}
    	\caption{\label{fig:thermal_performance} Focal plane environmental measurements over the initial 22 months of operations, including both commissioning and regular survey ops, which started in May, 2021. Grayed zones are seasonal,  maintenance, and Covid-19 shutdown times, during which temperature and humidity are allowed to track exterior ambient conditions. Left: While exterior air temperature varies from -7 to +30\degree C, air inside the enclosure stays within $\sim$\,+10 to +15\degree C. Middle: While ambient dew point may drift as high as +17\degree C (much higher than the hardware), dew point inside is suppressed to $\sim$\,-50 to -20\degree C, preventing condensation. Right: The system controls temperature of the FPR (the large aluminum ring to which petals are mounted) to a set point of 11\degree C. Coolant is typically about 5 to 8\degree~lower, removing heat load of both the focal plane equipment itself, as well as parasitic heat absorbed along the 50\,m hose run from the chiller to the enclosure.}
    \end{figure}

    Clean, dry air is delivered to the enclosure at a rate of 5\,--\,10\,CFM, and is provided by a system installed in the Mayall facility. Humidity inside the thermal enclosure is not actively controlled. Rather, the enclosure is sufficiently well-sealed such that the dry air input drives dew point well below our operating threshold of -15\degree C. Readout of temperature and humidity sensors, control of fans, and temperature and flow set points of the liquid chiller are controlled by the FXC computer (see \S\ref{sec:electrical_system} and \S\ref{sec:environmental_control}). Environmental conditions and performance of the system are summarized in figure \ref{fig:thermal_performance}. 
    
\subsection{Electrical system}
    \label{sec:electrical_system}
     
\begin{figure} 
    	\includegraphics[width=\textwidth]{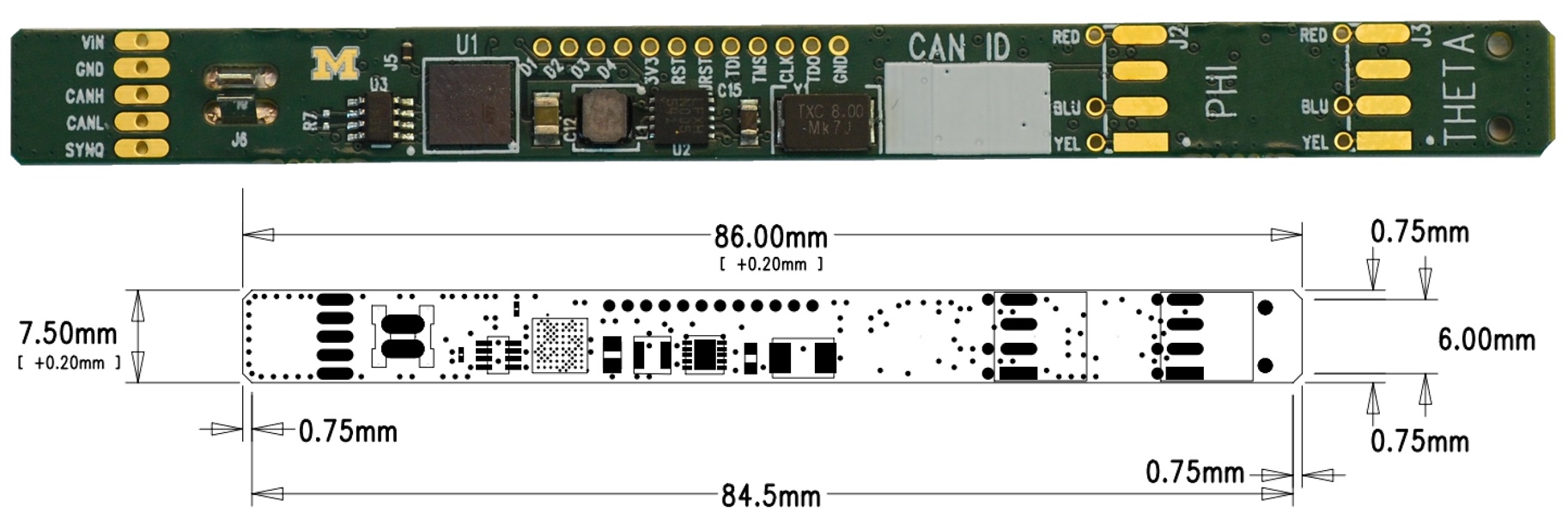}
    	\caption{\label{fig:fipos} Electronics board for the fiber positioner robots, from the University of Michigan (UM).}
    \end{figure}

    Each robot has an individual electronics board (FIPOS) screwed to the back of it \figref{fig:fipos}. These were built by UM. The boards can be powered with input voltage anywhere in the range 5\,--\,12\,VDC. The input voltage gets passed directly to the motor coils by CMOS switches (ADG1636), and is regulated down to 3.3\,VDC for the digital components. These include a 32-bit, 72\,MHz, single core, Cortex-M3 processor (STM32F103), and a CAN transceiver (MAX3051) capable of 1\,Mbps communication rates. A small resistor and amplifier (MAX9634) are connected to each of the two motors, inline with the input voltage. The signal is digitized with the microprocessor's built-in ADCs. Each board is 86\,mm long x 7.5\,mm wide. The narrow aspect ratio allows the board to pass through the 8.3\,mm mounting hole in the petal. The boards are made of 6 layers, and include a temperature sensor. A 400\,mm color-coded pigtail is soldered to the end of the board, and is terminated at the other end by a small 5-pin connector (Samtec S1SS-05-28). 

    Maximum power consumption of the robot is 3.23\,W with both motors spinning at 100\% duty cycle, but we typically operate motors at 70\% duty. During a typical reconfiguration sequence, each motor physically rotates for a total of only $\sim$\,5~seconds. The robot consumes 180\,mW when idle, which occurs primarily during communication and processing downtimes of the reconfiguration period ($\sim$\,2~minutes, see \S \ref{sec:reconfig_performance}) between exposures. Otherwise we put the FIPOS board into a lower power sleep mode, in which 15\,mW is consumed.
    
    Power is delivered to the robots by twenty 600\,W supplies (Mean Well HRPG-600-7.5). Two supplies are mounted to each petal, each of which delivers power to 250 or 252 robots (500 or 504 total motors per supply). The power supplies are fed 120\,VAC and output 7.5\,VDC. We tested the supplies early on to ensure they were capable of ramping current fast enough for the switching speed (18 kHz) at which we pulse-width modulate our motor coils. Given the constraints of move scheduling with anticollision (see \S\ref{sec:move_scheduling}) we typically operate $\sim$\,250\,--\,300 motors per supply at any given time during a fiber reconfiguration.
    
\begin{figure} 
    	\includegraphics[width=\textwidth]{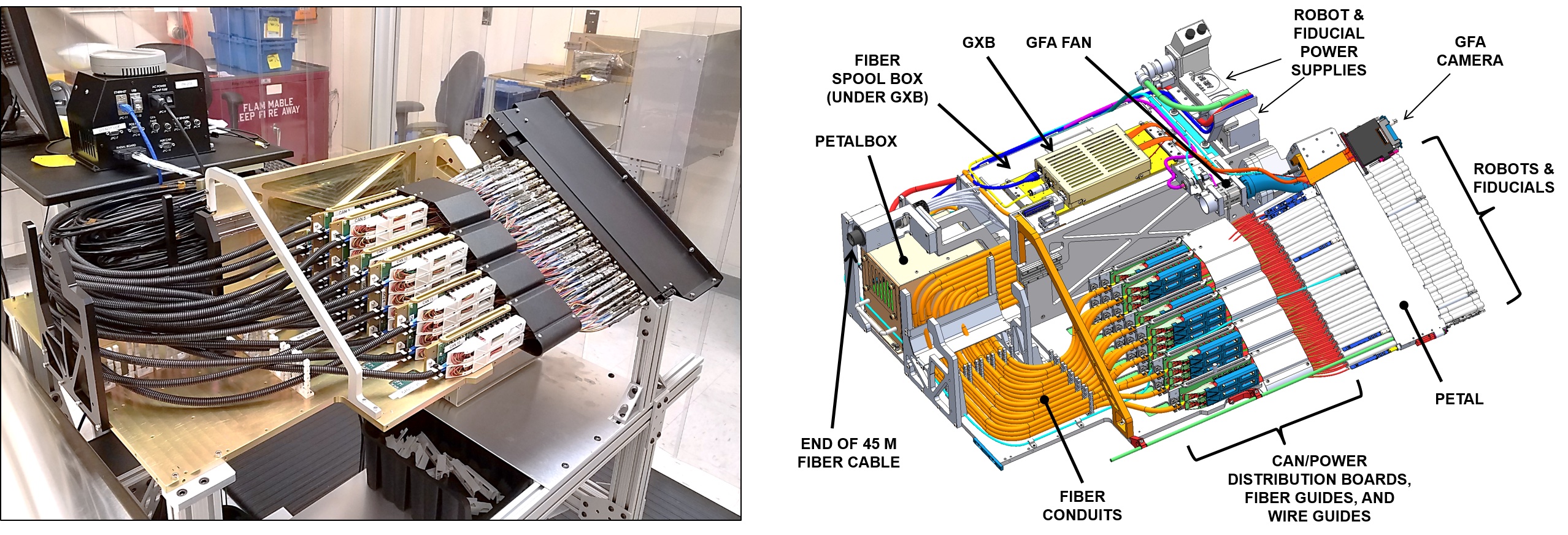}
    	\caption{\label{fig:petal_assembly} Key components of the petal assembly are illustrated at right. A petal is shown during splicing at left.}
    \end{figure}

    A large printed circuit board runs radially from the power supplies toward the tip of the petal, carrying power, CAN, and SYNC bus traces. From this radial board, seven transverse circuit boards project orthogonally across the robot array \figref{fig:petal_assembly}. Each petal has 10 CAN communication buses, which are carried on these seven boards. The boards are mounted to stiff aluminum panels, making them robust for connector mating and routing of fibers. 
    
    At the back ends of the robots, fibers and electrical pigtails are gathered in bundles of 12\,--\,14 units. The pigtails plug into the transverse boards in patches of 14 Samtec connectors. Between these connector patches, we mount plastic wire and fiber guides. The fibers bypass these connector patches through the guides, and are gathered into conduits of 25 fibers each. The 20 conduits then transmit the fibers to the spool box. Fusion splices to the 45\,m fiber cable are protected in the spool box. Inside the box the fibers are arranged in combs, with the fibers sandwiched by soft foam at the junction, and provided with $\sim 1$\,m of strain-relieving take-up length on either side \citep{poppett20}.
    
    Most hardware on each petal is controlled by a small Linux computer (Beaglebone Black Industrial). This computer is packaged together with two CAN control boards (sysWORXX USB-CANmodul8\footnote{\url{https://www2.systec-electronic.com}}) and three 12\,VDC power supplies in the `petalbox'. The petalbox controls CAN communications to robots and fiducials, state of robot and GFA power supplies, readout of temperature sensors, GFA thermoelectric cooler enables, and the GFA fan. It does not control operation of the GFA camera itself (which has a direct ethernet connection to the ICS). The petalbox interfaces to the Instrument Control System (ICS) via ethernet, and is powered by 120\,VAC.
    
    The GFA camera on each petal is supplied with DC voltages and an ethernet pass-through by the `GXB' electronics box. The GXB additionally provides interlocks to shut off power to the GFA, and hardware for control of the thermoelectric cooler inside the camera. The bias voltages on the GFA camera's CCD are automatically powered down by its firmware when the telemetry (humidity, electronics temperature) drift outside of safe operating limits. The GXB hardware interlock shuts off the GFA power completely if the electronics overheat.
    
    Environmental management is handled by the `FXC', an electronics box containing a similar Linux computer to those in the petalboxes. The FXC has interfaces to the main air circulation fans, as well as temperature, humidity, and smoke sensors. A key function of the FXC is control of the liquid chiller. Commands to the chiller are sent via ethernet. The computer in the FXC is powered by a 24\,VDC line, independent from the other interlocked components. In the event of tripping an environmental interlock (shutting off nearly everything else inside the focal plane) the FXC computer maintains power (see \S\ref{sec:hw_interlocks}).
    
    The 120\,VAC lines to each of the petalboxes and the fan power supplies are on independent, ethernet-switchable outlets. We used two rack power distribution units (Raritan PX3-5219-N1), both mounted inside the thermal enclosure. For each petal, 120\,VAC is delivered to the two robot power supplies on a single line. All of these 120\,VAC lines ultimately terminate at a relay box on the telescope facility main floor (``M-floor''). All 12 of the focal plane power lines (10 for the robot power supplies and 2 for the Raritan PDUs) are connected to a 220\,VAC power panel with UPS backup.

\subsection{Hardware interlocks}
    \label{sec:hw_interlocks}
    The focal plane has numerous automated safety features. We broadly classify these as either \textit{hardware interlocks} or \textit{fault management}. They are designed primarily to prevent damage to hardware, and secondarily to prevent unnecessary loss of survey time.
    
    \begin{description}
        \item[Hardware Interlocks] Intended to be fail safe and are implemented in hardware (i.e., not software that could easily be changed). `Hardware' can include single-purpose, low level processors or FPGA-based logic implementation.
        
        \item[Fault Management] Implemented in software and responds to an error condition with more specific actions, triggered before any interlock would come into play. This is discussed in \S\ref{sec:fault_management}.
    \end{description}

    The focal plane has an environmental Power Interlock. This is a hardwired circuit to a set of relays on the M-floor of the telescope facility. Interlock thresholds are shown in Table \ref{table:power_interlock}. When smoke is sensed or dewpoint exceeds its threshold, power is immediately killed to nearly all components of the focal plane system: fans, petalboxes, GFA cameras, robots, and fiducials. This condition also triggers an inhibit line on the Novec chiller, which causes the chiller to cease operation. When an over-temperature event occurs, power is shut off to the FIPOS power supplies and petalboxes. The relays can also be tripped by manually pressing an emergency stop button on the electrical panel on the M-floor. This disables power to the FXC as well, which means that while off, there is no information about the internal environment in the FPE. 
    
    \begin{table} 
        \centering
        \caption{Sensors and thresholds to trigger hardware Power Interlock.}
        \label{table:power_interlock}
        \begin{tabular}{|l|c|c|c|} 
            \hline
            \textbf{Sensor} & \textbf{Units} & \textbf{Lower Limit} & \textbf{Upper Limit} \\ \hline
            Internal air temperature & \degree C & 0 & +35 \\ \hline
            Internal humidity & RH\% & 0 & 85 \\ \hline
            Dewpoint & \degree C & -128 & -2 \\ \hline
            Smoke detected & 0 or 1 & n/a & 1 \\ \hline
            HXC or FPD fan failure & rpm & $< 546$ for $\ge 2$\,s& n/a \\ \hline
        \end{tabular}
    \end{table}
    
    Each GFA CCD camera has two additional levels of hardware interlock. Internally, an FPGA monitors temperature of itself, the CCD, and air in proximity to the CCD. If a limit is exceeded (50\degree C for the hardware or 45\degree C for the air in the camera), the FPGA turns off CCD biases and sends a request via ethernet to the petalbox, for the camera's power supply (GXB) to be shut down. Another temperature sensor is mounted on the GFA's hottest component, the A/D converter. This sensor is monitored by hardware logic in the GXB. If the sensor exceeds 75\degree C, the GXB turns off GFA power immediately.

\section{Software}\label{SOFTWARE}

    Software for DESI's focal plane is distributed among several independent applications. The entire system contains 22 hardware controllers, 21 software applications running on the DESI computer cluster, as well as custom firmware operating on the FIPOS boards that drive the fiber positioner robots and fiducials. The applications largely fall into four categories: petal and positioner control, GFA control, FVC control, and environmental control. Most of these categories have one hardware controller for fault management and low level operations and one application for higher level functions like move scheduling and image processing.

    These pieces work together with the Instrument Control System (ICS) to perform the \emph{positioning loop} (DESI Collaboration et al. 2022 in prep.). This process begins with the GFAs imaging a field, allowing ICS to guide and focus DESI. At the same time, fiber positioner robots move to pre-determined positions corresponding to targets on the sky. The FVC then measures the positions of the robots. ICS uses these measurements to issue an additional correction move to the robots. Finally the FVC measures the fibers a second time, to determine their final positions.

    ICS provides many important pieces of this process which go beyond the scope of this paper. This includes PlateMaker, which is responsible for astrometry calculations as well as coordinate transformations between the focal plane, sky and fiber view camera. ICS also interfaces with the telescope control system (TCS) and hexapod for slewing and focusing on target fields. Finally, ICS controls LEDs inside DESI's spectrographs which back illuminate the fiber positioners for measurement by the fiber view camera (DESI Collaboration et al. 2022 in prep.).

\subsection{Control of petal hardware}
    \label{sec:petal_control}
    All petal hardware, other than the GFA camera, is controlled by software on the petalbox computer. This software, called PetalController, is written in Python and based on a framework provided by the ICS. PetalController handles CAN communications with fiber positioner robots and fiducials. It receives move schedule tables as Python dictionaries, converts them to CAN messages, and sends them to the robots. PetalController reads temperatures and voltages, monitors for error conditions, and publishes telemetry data to the ICS. It controls the state of power supplies for the robots, fiducials, GFA camera and thermoelectric cooler, and the speed of the cooling fan for the GFA camera.
    
    Commands to PetalController are made by a single Petal application, running on the instrument control Linux cluster. Support for alarms and notifications, telemetry archiving, message passing, and command interfaces are all provided by standard ICS libraries.

\subsection{Positioner firmware}
    The fiber positioner microcontroller is loaded with custom firmware, written at UM, LBNL, and EPFL. The firmware controls all aspects of robot operation, including motor control, interpretation and response to CAN messages, thermistor readout, and motor coil current sensing.
    
    Motors are commutated open-loop, with pulse-width modulation at 18 kHz, corresponding to 4,000 clock cycles per second at the microcontroller's 72 MHz speed. The motors have 3 coils, spaced 120\degree~apart. At each 55\,$\mu$s timer update interval, each coil is assigned a duration for which its switch should be closed. The durations are calculated by a cosine lookup table, with the coils phase-shifted by 120\degree. Frequency of the cosine table determines rotation speed of the magnetic field. The pulse duration is multiplied by a user-configurable `duty cycle' parameter (0\,--\,100\%), thus controlling the net current.
    
    As of December 2021, the firmware currently supports 2 hard-coded speeds, \textit{cruise} (9,900 rpm at rotor = $176.07~deg/s$ at the output shaft) and \textit{creep} (150 rpm at rotor = $2.67~deg/s$ at the output shaft). We anticipate future firmware upgrades to allow selection of speeds in-between.
    
    Upon power-up, a bootloader routine waits 2 seconds. During this period, the unit may be signaled by CAN message to go into a reprogramming mode. In that mode, new firmware may be sent to each robot via CAN.
    
\subsection{Move scheduling and anticollision}
    \label{sec:move_scheduling}
\begin{figure} 
    	\includegraphics[width=\textwidth]{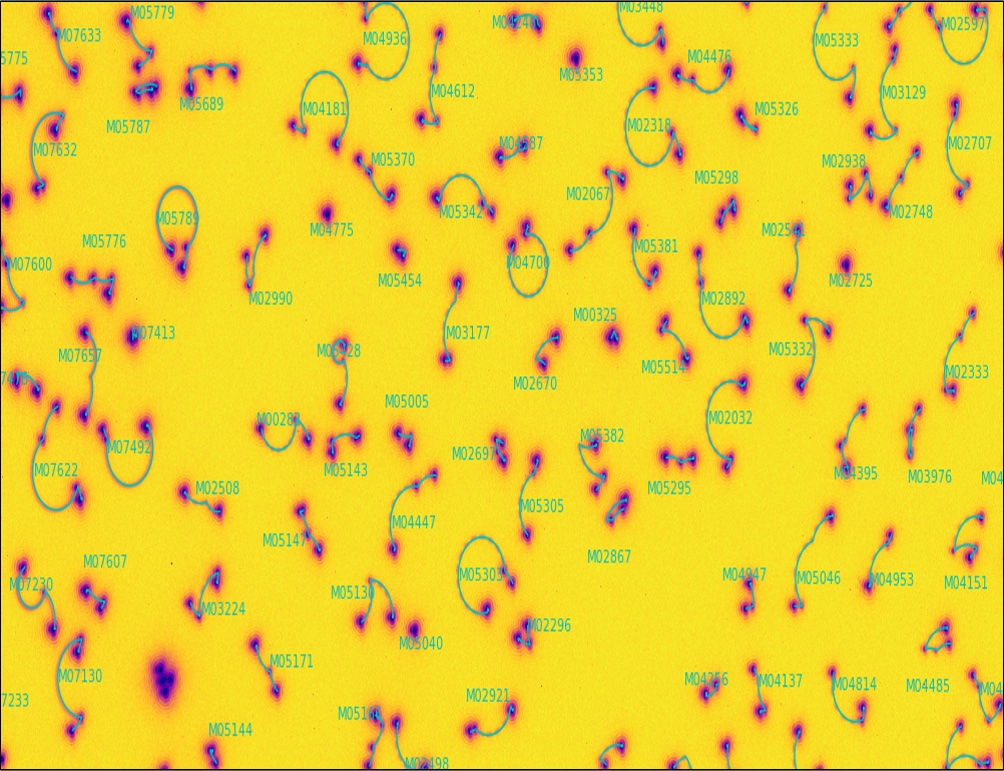}
    	\caption{\label{fig:move_paths} Motion paths of positioners, imaged by leaving the FVC shutter open and the fiber backlit during the move. Higher image saturation occurs at pause points during the path. Arcs of motion are naturally dashed due to pulse width modulation of the fiber backlight source. Positioner id numbers and motion paths (thin solid lines) are overlaid.}
    \end{figure}

    The fiber robots run open-loop and have overlapping patrol regions. They therefore require precalculated motion paths \figref{fig:move_paths}, in order to avoid colliding with one another. These `move schedules' are timed sequences of motor rotations and pauses, such that each robot can get to its target while avoiding its neighbors.
    
    The move schedule calculation is performed off-instrument, in a modern multi-core computer. Our computation takes advantage of the natural parallelism of the petal architecture: the move tables for the 10 petals can be calculated simultaneously in 10 separate cores, with the constraint that we don't allow robots along the edges to exceed their respective petal boundaries. Calculation time is typically 5\,--\,6~sec for blind moves and 2\,--\,3~sec for corrections.
    
    Prior to moving, each robot receives its particular move table and stores it in local memory. Then upon receiving a synchronized start signal, all robots execute their scheduled motions on their own local clocks. The synchronization requirement between positioners' motion start times is $\sim$\,2\,ms. Our electronics include the capability for synchronization signals via either a dedicated hardware line or with broadcast commands sent in close succession on the CAN buses. In practice we've found that either method works.
    
\begin{figure} 
	\includegraphics[width=\textwidth]{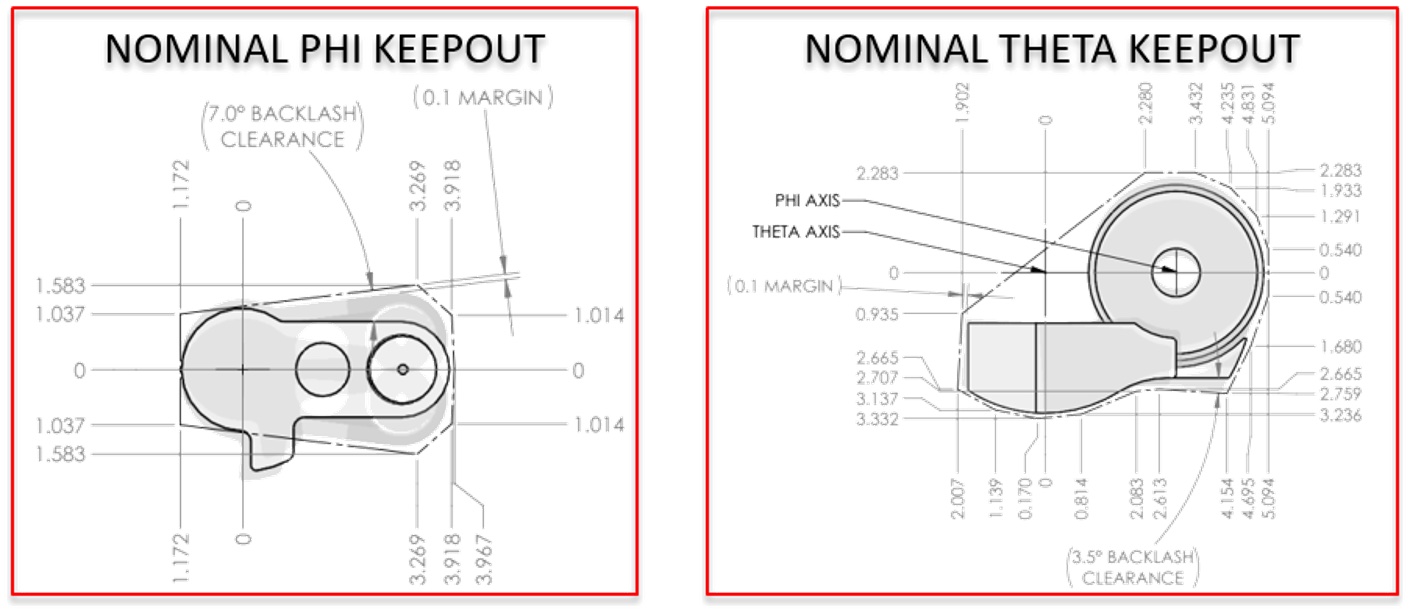}
	\caption{\label{fig:nominal_keepouts} 2D polygon keep-out zones drawn around the two moving bodies of each positioner. Additional polygons (not shown) gracefully provide the system with fixed boundaries at the petal-to-petal edges and around the GFA cameras.}
    \end{figure}

    In software we model each positioner as a pair of 2D polygons, encompassing the nominal mechanical outlines of the eccentric and central bodies \figref{fig:nominal_keepouts}. The \textit{cruise} motion of every positioner is simulated discretely, quantized at 0.02~second time steps. The polygons are slightly expanded, to include margin for variations in mechanical build tolerances, start signal synchronization, quantization of the simulator, and small final anti-backlash motions.

    We've coined our algorithm `Retract, Rotate, Extend'. All robots are first retracted ($\phi$ motion only) to within non-colliding circular envelopes, then rotated about the central $\theta$ axis, and finally extended to their target positions ($\phi$). This avoids the majority of would-be collisions. Each of these three stages of motion are then `annealed' (spreading out the different positioners' moves in time), to reduce instantaneous power. Annealing also further reduces the number of would-be collisions. Finally, to deal with any remaining collision cases, small off-path adjustment moves and delays are iteratively tested and incorporated into the move schedules as necessary. In rare cases where the algorithm cannot find any safe path to a given target (for example if there is an interfering, disabled neighboring positioner), the robot in question is labeled `frozen' in the log for that move. Similarly, in cases where a requested target position overlaps a neighboring positioner or boundary, or where a request is made of a disabled positioner, that request is rejected and labeled in the log.

    The collision avoidance algorithm is an interesting (and well-bounded) problem, and is also a relatively easy problem to provably solve in simulation. In practice we have found that the most important design criterion for the algorithm is architectural flexibility, to handle prosaic real-world constraints such as:
    
    \begin{itemize}
        \item Avoiding disabled neighbors, adjacent cameras, and petal-to-petal boundaries.
        \item Obtaining correct (to $\sim 5$\,\micron) calibrations of kinematic dimensions (i.e. center position, arm lengths). For DESI, this totals $> 40,000$ parameters.
        \item Avoiding positioners with broken fibers, whose precise locations are difficult to measure.
        \item Restricting extension of questionably-performing robots to a non-colliding smaller patrol zone.
        \item Limited selection of discrete motor speeds (e.g. `cruise' and `creep').
        \item Feedback is not real-time. The FVC system in DESI takes 5+~sec per measurement.
        \item Feedback is strictly 2D. Having one centroided spot per positioner gives Cartesian (x,y), but the transform from (x,y) to robot arm angles is not unique, because the $\theta$ and $\phi$ travel ranges exceed 360\degree~and 180\degree.\footnote{For example, on a robot with central axis travel range $-190\degree \le \theta \le +190\degree$ from hardstop to hardstop, there is a critical difference between asserting $\theta = +185\degree$ versus $\theta = -175\degree$, based on its singular measured fiber location in Cartesian space. Similarly, there is an important physical difference between asserting for example $(\theta, \phi) = (-5\degree, +5\degree)$ versus $(+5\degree, -5\degree)$, again based on a single Cartesian fiber location.}
        \item Accommodating unreliable communicators. Some robots are reliable enough on the CAN bus that we generally use them, however in rare cases they fail to properly communicate. In such cases we must be able to avoid their unplanned motion (or stasis).
        \item Variations in output gear speed, i.e. cases where $\frac{angle~ moved}{angle~commanded} < 1.0$ (see \S \ref{sec:pos_util}).
        \item Power supplies, though large (1200\,W available per petal), cannot support all 502 robots simultaneously moving both motors at 100\% duty cycle.
        \item FVC measurement error can make positioners that are very close to each other appear to move in and out of a collision state with their neighbors.
    \end{itemize}

    For ease of software integration and maintenance, an early goal was set in the DESI project to maximize usage of Python as the instrument control language. To achieve sufficient speed of the move scheduling calculation, we took three approaches. First, the polygon collision calculation code was isolated in a standalone module. It was prototyped in Python, then converted to Cython\footnote{\url{https://cython.org/}}, and finally optimized for speed. At this low level (translating and rotating polygons, checking for intersections), data structures are simple and easily statically typed. Second, the remainder of the move scheduling calculation suite was highly cached. At this higher level of abstraction (representation of hundreds of uniquely calibrated robots swinging around each other in a scheduled dance), more complex data structures were required. Data storage and retrieval in Python are fast, while loop executions are relatively slow; hence we put a premium here on never calculating the same thing twice. This policy produced reasonably fast code, but came at some cost of complexity, putting a premium on keeping all cached data up-to-date. Finally, and most importantly, we wrote ourselves a detailed test harness, with which to repeatably test and optimize in simulation every possible corner of the move scheduling system.

\subsection{Offline software tools}

    Every move and action commanded to the positioners is logged in a database which takes a snapshot of the current software state of the robots. This database can be queried offline to analyze positioners for calibration of parameters or positioning performance. \texttt{Desimeter}\footnote{\url{https://github.com/desihub/desimeter/}} is a software suite for processing positioner information offline. This includes making batch queries to the positioner database, processing raw FVC images, and transformations between sky, FVC, focal plane, and local robot coordinate systems. We wrote the desimeter transformation code in isolation from PlateMaker (used online during observing) so that we can cross-check between the two.
    
    Positioner calibration parameters (kinematic `arm lengths', central axis location, angular zero points, effective gear ratios) are calculated by making best-fits between commanded angular motions of the motors and measured locations of the fibers. This can be done either with a random scattering of points, or (often more efficiently) by independent arcs of points on the $\theta$ and $\phi$ axes. This optimization code is also a part of the desimeter suite.
    
    Many of these tools are also used by DESI's \texttt{fiberassign}\footnote{\url{https://fiberassign.readthedocs.io}} software, which generates configurations of the positioners, called tiles, to observe targets on the sky. There is a daily one-way sync from the instrument control model of the positioners to the model used by fiberassign. This enables tiles generated on the fly during the night to remain aware of potential changes in the instrument model of the positioners, such as changes in the usability of robots, calibration parameters, or synthesized parameters like keep-out polygons. Letting these parameters go out of sync causes tiles to include unreachable target requests, which the online control system must then reject. Therefore keeping this information up-to-date is important for aligning the maximum number of fibers with targets and thus maintaining survey homogeneity.

\subsection{FVC centroiding and spot matching}
    \label{sec:centroiding_spot_matching}
    
    As described in \S \ref{sec:fvc}, the FVC is used to close the control loop with the positioners. Centroids of spots in the FVC image provide a measure, in pixels, of the relative locations of the fibers and the fiducial sources on the focal surface. The transformation of these measurements from pixels at the FVC image plane to physical coordinates at the focal surface is done by a software module called PlateMaker \citep{honscheid16}, which is not discussed in detail here.
    
    The FVC software is separated into two parts: `fiberview' and `spotmatch'.  Fiberview records FVC images, detects spots above a fixed brightness threshold with intensity profiles matching the diffraction-limited PSF (this filters cosmic ray hits and hot pixels), and precisely measures pixel centroids of each source. Spotmatch associates each fiber spot to its respective location as predicted by PlateMaker. Because the predicted locations are sometimes far from their actual locations, due to robot positioning errors, and because of intermittent errors illuminating fiducials, spot matching is not always trivial. To ensure correct associations, PlateMaker also provides a prediction for the location in the FVC image of each robot's center of rotation. For spots that inadvertently appear far from their expected location, their proximity to one of the robot centers is usually enough to make a secure match. For normal cases, proximity to the nearest predicted source is usually the correct association (after correcting for small errors in scale, translation, and rotation of the PlateMaker model).  Once each centroid is matched with its predicted source, this information is fed back to PlateMaker, which translates the measured pixel locations to the reference frame of the focal surface and determines any correction moves required to align the fibers with their sky targets. 

\subsection{GFA camera operation}
    To fit within a small form factor of each GFA housing, the electronics are distributed across several stacked printed-circuit boards (see \S \ref{sec:GFA_hardware}). A hybrid field programmable gate array (FPGA) / computer processing unit (CPU) ties all operations together. The CPU runs a Linux operating system supporting high-level functionality, such as standard protocols for network communication, data transfer, and command processing. At a lower level, the FPGA logic is programmed for hardware-level operations, like control of the CCD, environmental sensors, a thermoelectric cooler, and the analog-to-digital converter required for signal digitization. With this system, the masked area of the CCD (2K x 1K pixels) can be read out to  memory by four amplifiers in 2.5~sec, with the option of exposing a new image simultaneously. Noise and dark current depend on the CCD operating temperature, but levels acceptable for telescope guiding (at a duty cycle of $\sim$\,5~sec) are achieved without any additional cooling, beyond that which is normally maintained in the focal plane enclosure ($\sim$\,10\,--\,15\degree C, as shown in fig. \ref{fig:thermal_performance}). 

    Typically each GFA is operated independently and remotely. A dedicated client application (`camera.py') sends commands over the network to each GFA, where a dedicated command server (`gfaserver') receives and processes the commands. There are two channels for communication: one for command and status, another for asynchronous data transfer. Except during image readout, telemetry data are continuously updated.

    Each client program is responsible for: (1) configuring  the respective GFA at power up (setting bias levels and clocking options); (2)  reconfiguring the GFA as necessary for each exposure (setting exposure time and readout mode); (3) monitoring the telemetry; (4) starting each exposure; and (5) receiving, repackaging, and saving the image data when they are sent back by the GFA. Each client program also acts as a command server to a higher level controlling process, executing GFA operations as requested. In normal operation while guiding the telescope, $\sim$\,5~second exposures are continuously recorded. If there is an occasional readout error (e.g. dropped data), the client can automatically recover by reconfiguring the GFA before the next exposure command is received.
    
\subsection{Environmental control}
    \label{sec:environmental_control}
    Environmental control software runs on the FXC, a small Linux computer housed within the thermal enclosure (see \S\ref{sec:electrical_system}). The FXC reads temperature and humidity sensors positioned at numerous points inside and just outside the thermal enclosure. The FXC controls the fans blowing air through the heat exchanger (HXA), into the FPD volume, and a small exhaust fan for driving air out of the system, a feature which in practice we have never used.  
    
    Every 60 seconds, the FXC updates the temperature set point for the liquid chiller, which is physically located some 50\,m away from the focal plane `as the coolant flows', on the telescope's C-floor. The set point is calculated with a PID loop, controlling on the measured temperature of the FPR (the large aluminum ring to which the 10 petals are mounted). Separately, the chiller runs its own internal PID loop, controlling its heating and refrigeration components to achieve the requested coolant set point. This second PID loop is built into the chiller vendor's firmware. The FXC can also set other key chiller parameters, such as target flow rate. These parameters are stored in chiller non-volatile RAM so they are typically only set when the chiller is initially configured.

\subsection{Fault management}
    \label{sec:fault_management}
    In addition to hardware interlocks (see \S\ref{sec:hw_interlocks}), we implement numerous fault management actions in software. Some actions only raise warnings to the operator, while others may selectively disable components. Whereas the tripping of a hardware interlock may incur significant instrument down-time, fault management responses tend to be quicker to recover from.
    
    In \S\ref{sec:hw_interlocks} we described the global Power Interlock relays, which can be triggered by hardware logic. The FXC computer is also provided with the ability to trip these same relays, based on information known only in software,  including chiller and dry air supply flow rates, fan speeds, differential air pressure across the FPD boundary (ensuring sufficient pressure exists to push air through the robots), structure temperature, air temperature, humidity, and dew point.
    
    Software limits may be quite restrictive, guaranteeing the instrument is not only in a safe envelope, but in our intended operational mode. For example, in the hardware interlock, the lower limit on air temperature and upper limit on dew point are set respectively at 0\degree C and -2\degree C  (see Table \ref{table:power_interlock}). In software however, we trigger a violation at +5\degree C air or -15\degree C dew point. The hardware limit is a final fail safe against frost conditions; the software limit allows additional protective actions prior to reaching that point.
    
    The FXC software requires any violation to persist $> 20$ seconds before triggering, to allow time for the relevant components to self-correct. If conditions have not improved, it alerts the ICS, so that other actions can be taken (for example shutting down GFA CCDs). After a 30~second window, the FXC then trips the relays.
    
    PetalController (see \S \ref{sec:petal_control}) provides the primary fault management on each petal. It monitors numerous temperatures, voltages, fan speeds, power supply and interlock states, and the responsiveness of all robots and fiducials on the CAN bus. It gathers regular temperature telemetry from all responsive robots, giving us a granular thermal map of the array. In response to these conditions, the petalbox can enable or disable power to any of these components. We also put a software rate limiter on the sending of move tables to positioners. The rate-limiter prevents unnecessary build-up of heat or mechanical cycles, for example if an erroneous external script were to repeatedly request moves.
    
\section{Assembly and Test}
\label{ASSEMBLY_AND_TEST}

We built key sub-assemblies at several institutions. These included the FPD at FNAL, the GFAs by the Spanish Consortium, machined petals provided by BU, FVC and fiducials from Yale, and positioner robots from UM and EPFL. These items were tested and sent to LBNL for system level integration. At LBNL we installed short lengths of ferrulized, anti-reflection coated fiber (PFAs) into the robots. Then we installed the robots into each petal, and finally spliced the PFAs to the fiber cables (Poppett et al. 2022 in prep), which were provided by Durham University.

When the petals were fully assembled, they underwent precise metrology, measuring the locations of the GFAs with respect to the fiducials and petal structure. We required that the fiducials be measured to 10\,\micron~accuracy with respect to the focal surface, and to 5\,\micron~accuracy with respect to the GFAs. Furthermore, we required that the fiber tips be located within $\pm$100\,\micron~of an ideal focal surface. 

The fully-assembled petals each underwent a formal acceptance process before being shipped individually to Kitt Peak for installation. The subsystem acceptance package included demonstration of compliance with key requirements, the compilation of a data set including the metrology measurements, and the completion of a full functional test. 

\subsection{Petal alignment to mounting ring}
    \label{sec:petal_align}

    Practical machining tolerances would have been insufficient to align the petals to our desired level of accuracy in the telescope. We therefore included alignment features at the petal's mounting interfaces, consisting of captive shims and gauge blocks.

    Each petal has three precision tooling balls, permanently mounted via pins and epoxy. Upon completion of machining each petal, BU contracted with ZEISS to measure all robot/fiducial mounting holes and spotfaces with respect to these balls. DESI's nominal aspheric focal surface was best-fit to the measured hole array, as projected 86.5\,mm ahead, the length of a robot from mounting flange to fiber tip. This complex measurement was done both as a quality control test and to establish the exact location of the focal surface on each petal, with respect to a simple set of datums (i.e. the tooling balls).
    
    Separately, the FPR and FPD were aligned by FNAL with respect to the Corrector Barrel. An array of 11 tooling balls around the edge of the FPR could thus be tied to the expected final position of the Corrector. At LBNL, we mounted the petals into the FPR, established a global coordinate system with the FPR tooling balls, and then aligned each petal to that system using its 3 local tooling balls.
    
    Alignment was done via a system of shims (see \S\ref{sec:fp_structure}). We achieved a focus alignment of 12\,\micron\,RMS of the tooling balls, with worst case error 33\,\micron~at the tip of one petal. For logistical flexibility, we also checked whether petals could be satisfactorily installed in alternate positions about the ring. We dismounted the petals, shuffled their locations, and re-installed them. In the shuffled positions, RMS error was 16\,\micron, and worst case error was 59\,\micron~(on the same petal as before). After completing the alignment with the tooling balls, we also back-calculated the overall error of all 5,020 holes with respect to the nominal focal surface. The overall alignment error was $\pm 15$\,\micron\,RMS by geometric metrics, translating to $\sim 0.12\%$ reduction in throughput \citep{duan18}.

\subsection{Positioner assembly and test}
\label{sec:pos_assy_test}

    The robots were designed at LBNL, SSL, and UM, as discussed in \S\ref{sec:robots_hardware}. Early prototypes were built at LBNL, and volume production was completed at UM. EPFL provided key precision parts, consultation, and independent testing. Production of gearmotors and bearings was by Namiki Precision Jewel Co.

\begin{figure} 
    	\includegraphics[width=\textwidth]{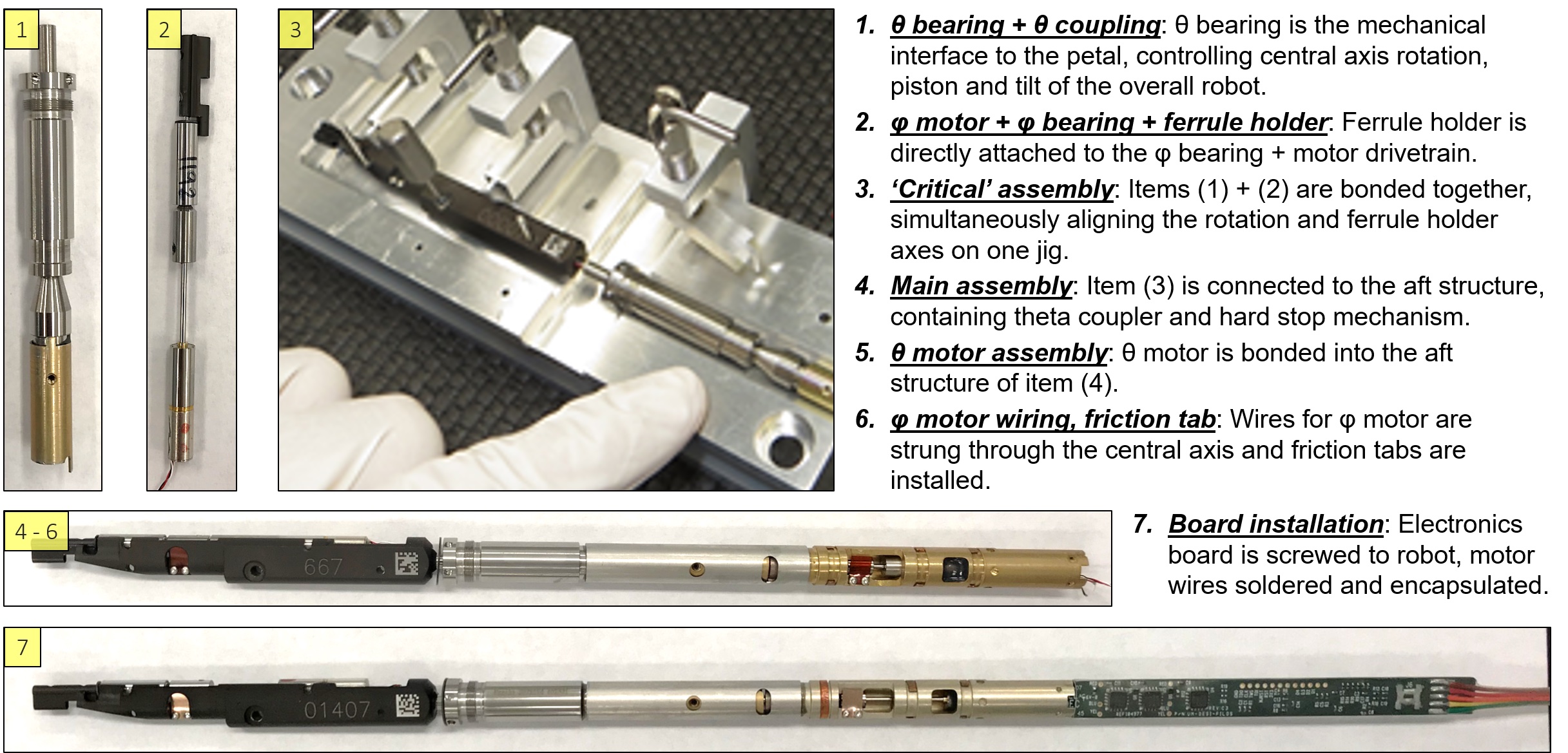}
    	\caption{\label{fig:subassemblies} We built the positioners in seven sequential sub-assemblies, with quality control checks at each step.}
    \end{figure}

    A fiber positioner robot is composed of seven sequential sub-assemblies, illustrated in figure \ref{fig:subassemblies}. We established quality control checks at each step, to minimize loss of parts and time (see \S\ref{sec:inventory_and_qc}). Upon completion of the robot, we installed a temporary, 1\,m long, cleaved fiber, and mounted the unit horizontally on a test stand with a calibrated metrology camera (SBIG STF8300, monochrome CCD). Each test stand could hold 40 positioners simultaneously. To ensure good centroids, the fibers were diffusely illuminated in a custom integrating chamber, 3D printed in white nylon, holding a board of blue LEDs. The positioning robots were tested for accuracy and reliability in a standard grid of 192 points. To ensure rejection of early-failing robots, the test sequence included $> 6,000$ total retargetings, and running to the hard stops 50 times. From these tests, we were able to give each positioner a grade (see Table \ref{table:pos_grading}). Only those positioners that received A and B grades were shipped to LBNL for further integration.

    \begin{table} 
        \centering
        \caption{Grading scheme for positioners, and measured yield for the first 4,000 units we produced.}
        \label{table:pos_grading}
        \begin{tabular}{|l|c|c|c|c|} 
            \hline
            \textbf{Grade} & \textbf{Max. blind move error} & \textbf{Max. corrected move error} & \textbf{RMS corrected move error} & \textbf{Yield} \\ \hline
            A & 100\% $\leq 100\micron$ & 100\% $\leq 15\micron$ & 100\% $\leq 5\micron$ & 96.1\%\\ \hline
            B & 100\% $\leq 250\micron$ & 100\% $\leq 25\micron$, 95\% $\leq 15\micron$ & 100\% $\leq 10\micron$, 95\% $\leq 5\micron$ & 0.8\%\\ \hline
            C & 100\% $\leq 250\micron$ & 100\% $\leq 50\micron$, 95\% $\leq 25\micron$  & 100\% $\leq 20\micron$, 95\% $\leq 10\micron$ & 0.6\%\\ \hline
            D & 100\% $\leq 500\micron$ & 100\% $\leq 50\micron$, 95\% $\leq 25\micron$  & 100\% $\leq 20\micron$, 95\% $\leq 10\micron$ & 0\%\\ \hline
            F & \multicolumn{3}{c|}{Does not meet any of the above} & 2.4\%\\ \hline
        \end{tabular}
    \end{table}
    
\begin{figure} 
    	\includegraphics[width=\textwidth]{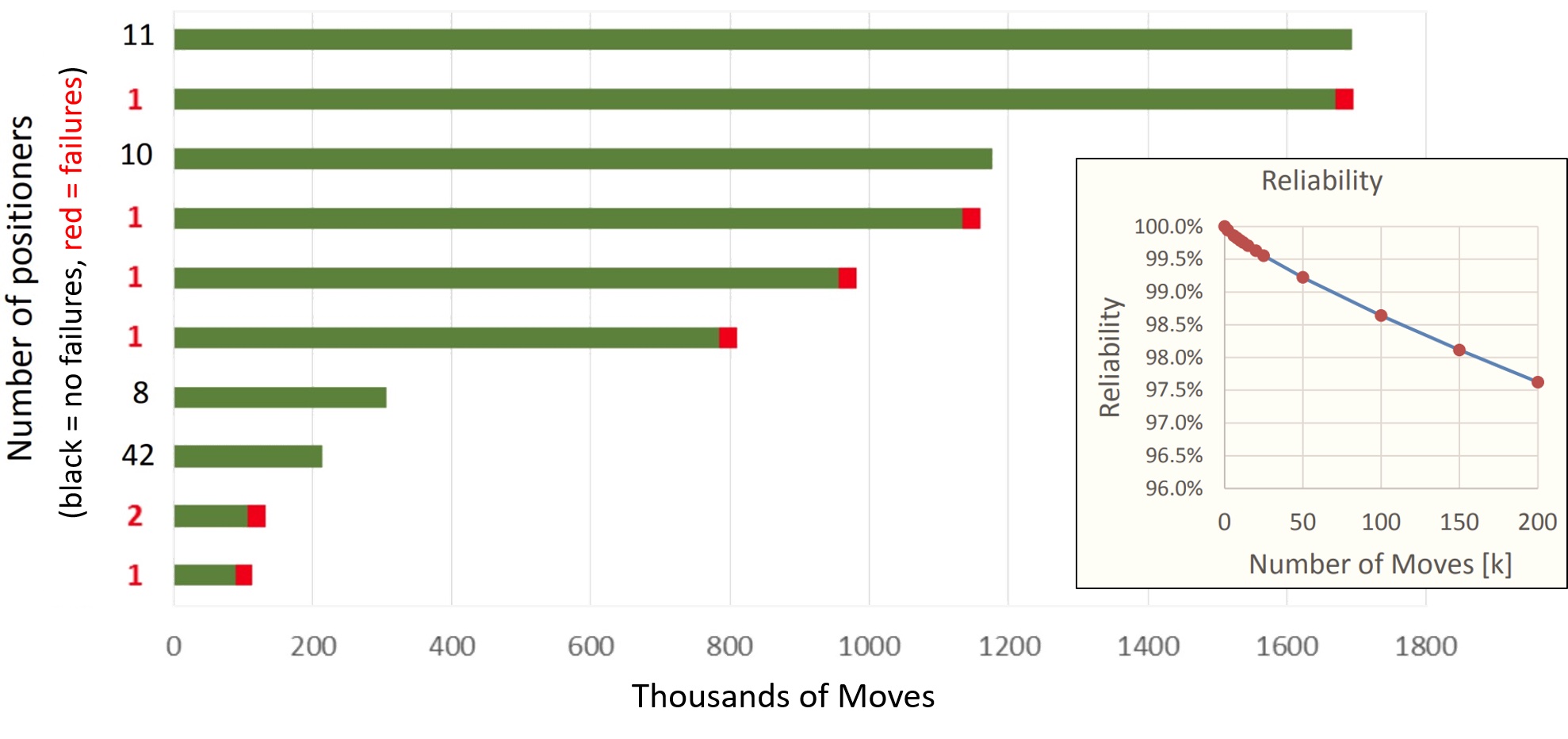}
    	\caption{\label{fig:lifetime_summary} Summary of robot lifetime tests during design qualification. Each cycle is a move at cruise speed to a random position within the robot's patrol radius. Failure was defined by a positioner becoming unable to reach its commanded targets, as measured by the test stand FVC.}
    \end{figure}

    Lifetime tests of $\ge$~100,000 moves were done on 78 positioners. Our requirement was a 90\% survival rate after 107,000 moves. In these sequences, we would move the positioner open-loop to a series of 2,000\,--\,10,000 random points, followed by a closed-loop FVC test grid of 192 measurement points. Of the 78 units life-tested, 23 were operated past 1,000,000 cycles, to improve the statistical power of our distribution \figref{fig:lifetime_summary}. A Weibull reliability fit indicated 98\% survival after 107,000 moves and 90\% after 1,200,000 moves \citep{schubnell18}.
    
    These lifetime tests did not reveal a significant class of motor failures, which we discovered only later when the positioners were installed on the telescope (see \S\ref{sec:pos_util}). We currently hypothesize that the tests missed this because the positioners were oriented horizontally when tested in the lab, rather than pointing downward as they do in the telescope.

    Robot performance at thermal extremes was qualified in a dry, thermal chamber, at temperatures ranging from -20 to +35\degree C. This testing was done on 83 robot units and several fiducials, all of which survived the thermal extremes.
    
    An array of 15 robots and a fiducial was tested for stability under the influence of air flows of several speeds, in horizontal, vertical, and face-on directions. This was done in consideration of the cooling air which we pass through the focal plate \citep{zhang18}.

\subsection{Fiber integration to positioners}
    \label{sec:fiber_integration}
    
    Our design required integration of fibers to robots prior to installation into petals. It also required installation of the robot to the petal from the front side. We therefore installed a short section of fiber (the Positioner Fiber Assembly, or `PFA') into each robot, and subsequently fusion-spliced the PFAs to the $\sim$\,45\,m cables. A significant logistical benefit of splicing was to decouple cable and slit fabrication in Durham, UK \citep{schmoll18} from petal assembly in Berkeley, California. The PFAs were spliced to cables only after all positioners had been installed and tested in a given petal.
    
    Each PFA is composed of a single $\sim 3.1$\,m strand of Polymicro FBP optical fiber with a core diameter of 107\,\micron~and a thin polyimide coating. This is the same fiber as used in the fiber cables. One end of the fiber was cleaved and bonded into a custom \o\,1.25\,mm borosilicate glass ferrule, then shipped to Infinite Optics for anti-reflection coating. For strain relief and protection through the working mechanisms of the robot, we bonded a \o\,0.4\,mm polyimide sleeve at the end of each ferrule. In addition to general robustness, the sleeve reduces strain-induced Focal Ratio Degradation (FRD), by preventing direct contact points to the fiber within the fiber positioner \citep{poppett18}. 
    
    We installed the PFAs into robots from the front side. To ease the installation, we would first install a temporary length of Hytrel furcation tube from the rear. With this guide tube in place, the fiber was smoothly pushed through the positioner.
    
\begin{figure} 
    	\includegraphics[width=\textwidth]{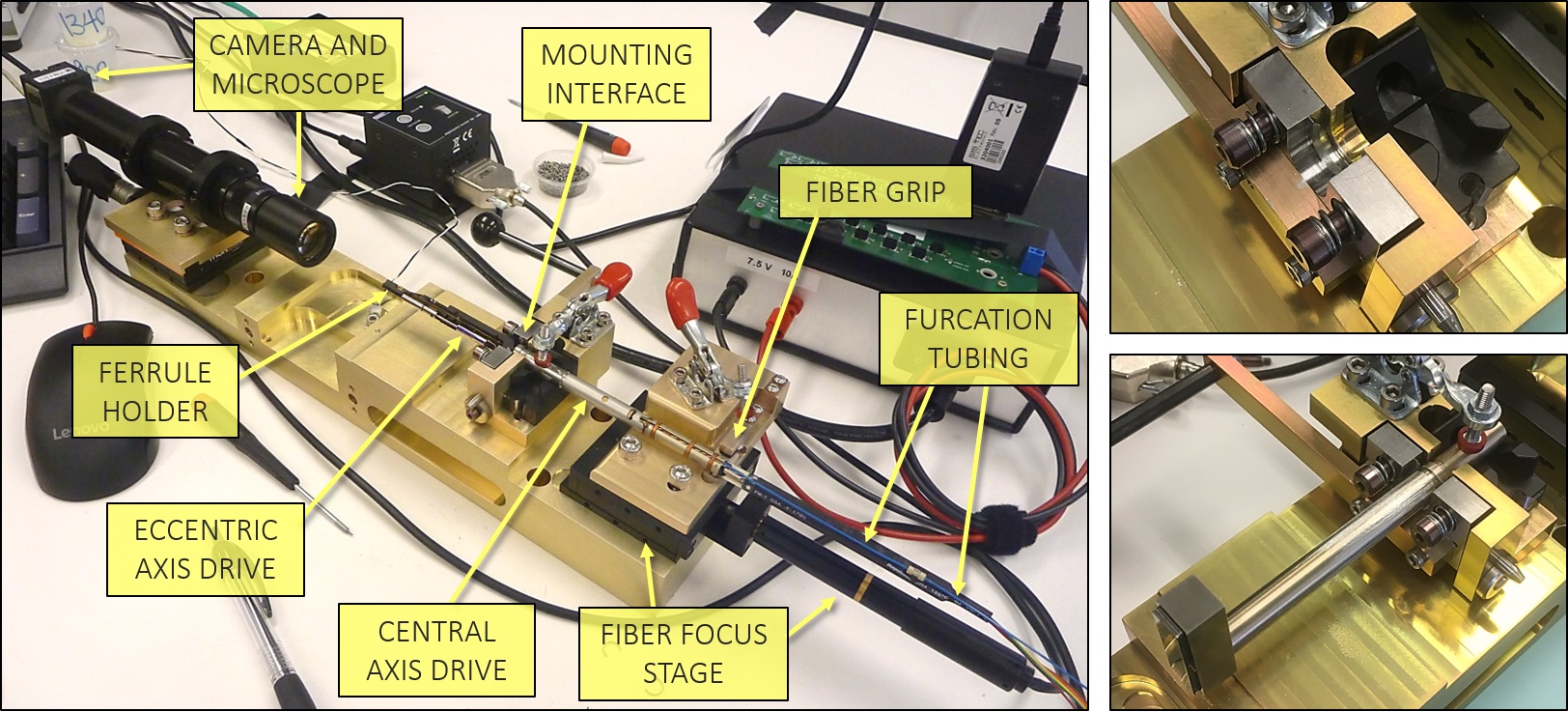}
    	\caption{\label{fig:fiber_alignment} The fiber alignment fixture (left) has a camera and microscope mounted at one end on a manual micrometer stage and a motorized fiber adjustment stage at the other. Between them, the robot mounts in a vee-and-flange, which we designed for functional similarity to the true focal plate interface, while allowing installation and removal using a simple toggle clamp, with no rotation of the assembly required along its screw threads. At the motorized stage, the fiber is gripped by another toggle clamp actuating a pivot arm with a soft foam grip. Temporary \o\,0.9\,mm furcation tubing is used to assist us in sliding the fiber through the mechanical assembly. At upper right is a closer view of the spring-loaded vee-and-flange clamping interface. At lower right, a calibration rod is loaded in the clamp.}
    \end{figure}

    The nominal distance from the robot mount flange to the focal surface is 86.5\,mm. We chose to set the axial position of the fibers optically rather than by a mechanical stop. This relieved us of any concerns about damaging the fiber tips. It also naturally built traceable test data (automated optical focus scans) directly into the alignment process. After inserting the PFA, we clamped the robot into alignment fixtures like that shown in figure \ref{fig:fiber_alignment}. The ferrule was manually adjusted to an approximate focus location. Then, at the rear, the fiber was lightly clamped by soft foam to a motorized stage (Thorlabs MT1/M-Z8), which was axially independent of the positioner. The fiber was backlit with an LED, and the tip was imaged by an Edmond Optics CMOS Color USB Camera (EO-3112C), with a Mitutoyo 10x Objective lens.
    
    An automated script stepped the fiber through several axial positions, imaging the fiber at each step. A parabola was fit to the imaged spot widths, with its minimum indicating best focus. The stage returned the fiber to the best focus position, completing the program. We used epoxy to retain the ferrule at the front of the robot. A small set screw inside the robot's eccentric axis arm held the ferrule in place while the glue cured.\footnote{When the set screw was over-tightened, it would crack the glass ferrule. Even with calibrated wrenches, torque can be challenging to finely regulate when fastening such small screws (size 0\,--\,80). We lost several fibers to this failure mode.}
    
    In the aft section of the robot, where the fiber emerges, we loosely retained it in an axial groove, alongside the $\phi$ motor wires, using three phosphor-bronze clips. At the rear of the electronics board, we transitioned from the \o\,0.4\,mm polyimide sleeve to \o\,0.9\,mm Hytrel furcation tube. We made the transition between these tubes by gluing them into either side of a PEEK `hardpoint', which was snapped into a clip on the electronics board. The fiber was free to slide there; we axially restrained it only at the ferrule and at the far end of the PFA (inside the spool box, where there is 0.5\,--\,1\,m of free slack prior to the soft foam mount pads that retain the splice joint).
    
    The remaining $\sim$\,2.9\,m of fiber, inside its furcation tube, was then spooled on to a custom injection-molded plastic holster. The holster had an integrated trough designed to hold the robot snugly, and a clamshell cover. In this state the robot and fiber assembly was safe and convenient for handling and storage.
    
    Calibration of the focus alignment fixture was done with an undersized, machined aluminum rod, to which we wrung gauge blocks as necessary to achieve the exact final length. The rod had mounting features and center of gravity to match a fiber positioner. Aluminum was chosen for thermal expansion matching to the robot's upper housing. To calibrate, we installed the rod in the fixture and focused the camera on the face of the front gauge block.
    
    We found it challenging at times to maintain perfect calibration of the installation fixture's focus position. In order to match the mounting of fiber positioner robots, we had designed the calibration rod with a similarly small flange (which sets focus position when the robot is screwed into the focal plate). However, a single positioner only undergoes one mechanical cycle in and out of the fixture, whereas the calibration rod experienced hundreds. Furthermore, we made the rod out of aluminum, to match the CTE of the robot's upper housing. The result was wear damage on the flange feature of the rod. Asymmetries in the damage profile on the rod's flange caused the focus point to vary slightly from day to day, depending on what rotational angle the rod had been installed in the fixture. These problems were identified midway through production, tracked, and mitigated (see \S\ref{sec:fiber_alignment_performance}).

\subsection{Positioner integration to petals}

Prior to installation into a petal, we visually inspected each positioner, cleaned it with ionized air, and commanded it to move. The motion was observed by eye as a check of basic function. After checking that the fiber had not been broken and could transmit light, the fiber was unspooled from its custom holster and fed through a given mounting hole in the petal. The fiber was laid as straight as possible on a long table behind the petal and then the electrical connector and pigtail wires were fed through the hole. Low-strength anaerobic threadlocker (Loctite 222) was applied to the M8.7 threads below the flat mating surface \figref{fig:sparkplug} and the positioner was lightly screwed in by hand. Prior to cure of the threadlocker, the unit was torqued in to 0.8 N-m, using a custom pin spanner wrench. We had designed 5 radial holes, \o\,1.6 x 1.0\,mm deep, into the flange of the theta bearing cartridge. The pin wrench reached around the fiber arm and upper housing of the robot to these holes, engaging them with a pin feature. Some complicated geometries were needed to ensure reliable engagement and capture of the flange, so we had the wrenches 3D printed in 17-4PH steel, hardened to H900, using the 20\,\micron~direct-metal laser sintering process from Protolabs, Inc. This proved strong and effective.

The need to access each positioner first by hand and then with the torque wrench meant that the positioners had to be installed serially, working across the petal. Each step of installation was tracked in a Google Sheets ``traveler", where we recorded information including positioner ID number, installation hole number, and confirmation of electrical function. Both motors ($\phi$ and $\theta$) were moved to confirm basic mechanical function. The positioner was then moved to a `parked' location, with the $\phi$ motor fully retracted and the $\theta$ motor in a neutral position, about halfway between its hard limits. Fiducials were also installed in this sequence, in a similar way to the positioners. We were able to install up to 50 positioners into a petal per day.

At the transverse electronics boards, fibers were laid in groups of 12 or 13 into plastic guides, which separated them at this point from their robots' corresponding electrical pigtails \figref{fig:petal1}. The 12 or 13 pigtails were secured in guides which snapped on top of the fiber guides. These guide stacks were mounted to the transverse board, adjacent to patches of 14 right-angle electrical connectors. After 25 positioners had been installed and connected, their fibers were gathered to the rear and fed into a black plastic conduit for protection. This conduit was eventually secured to the large radial aluminum plate which carries the radial electronics board. Once all positioners were installed, the PFAs were ready to be spliced to the fiber cables, as discussed in Poppett et al. 2022 (in prep).

\subsection{Inventory and quality control}
\label{sec:inventory_and_qc}

\begin{figure} 
    	\includegraphics[width=\textwidth]{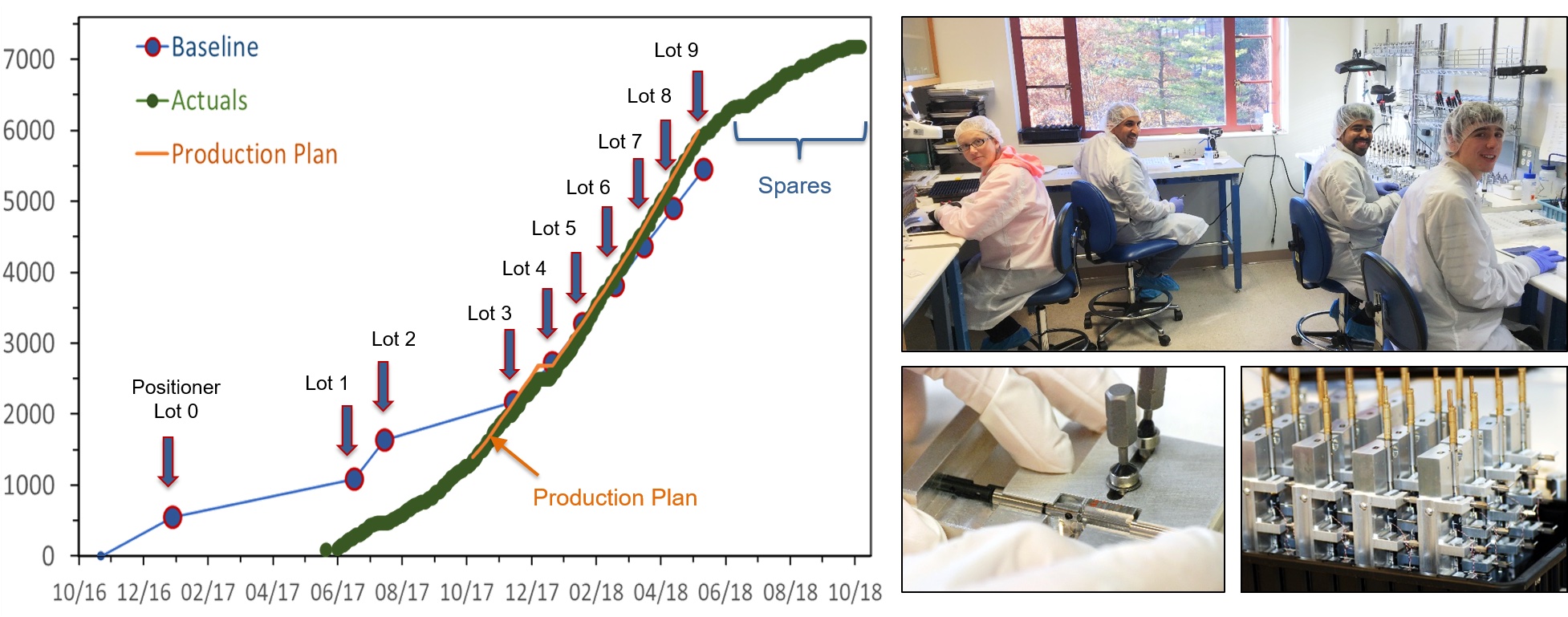}
    	\caption{\label{fig:positioner_production} Mass production of fiber positioners is plotted at left. We produced a total of 7,148 robots. A pre-production run of 440 units (not shown here) lasted from December 2016\,--\,March 2017. Tooling and procedures were finalized by the time of a May 2017 manufacturing readiness review. Full production began in June 2017. By fall 2017, all manufacturing startup issues had been resolved, and production accelerated to eventually exceed the planned cadence (500 units per month) by mid-2018. Upper right: Members of the production team at UM in the positioner assembly clean room. Lower right: Some subassembly bonding fixtures.}
    \end{figure}

We had a pre-production phase for robot assembly, during which 440 positioners were built at UM. During this period, we developed and refined our quality control (QC) protocols, procedures, tooling, and work spaces. We built and tested a SQL database for parts QC and inventory and to provide continuous production statistics. The database allowed us to identify production problems in a timely manner, and to mitigate these issues with the supplier or the assembly teams as soon as possible. With 32 parts per positioner (22 custom parts + 10 fasteners), and seven subassembly steps, the database helped maintain production rates with a reliable part count and clear QC interfaces. In addition to the SQL database, we made extensive use of Apple iOS applications and Google Sheets. These provided flexible and accessible real-time data management. A similar approach was used to track positioners as they were shipped, stored, and moved through the final integration steps and QC, including PFA installation and splicing. By the conclusion of the pre-production phase, given sufficient parts supply, the UM team was capable of a build rate of 50 positioners per day \figref{fig:positioner_production}. 

The tolerances required for the positioners and the irreversibility of glued assemblies necessitated 100\% QC on each part, and regular evaluation of tooling accuracy. We put considerable effort into developing QC processes that were satisfactorily prognostic while remaining as simple as possible. Typical inspection tools were dial indicators, granite surfaces, machined gauges, and V-blocks. Visual inspections were performed on each part for burrs and defects, on each subassembly for adhesive under- or over-fill, and for screw engagement. Each bearing was manually turned to feel for friction, and a microscope was used for a final inspection \citep{leitner18}. Each subassembly was individually qualified to minimize part loss.

In addition, the subassembly sequence was optimized to minimize loss of the most expensive and long lead time components. During the pre-production phase we determined our real-world production yields, and this defined the number of parts required to yield at least 6000 production positioners (sufficient for 10 production petals and 2 spares). Our modular assembly procedure (having clearly defined, discrete subassemblies, and continuous training of staff) allowed us to continue steady production in circumstances when particular component deliveries were delayed or available parts were non-compliant. We would flexibly shift our workers to produce more subassemblies of the parts we had in stock.

During the pre-production phase, we identified an issue where motor gear trains could be compressed to a state of inoperability by a slight, nearly zero force, axial motion of the output shaft. This was found by inspection of CT scans of the positioner done at Los Alamos National Laboratory. Working closely with our supplier we modified the motor assembly specifications and adjusted our assembly technique to bias the assembly in order to reliably achieve the required gear spacing of the motor assembly.

For a one-time mass production run, the rate of assembly and the yield were excellent. UM's diverse team of students, technicians, and engineers produced 150\,--\,200 units per week. Yields on the seven sub-assemblies ranged between 96\,--\,100\%. For complete positioners, the yield was $98\%$. The biggest challenges were (1) resolving the early problem with the motor gears that impeded our pre-production run, and (2) maintaining sufficient parts inventory to keep the production rolling.

\subsection{GFA integration to petals}
\label{sec:GFA_integration}
The GFA is mounted such that the planar active sensor area intersects the curved focal surface (as shifted by the GFA filter), with 1/2 of the sensor area above and 1/2 below focus. In the nominal mounting position, deviations from focus due to this mismatch between the planar CCD sensor and the aspheric DESI focal surface are $\sim$\,48\,\micron~peak-to-valley across the sensor. We aimed to mount all sensors to within 50\,\micron~of this nominal position.

In each camera, IFAE measured the CCD sensor plane optically, and aligned it to within 20\,\micron~with respect to the mechanical base. Between the petal and the camera, we placed an intermediate mounting plate, to which the camera would interface. The mounting plate was attached to the petal with three M3 screws. A shim stack beneath each screw provided the necessary adjustment of piston, tip, and tilt.

We aligned the mount plates at LBNL, long before any actual GFA cameras had been built. The locations of the plates were measured relative to the three tooling balls on each petal with a touch CMM (see \S\ref{sec:final_petal_metrology} and figure \ref{fig:structure2}). Shims were iteratively changed to achieve correct spatial alignment. It typically took 2\,--\,3 measurement and adjustment iterations. The worst case error among the mount plates was 7\,\micron.

Upon receipt of cameras at LBNL, we installed two GIFs (fiducials with mount features specific to the GFA housing). A shim stack at each GIF mounting location allowed us to independently align them to the focal surface. Prior to installing cameras on petals, we measured the GIF dot positions relative to the camera sensor. For this measurement we built a custom metrology rig, consisting of a spot projector mounted on X-Y motorized stages, together with a beam splitter and a Thorlabs DCU223M camera. First we would image the GIF pinholes, then translate the stage to a pattern of positions over the CCD. The spot would be projected through the GFA filter and read out with the camera. Using this rig, the locations of the GFA sensors were measured relative to the GIF pinholes with a lateral accuracy of 5\,\micron.

The metrology rig was also capable of imaging the reflection of its own projected spot. We mounted the GFA on a separate, orthogonal stage with a manually operated micrometer, to adjust the focal distance between the GFA and spot projector. Around the GFA, we mounted and precisely measured three gauge blocks, to establish a datum plane. Translating the camera with this stage, we could determine the relative Z position of key features of the camera as determined either by the image of GIF dots, the spot as read out on the sensor, or as reflected off of exterior mechanical datum features.

After integration of robots, fibers, and electronics, the GFA was mounted on the petal. The cooling fan and duct were installed, the camera was connected to the GXB and Petal Controller, and a grounding wire was attached between the GFA and the petal structure. Dark current and noise measurements were repeated after installation to confirm that no grounding issues were present, keeping care that a cover was always on the GFA sensor to avoid any saturation events. We measured an average read noise of 18 e$^{-}$ RMS with a gain of 3.7 e$^{-}$/ADU, within the requirement of 25 e$^{-}$/pixel.

\subsection{Metrology of completed petals}
\label{sec:final_petal_metrology}

The DESI operating model relies on being able to link the reference frames of the GFA cameras and the fiber positioners, enabling alignment of the fibers with targets based on their RA/DEC location \citep{honscheid18}. The illuminated fiducials (FIFs and GIFs) provide the mutual reference points between them. In the previous section we described how we linked the location of the GFA sensors to the GIFs. Once all positioners and the GFA were installed on a petal, we measured the locations of the FIFs and GIFs relative to each other.

For this measurement, we used the ZEISS CMM at LBNL, which has an optical probe in addition to the standard touch probe. The lateral position of the optical probe was cross-calibrated to the touch probe by both imaging and touching a reference ball. To cross-calibrate the focal position of the optical probe, we used a flat surface on the GFA housing that could both be touched and optically imaged. Due to possible tilts in the GFA, calibration at this point induced an error in the focus measurements of $\pm$10\,\micron.

Our operator guided the CMM to each of 4 pinholes on each fiducial (FIFs and GIFs), and used the optical probe, to measure their X, Y and Z locations. The four corners of the GFA were measured with both the touch and optical probes, and the face of one fiducial was additionally checked with the touch probe. These measurements were repeated four times on each petal. To get statistics on fiber focus, we also backlit all fibers and used the optical probe to measure the locations of $\sim$\,50 fibers, repeating for a total of 3 measurements each.

\subsection{Full petal tests in the lab}
\label{sec:labtests}

Following fiducial and fiber metrology, final functional tests and positioner accuracy tests were completed in the lab at LBNL before shipping to Kitt Peak. The main performance metric we wanted to confirm was that we could place the fiber tips within 10\,\micron\,RMS of their nominal lateral target positions. This was the first time that we were able to test much of the system software, relying on interaction between the petal controller, move scheduling, PlateMaker, FVC centroiding and spot matching, and fault management.

\begin{figure} 
    	\includegraphics[width=\textwidth]{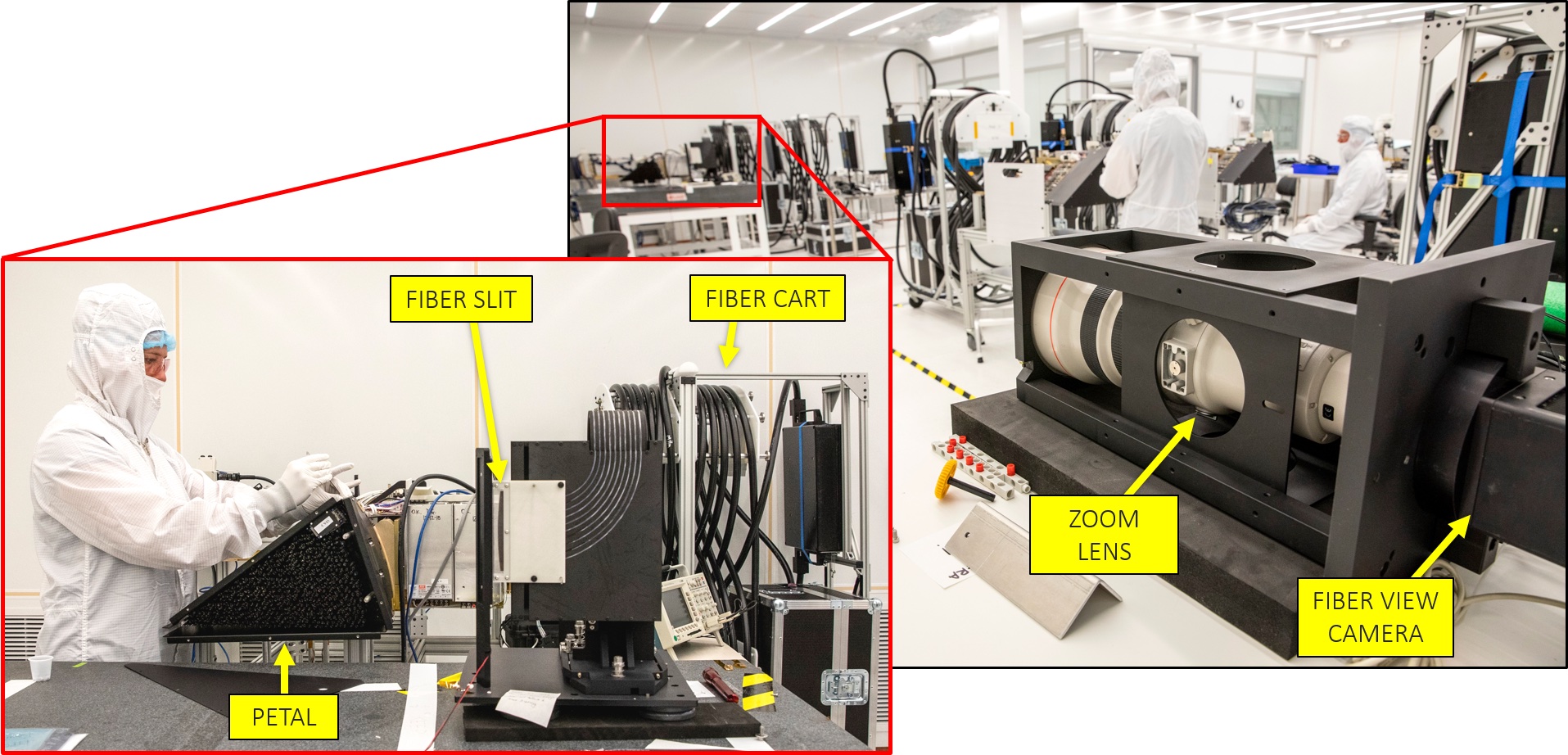}
    	\caption{\label{fig:petaltest_layout} Petal testing in a clean room at LBNL. The metrology camera sits $\sim$\,12.5\,m from the petal, which is connected to a fiber cable on an adjacent cart. The fiber slit is illuminated by an LED. The positioners lay horizontally, pointing at the FVC metrology camera. Credit: Marilyn Chung/LBNL}
    \end{figure}

The tests took place in a Class 1,000 clean room at LBNL. The petal was set on an integration cart with the positioners oriented horizontally \figref{fig:petaltest_layout}. The fiber slit was installed in a temporary illumination system and an FVC was positioned at the other end of the room, approximately 12.5 meters away. When a petal was first set up, the fiducials were used by the FVC and PlateMaker to establish a distortion map of the FVC lens and solve for the focal plate coordinate system. When the positioners were backlit with the illuminator, the FVC could measure the centroid of each positioner. Given the solved focal plate coordinate system, the positioners were identified by Spotmatch. A challenge in laboratory testing was that without the DESI Corrector, the chief rays of the fibers along the curved focal surface did not point directly at the camera. Depending on how the petal was oriented relative to the FVC, some fibers and fiducials appeared dimmer than others. Once all fibers were correctly identified by Spotmatch and the plate scale had been determined, the positioners were calibrated by measuring arcs of points about their $\theta$ and $\phi$ axes.

During the lab tests, the anticollision software was not yet mature enough to incorporate. We kept each positioner within a restricted patrol envelope, to ensure that it could not overlap with its neighbors. We tested as-installed positioner performance by moving all robots along square grids of 24 targets, 5 x 5 with no center target. The grids were $\sim$\,3 x 3\,mm in size. We tested up to 3 move iterations.

These were our first tests moving all 502 robots in a petal simultaneously. We found several positioners with communication errors that would bring down the CAN network for all devices on their bus. These were identified and unplugged from their respective buses. The positioners remained mechanically in place, and their fibers were left intact. Several robots started to show performance issues that hadn't been identified during production QC testing. Some of these motors, mostly $\phi$, were found to be ``sticky'', meaning that their motion was slow and lurching. Other motors were seen to not move at all. Only grade A and B positioners (see \S\ref{table:pos_grading}) had been installed on the petals, therefore some deterioration had occurred. The root cause, discussed in section \ref{sec:pos_util}, was not uncovered until 2021, two years after the petals had been installed on the telescope.

Despite these issues, overall each petal performed excellently during the lab tests, with $>$~97\% of all installed positioners meeting requirements. In total, prior to shipment, we found 17 broken fibers. We disconnected 32 positioners from the petal controller, mostly due to being unresponsive or causing CAN bus issues. There were 96 positioners that were disabled but not disconnected because they did not meet performance requirements. For the positioners that weren't disabled, their maximum blind move accuracy was $\leq$200\,\micron~and their RMS error after the third correction move was $\leq$5\,\micron~over the 24 grid moves.

Prior to shipment, each petal underwent functional testing. This test set a baseline for the expected functionality of the petal and was repeated several times: pre-installation on the shipping crate, post-installation on the shipping crate, post-shipment to Kitt Peak, and post-installation on the telescope. The test included a grounding checkout, confirmation that all telemetry was transferred, and that the fans were fully functional. To test functionality of the GFAs, we took a zero second exposure to measure the noise levels. During the test, all positioners and fiducials were powered on, and the positioners were moved in both axes. No positioner accuracy tests were performed. 

\subsection{FVC testing}
We tested the FVC feedback system in several phases, prior to the DESI focal plane installation (summer 2019). Early on, the camera was tested in the lab at Yale, pointing the camera at a well-measured plate of fixed, backlit fibers from a representative distance \citep{baltay19}. These tests demonstrated feasibility. In 2016, we successfully tested the system in-situ at the Mayall Telescope, as part of ProtoDESI. In this test, the FVC imaged a set of three fiber positioners and 13 fiducials, looking through the pre-DESI prime focus corrector \citep{fagrelius18a}. In spring 2019, the system was again tested, but now looking through the DESI corrector, at the Commissioning Instrument and its array of fixed fiducials \citep{ross18}.

In parallel with these campaigns, we built significant experience with smaller-scale FVC systems on the robot test stands, as part of our robot development and QC testing. In these stands we tested fiducial reliability and debugged the feedback side of the positioner control interface.

\section{Installation}\label{INSTALLATION}

We shipped petals individually from Berkeley to Kitt Peak, and installed them one-by-one directly into the FPR on the telescope. We decided upon this approach (rather than integrating petals to the FPR in the lab, then shipping and installing them as a single massive unit) to reduce both schedule and technical risks. As opposed to a more monolithic installation strategy, our approach entailed a higher total number of critical lifts and shipments, but allowed simpler fixtures, repetition of procedures, elimination of critical flipping manuevers, reduced risk per shipment, and created logistical flexibility in scheduling of deliveries, functional tests, and integration.

During the shipping and installation phase of the project, careful tracking of high-value subsystems and tooling was critical. The success of this approach relied on significant travel of the FPS engineers and scientists to Kitt Peak and thoroughly-reviewed procedures for handling the hardware.

\begin{figure} 
	\includegraphics[width=\textwidth]{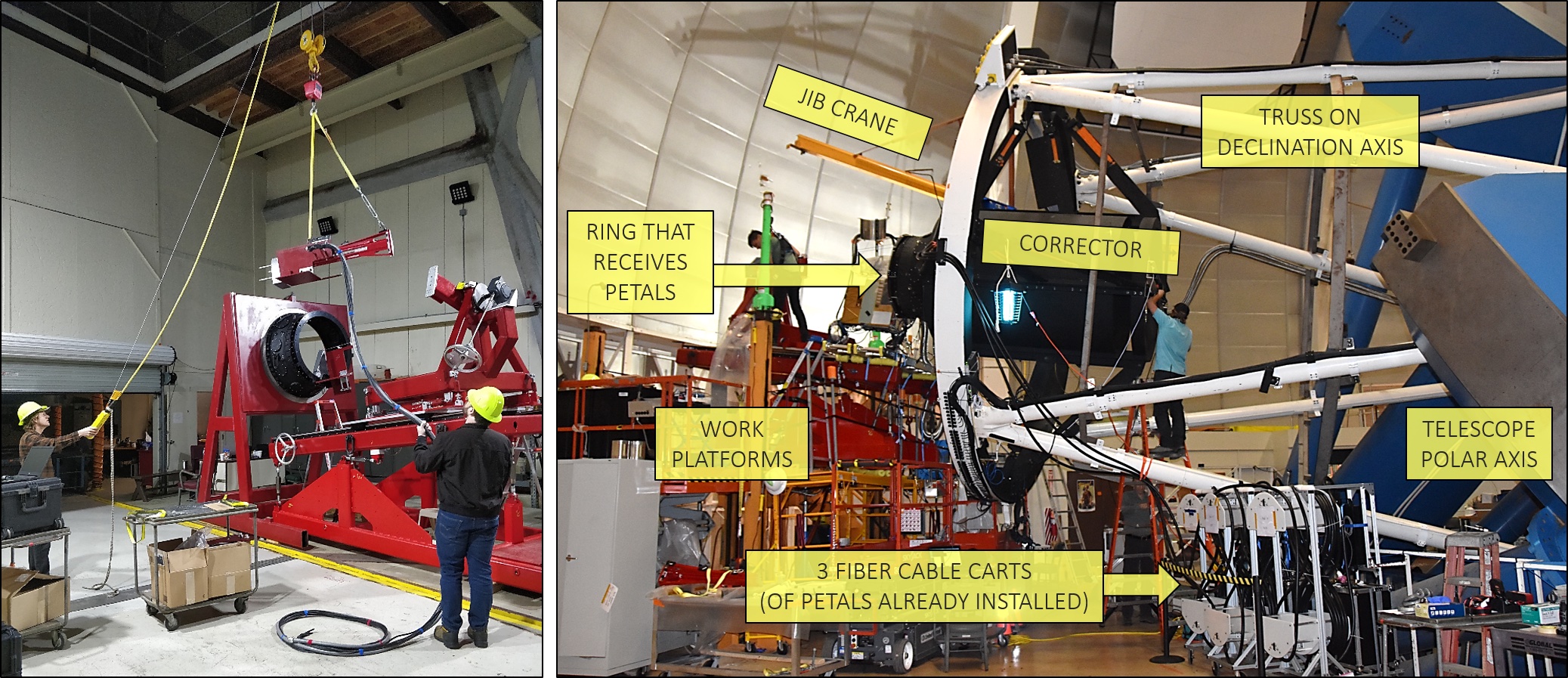}
	\caption{\label{fig:install_overview} At left is a mock installation trial in the Mayall Garage. The FPD (black ring) is mounted on a steel A-frame with a height, inclination, and bolt interface matching the corrector barrel on the telescope (when mounted on the southeast work platform). At right is an overview picture of the real installation, taken during insertion of our third petal to the telescope. The installation is done with the telescope parked at the South-East (SE) platform of the dome.}
\end{figure}

We shipped petals to Kitt Peak beginning in spring 2019. At the same time, we practiced mock test installations of the FPD and petals with Kitt Peak staff in the Mayall garage. The FPD was installed to the telescope in June 2019, followed by the ten petals, complete by the end of July \figref{fig:install_overview}. The thermal system was subsequently installed in early August \citep{besuner20}.

\subsection{Shipping to Kitt Peak}

\begin{figure} 
	\includegraphics[width=\textwidth]{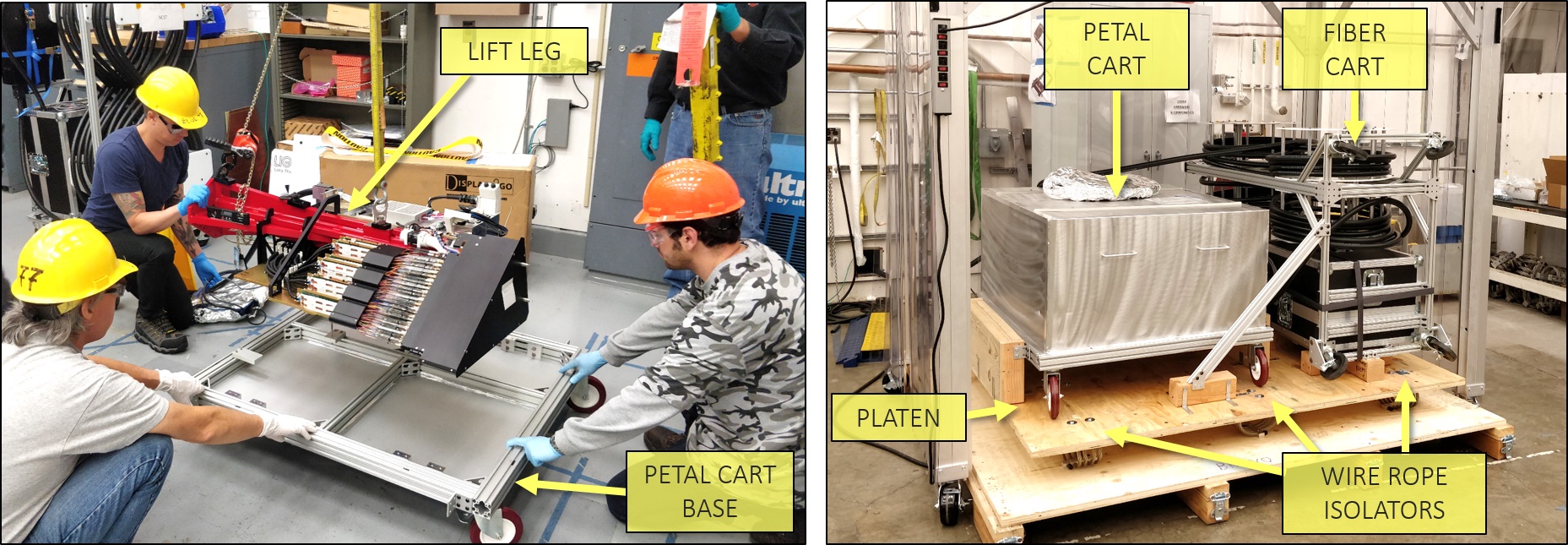}
	\caption{\label{fig:petal_shipping} Petals are transferred from their assembly carts to lower profile shipping carts (left) via crane procedures. At right, the wood base and vibration-isolated internal platen of a petal shipping crate are shown, with a petal cart and its fiber cart mounted.}
\end{figure}

We shipped petals in individual crates from LBNL direct to Kitt Peak. Each crate contained two carts \figref{fig:petal_shipping}, one for the petal and the other for that unit's permanently-attached 45\,m fiber cable and slithead. After completing the acceptance process for a given petal assembly, we transferred it from its integration cart to a shipping cart with stiffer members, a protective sheet metal enclosure, and a lower center of gravity. The permanently-attached fiber cable and slit remained on their separate fiber cart. We then wheeled the two carts up a ramp onto a vibration-isolated platen on the petal shipping crate. We rotated the fiber cart 90\degree~for a lower center of gravity during shipping. All lifts, rotations, and transfers were done according to well-rehearsed, detailed procedures. Both carts were screwed down to an internal wood platen, mounted to the crate base via 12 wire rope isolators (IDC SB16-610-04-A). The crates were sized to fit 2 at a time in an enclosed air-ride truck.

Once the shipping crates arrived at Kitt Peak, the petal and attached fiber cable assemblies were removed from the crates, and remained on their respective pairs of carts until installation to the telescope. In this configuration, the petals all underwent functional testing, as described in section \ref{sec:labtests}, which confirmed that no items were damaged during shipment.

\subsection{Installation to telescope}

\begin{figure} 
	\includegraphics[width=\textwidth]{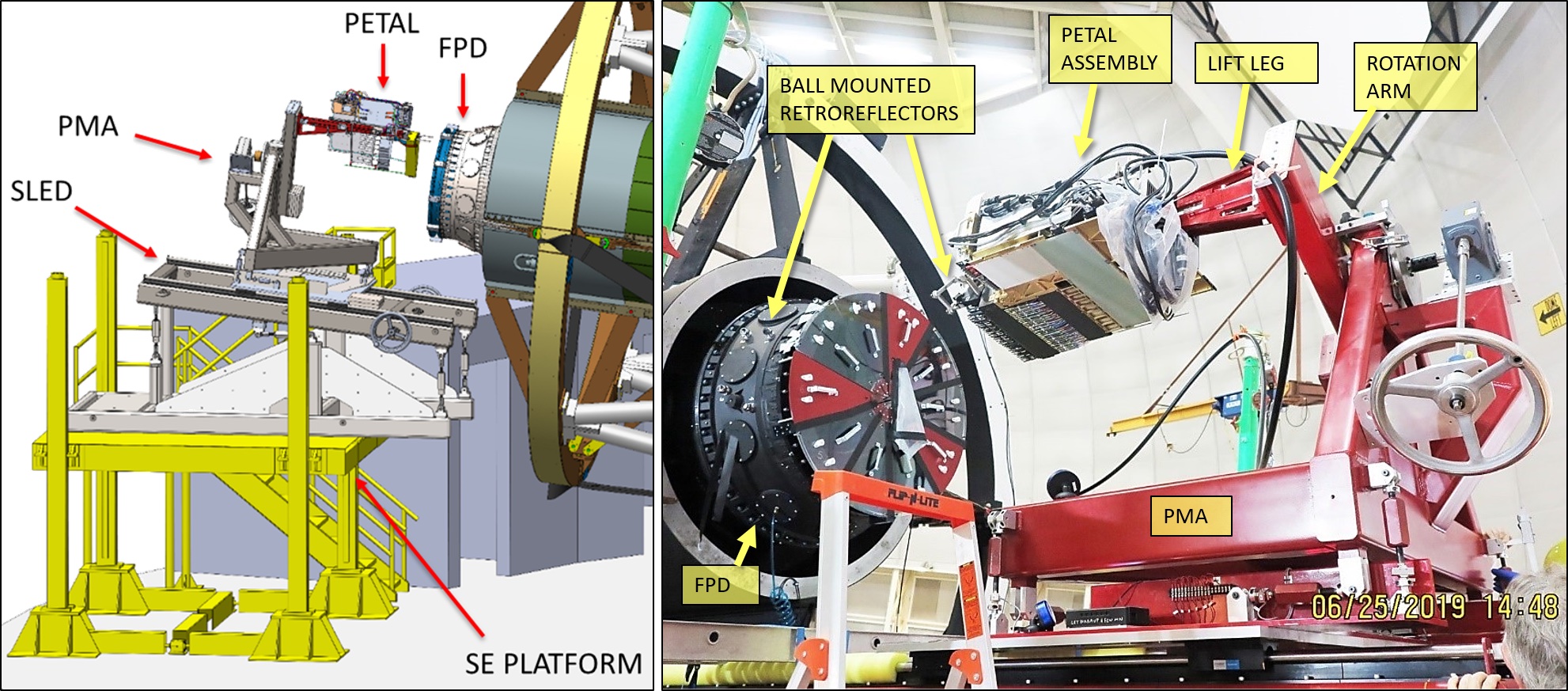}
	\caption{\label{fig:install_fixtures} Illustration and photograph of the system with which we aligned and installed petals at the Mayall Telescope. The system is also used for removal of petals, which was done in summer 2021 to install some electrical upgrades.}
\end{figure}

We inserted  the petal assemblies one at a time into the telescope on a large steel fixture with 16 degrees of freedom. The system had three major components \figref{fig:install_fixtures}: the South East (SE) Platform, Sled, and the Petal Mounting Assembly (PMA). The SE Platform is an existing platform at the Mayall Telescope's south east annex and its height can be adjusted coarsely. The Sled includes one pair of linear rails and an acme drive screw. The rails platform is mounted to six manually-adjustable struts, used for alignment of the rails to the Corrector Barrel's central axis. The PMA includes the rotation arm used to address the 10 petal positions. The arm has a main rotation axis and a rotary trim stage at the end where the petal mounts. The PMA rides on the Sled rails, and is mounted on six manually-adjustable struts, for alignment of the central PMA axis to Barrel and Sled.

Alignment measurements were done using a laser tracker \citep{shourt20}. Ball mounted retroreflectors (BMRs) were magnetically attached to kinematic nests, permanently glued on the FPD. We had previously measured these ball nests accurately with respect to the Barrel central axis on a CMM in the lab. This established the Corrector central axis. We then aligned the Sled and PMA to this axis.

\begin{figure} 
	\includegraphics[width=\textwidth]{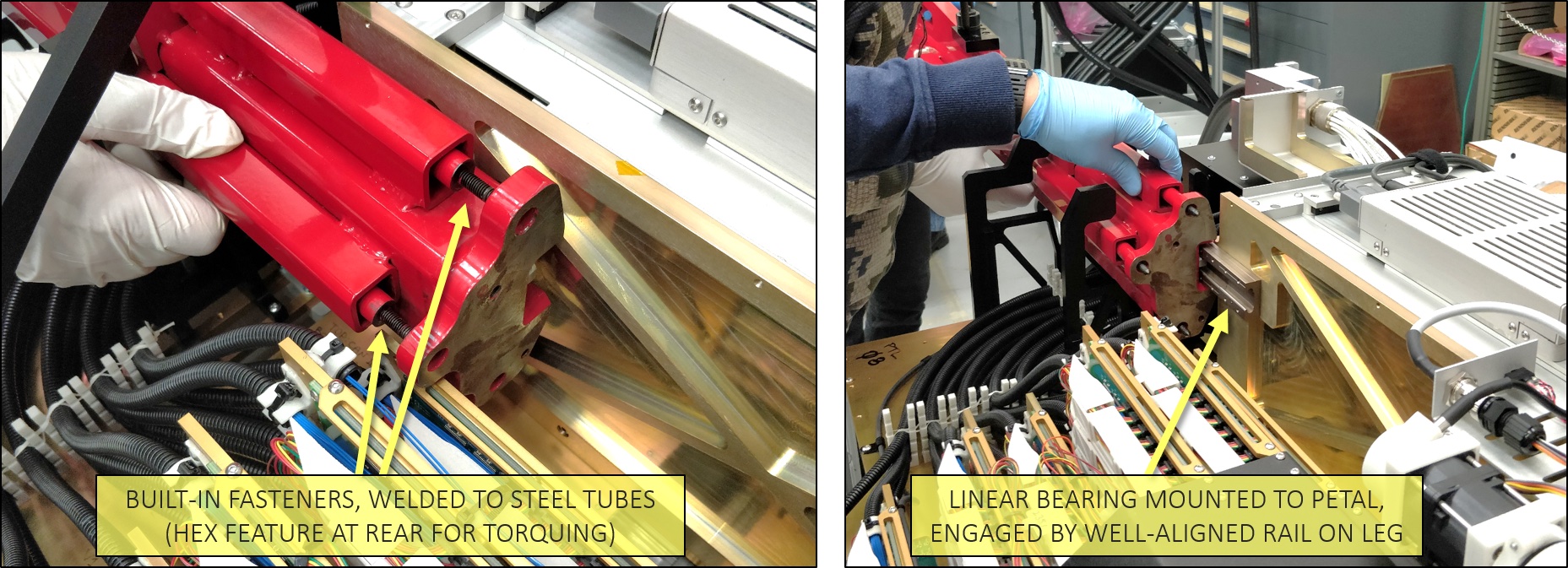}
	\caption{\label{fig:petal_leg} Petals are handled with a steel lift leg, with attachment points for horizontal (crane) and cantilevered (insertion fixture) attachment.}
\end{figure}

The petal assembly is lifted by a steel `leg' \figref{fig:petal_leg}. The leg inserts from the rear of the petal assembly, guided by a rail/bearing, and attaches rigidly to the petal structure. The leg has lift points at both its own center of gravity and that of the petal assembly. At its base, the leg is bolted to the end of the PMA rotator arm. We made a special holder for small ball mounts which we could kinematically attach to the two insertion guide spikes that project from the front of each petal \figref{fig:structure1}. Measuring the petal BMRs with respect to those on the FPD, we could adjust the struts such that the items were aligned within 25\,--\,50\,\micron. We then drove the PMA down the Sled bearings with a manual hand wheel, right-angle geared to an acme screw.

Once the petal had been aligned via laser tracker measurements, we removed the BMRs from the guide spikes and drove it forward into the FPR. The first parts of the petal to contact the FPR are the stainless steel guide spikes. We sized these relatively loosely: 0.2\,mm smaller on the diameter than the corresponding steel bushings in the FPR. This gave us some play for misalignments, in particular the dynamic shifts that occur as loads move across the SE work platform, and as the whole platform moves with respect to the telescope. We included load cells (Omega LC703-1K, read out with a Graphtec GL220) in series at each of the six PMA struts, so that we could immediately identify hard contact conditions. We would drive the petal forward until encountering a full-stop condition against the FPR ring. This we could quantitatively determine by the axially-oriented load cell. For fine adjustments of the PMA struts during this final insertion step, our feedbacks were the load cell measurements, feeler gauges, and visual inspection. 

\begin{figure} 
	\includegraphics[width=\textwidth]{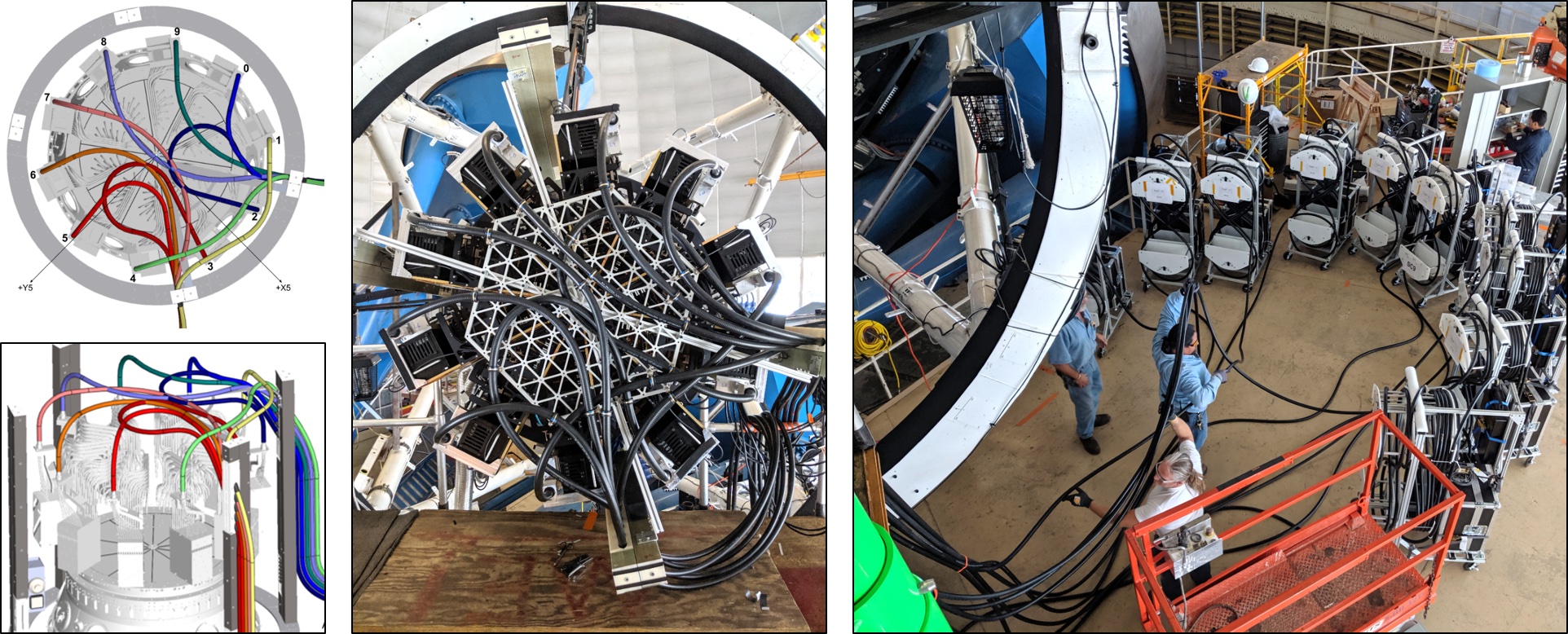}
	\caption{\label{fig:fiber_cables} Fiber cables are lashed with hook-and-loop ties to an aluminum hexagonal grid at the rear of the focal plane assembly. The grid is supported by the thermal enclosure, and sufficient slack was designed into the cable routes to ensure free hexapod motion. At right, a view of the 10 cable carts on the C-floor, after completing installation of petals, but prior to routing cables along the truss and lowering each slithead to the spectrograph enclosure.}
\end{figure}

At all times during shipping, lifts, and insertion, the 45\,m fiber cable and slithead (the component at the other end of the cable, which mounts and linearly aligns the fibers in each spectrograph) are permanently connected to each petal assembly. Cable routing was carefully planned in CAD, including both the final configuration when installed, as well as how to manage the fiber carts on the C-floor during the installation process \figref{fig:fiber_cables}.

\subsection{Functional Verification}
After all ten petals were installed, we completed functional tests on each petal, confirming no grounding issues were impacting the function of the GFAs or petal electronics. Then began the work of getting Spotmatch to work with the PlateMaker software to correctly identify each fiducial and fiber, given the distortion pattern of the FVC lens and corrector. For this to work, fiducials from all 10 petals had to be illuminated, but we backlit only one petal at a time to avoid confusion between fibers on adjacent petals. We tuned the power to each fiducial and the backlight LED so that all points of light had a similar signal-to-noise on the FVC.

Once all fibers/positioners were correctly identified by Spotmatch, we repeated the calibration and performance tests that we had done in the laboratory prior to shipment (see \S\ref{sec:labtests}). For most of this testing, the telescope was set at zenith so that the positioners pointed downward, parallel to gravity. It was found that some positioners that had performed well in the laboratory were no longer performing well enough for use on-sky. All positioners that did not meet requirements were parked and identified in a database. The majority of the positioners ($>$~85\%) worked as expected and were tested in a variety of dome conditions. We measured significant in-dome turbulence due to temperature gradients during the day, exacerbated by various fans in the dome. The most stable conditions occurred in the evening with the dome open, after the telescope had come into equilibrium with the ambient air. Final characterization of the positioners was completed in this configuration. 

Since this was the first time that more than one petal was tested at the same time, the team was careful to incrementally test the capacity of the thermal system to deal with the heat load from the positioners. A minimum time between successive moves was established so that heat from the motors could be removed from the system before the subsequent move. While fault management was in place at this time, the thermal system hardware interlocks had not yet been fully installed, so vigilant supervision of the temperature of all positioners took place whenever there was power to the focal plane. 

The final step before integrating the focal plane into the larger instrument control framework was to complete a positioning accuracy test on all operable positioners simultaneously. The positioners that met accuracy requirements were then enabled to receive targets from the ICS. Excluding outlier targets (nominally defined as the the worst 0.7\%), the max blind move accuracy was 120.4\,\micron~and the RMS error was 4.7\,\micron, for the 87\% of installed positioners that were enabled at that time \citep{fagrelius20}. 

\subsection{Commissioning}
On-sky testing with the focal plane began first with the GFA cameras and stationary robots in late October 2019. Once the functionality of guiding from the GFAs was verified, the commissioning team was able to build upon work characterizing the telescope that had begun on the DESI Commissioning Instrument \citep{ross18}. During this period of observing, with some spectrographs enabled, we did coarse telescope dithering around bright targets to assess our initial mapping from focal plane to sky coordinates.

The first on-sky fiber positioning with the full control loop occurred in early December 2019, with the robots operating within a restricted patrol radius. We avoided any use of the anticollision algorithm until we had fully characterized the robots and debugged the feedback system. Initial sky to fiber mappings were offset by up to 10 arcsec and little flux made it down the fibers when targeting fainter objects. Operation of the robotic positioners enabled fine fiber dithers around bright objects to correct smaller scales of the focal plane to sky coordinate mapping. By March 2020 the mapping had been sufficiently characterized and verified, and the software was sufficiently robust, such that we made the FPS available to the operations team, to begin the spectroscopic measurements of the Survey Validation (SV) campaign.

We measured on-sky fiber positioning accuracy with a series of ``dither tiles'', in which astrometric standard stars are offset by varying, randomized distances between each exposure in a 13-tile sequence. Treating the spectrographs as photometers, these random offsets produce flux profiles of numerous bright stars simultaneously in the field of view. The peaks of the flux profiles give true positions of the stars as projected on the focal surface. We repeated these dithers at several telescope pointings, exploring most of the available parameter space. Using this method, errors in focal plane metrology and the FVC distortion model were identified. On-sky positioning accuracy, originally $\sim$\,10\,arcseconds for our first targeting attempts, improved to $\sim$\,0.1\,arcseconds (DESI Collaboration et al. 2022 in prep.) after addressing these error sources. 

Following DESI's shutdown for the COVID-19 global pandemic, which lasted from March 16 to November 20, 2020, the anticollision move scheduling code (\S \ref{sec:move_scheduling}) was tested on the full instrument, and demonstrated to perform as required. (We had tested it thoroughly on spare petals in the lab, prior to this.) We progressively released the petals for operation of robots over their full patrol envelopes. Petal 0 was allowed full reach starting on January 29, 2021, followed by petals 2, 4 and 5 on February 6, 2021, petals 7 and 8 on February 28 2021, and all petals on March 6, 2021. This marked the passing of a fully functional focal plane to the survey operations team.

\section{Performance}\label{PERFORMANCE}
Performance of the FPS is described in six sections below, covering positioner accuracy (\S\ref{sec:positioner_accuracy}), fiber alignment (\S\ref{sec:fiber_alignment_performance}), positioner utilization (\S\ref{sec:pos_util}), reconfiguration time (\S\ref{sec:reconfig_performance}), guider performance (\S\ref{sec:GFA_performance}) and fiber view camera performance (\S\ref{sec:fvc_performance}).

\subsection{Positioner accuracy}
\label{sec:positioner_accuracy}

\begin{figure} 
    	\includegraphics[width=\textwidth]{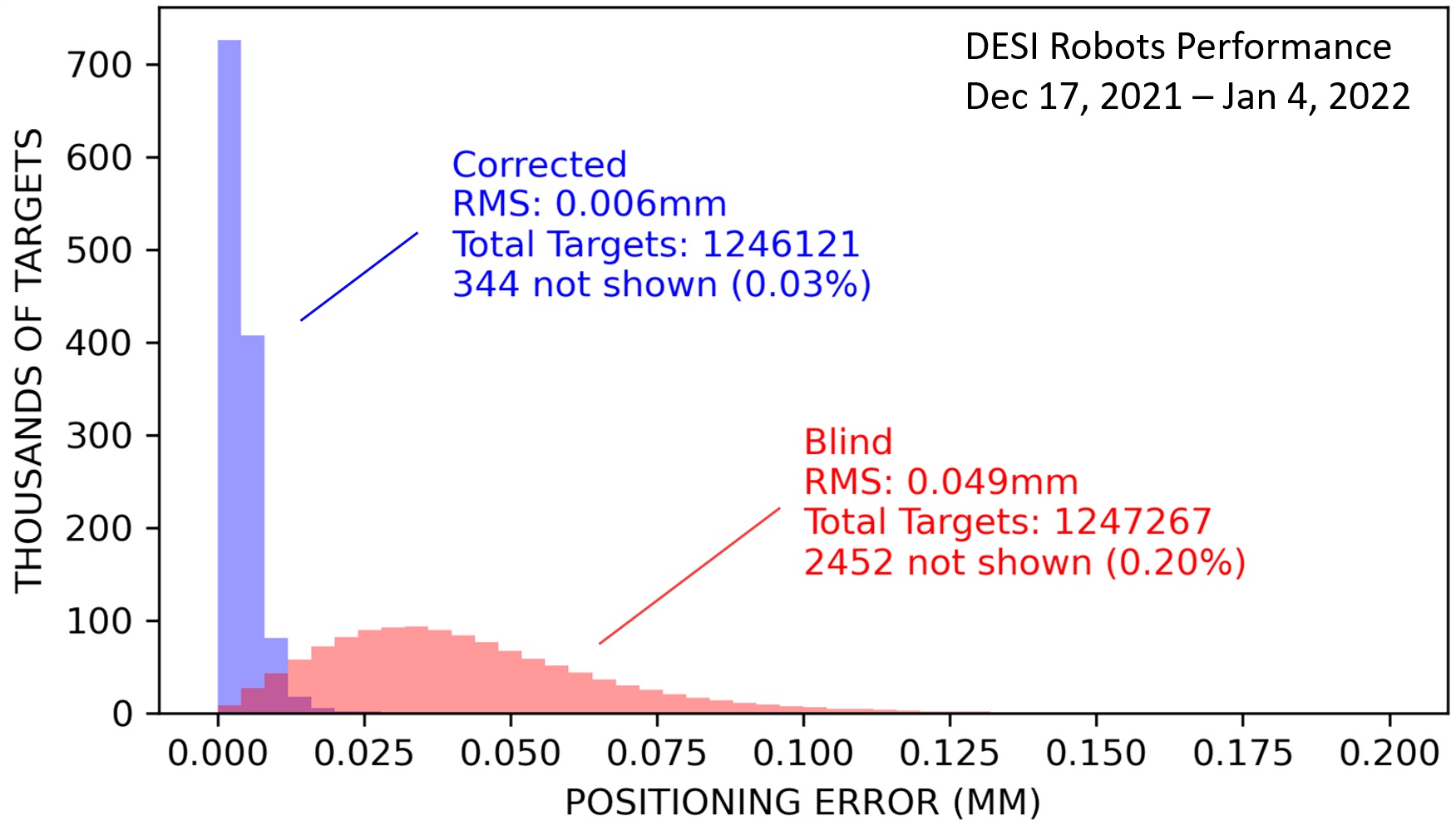}
    	\caption{\label{fig:positioner_accuracy_with_turbcorr} Targeting accuracy for DESI fiber robots over a 2.5 week period of survey operations, during which the Focal Plane System acquired more than 1.2 million targets. During this period, we operated with code that corrects some of the dome turbulence effect in the FVC measurement. This reduced the positioning error to 6\,\micron\,RMS (prior performance was 9\,\micron\,RMS, as shown in figure \ref{fig:positioner_accuracy_hist}).}
    \end{figure}

    The DESI requirement for positioner accuracy is that the fibers be aligned with their commanded target locations as-measured by the FVC within 10\,\micron\,RMS. Figure \ref{fig:positioner_accuracy_hist} plots fiber positioning accuracy over three months of survey operations, during which over 7.8 million targets were acquired. Accuracy of the system was 9\,\micron\,RMS, with most of the error due to measurement effects (spot centroiding $\sim 3$\,\micron~and dome turbulence $\sim$\,3\,--\,8\,\micron). Accuracy was subsequently improved to 6\,\micron\,RMS \figref{fig:positioner_accuracy_with_turbcorr} after we implemented code to subtract some of the turbulence out of the measurement (see \S \ref{sec:fvc_performance}).

    Dither tests, as described above, tell us the overall positioning accuracy, since they include corrector, telescope, and atmospheric effects. However, they consume significant telescope time, and so were largely a one-time commissioning task. Using the dither mode, we measured total fiber positioning accuracy on sky to be 0.19 arcsec (13\,\micron) 2-D RMS. These results were valid to high fidelity and were repeatable across several nights. This achieved the design goal for positioning accuracy, losing less than 3\% of the light for DESI Main Survey Emission Line Galaxies (ELGs) under nominal conditions of 1.1 arcsec FWHM atmospheric seeing.

\subsection{Fiber focal and angular alignment}
\label{sec:fiber_alignment_performance}

Over the full array, our alignment process for fiber tips (see \S\ref{sec:fiber_integration}) achieved a focus with a standard deviation of $\sigma = 14.3$\,\micron, corresponding to a throughput loss of $\sim 0.1\%$. In addition to this scatter, we saw a larger systematic offset, likely caused by wear on calibration rods (see \S \ref{sec:fiber_integration} and figure \ref{fig:fiber_alignment}) in our fiber integration fixtures. Measured over 5,526 units, the fiber tips had a mean position -48.6\,\micron~below the nominal focus depth. Fortunately, we had designed sufficient adjustability into our structure, and our QC measurements identified the problem early enough in the process so that we were able to remove this offset by shifting the mount position of the petals themselves. We made this adjustment at the captive shim stacks between the petals and forward lip of the FPR (see \S\ref{sec:fp_structure}, \ref{sec:petal_align}). Simultaneously we counter-shifted the GFA mount plate (\S\ref{sec:GFA_integration}) on each petal in the opposite direction, to keep the cameras in focus.

A study of the impact of fiber angular error on total throughput showed errors within $\pm0.5$\degree~would incur loss $<$~1\%, and within 1\degree~would incur loss $<$~3\%. In production, we measured the as-built angle of tilt between ferrules and $\phi$ bearing cartridges, which controls rotation of the fiber about the eccentric axis. For the 2,023 units we measured, the mean absolute error was 0.14\degree, well within our 0.3\degree requirement. The RMS error was 0.25\degree, corresponding to $\sim$\,0.2\% throughput loss. Six outlier units (0.3\%) had tilt error $>$~1.0\degree, with one fiber strongly misaligned at 6.48 deg. Over the other 2,017 units, the RMS error was 0.17\degree. 

\subsection{Positioner utilization}
\label{sec:pos_util}

Not all fibers are usable for survey operations. As of December 2021, DESI operates with 85.3\% of its fiber robots acquiring targets, 14.3\% disabled due to mechanical or electrical problems, and 0.4\% having damaged or broken fibers (see Table \ref{tab:positioner_utilization}).

\begin{table}
    \centering
    \begin{tabular}{l r r l}
        \toprule
        \textbf{Status} & \multicolumn{2}{l}{\textbf{Count}} \\
        \midrule
        Nominal & 4280 & (85.3\%) \\
        \midrule
        Improper motion & 649 & (12.9\%) \\
        \midrule
        Communication failure & 70 & (1.4\%) \\
        \midrule
        Damaged fiber & 21 & (0.4\%) \\
        \bottomrule
    \end{tabular}
    \caption{\label{tab:positioner_utilization} Fiber positioner real-world utilization. Counts are as of Dec 18, 2021. }
\end{table}

The original survey plan was to configure a minimum of 8\% of the fibers in each exposure on blank sky positions for spectroscopic sky calibration. Specifically, each petal is to have a minimum of 40 fibers with at least one fiber from each slitblock looking at sky (there are 25 slitblocks per petal). To mitigate the loss of function from disabled positioners, we use their fibers for this purpose whenever they fall on blank sky positions, which is typically 60\% of the time. In order to provide sufficient sky positions in each petal (and for each slitblock), it is often necessary to assign additional fibers from enabled positioners to blank sky. The net result is that for any given exposure, 9\,--\,13\% of all fibers are used for sky measurements, with 1\,--\,2\% of these being functional positioners.

A significant fraction of the positioners have motors which move at consistently smaller angles than commanded, often in a halting manner rather than smoothly and continuously. In early 2021 we investigated a number of root cause hypotheses, and eliminated those related to both electrical (such as grounded wire or failed solder joint) and externally-determinable mechanical failure (such as broken bond joint or slipped set screw). Our remaining hypotheses related to internal behavior in the gearmotor. During a summer 2021 shutdown period, we removed 18 failed robots from the focal plane for direct diagnostics.

\begin{figure} 
	\includegraphics[width=\textwidth]{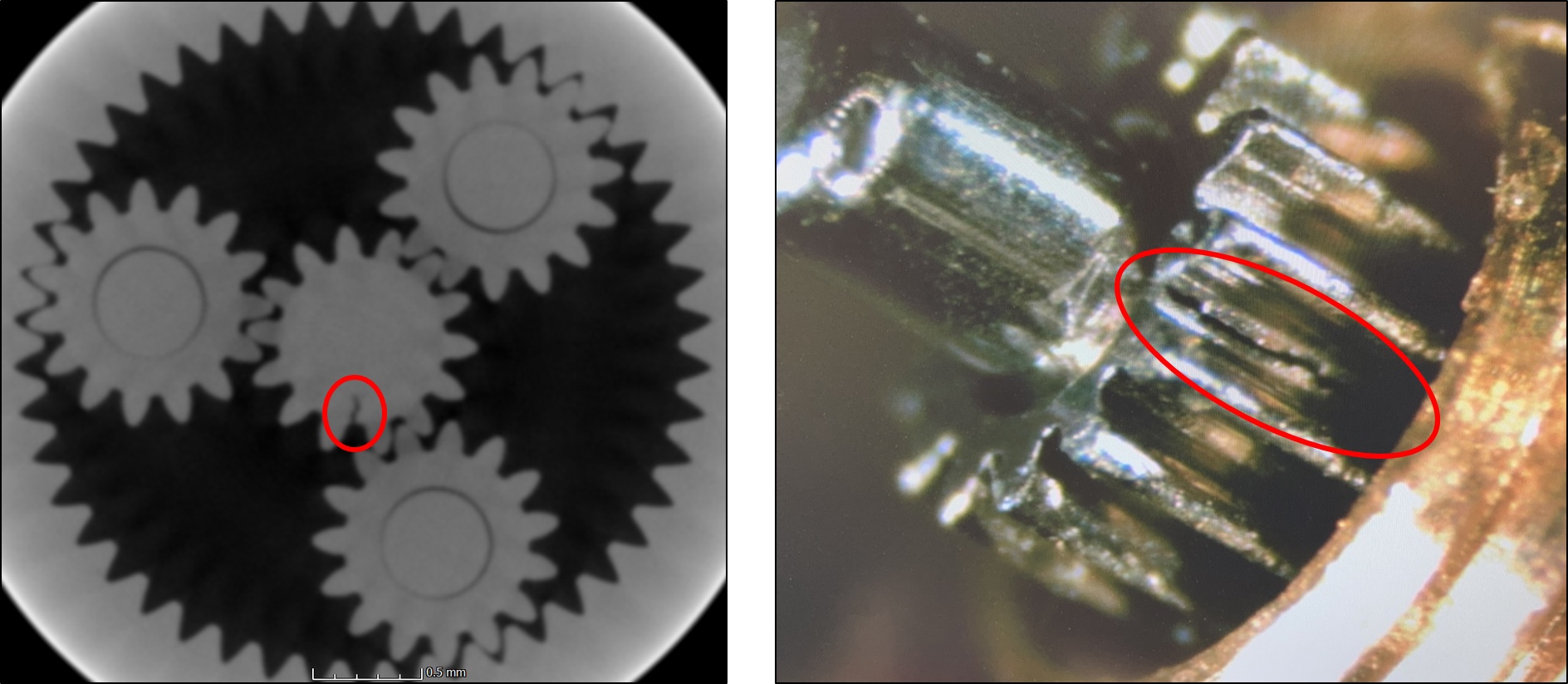}
	\caption{\label{fig:cracked_pinion} X-ray (left) and optical (right) views of a typical cracked pinion gear. In the right hand view, one can see how the pinion has slipped down the shaft (normally it would be flush to the shaft tip, in the upper corner left of the image).}
\end{figure}

The root cause was identified in October 2021 by high-resolution CT scans. Inside the failed gearmotors, torque is transmitted from the motor shaft to the gear train via a pinion gear. In the failed motors, this gear was found to be cracked radially through its wall thickness, from top to bottom of gear. In all items we inspected, the crack occurs at the root between two gear teeth. The crack can be seen by optical inspection, in those cases where the gear can be unscrewed from the motor \figref{fig:cracked_pinion}. We further inspected 13 spare motors that had never been operated since receipt from the factory and found 3 with cracked pinions (23\%).

The method of attachment of pinion gears to motor shafts was a pressure fit. Tolerances for such fits at the sub-millimeter size scale can be challenging to meet, and we surmise that this fraction of motors may statistically have been outside allowable tolerances, thereby over-stressing the gear.

Unfortunately, the failure mode has a subtle operational behavior, in that it typically only reveals itself under a combination of axial and azimuthal slippage, specifically when the gears are biased toward the output shaft direction, i.e. when the robot is oriented downward. Our QC testing in mass production was done with the robots always fixtured in a horizontal position. Hence we missed the failures until the petals were mounted in the telescope, where they often point downward. The failure rate has been a constant $\sim$\,1 unit per week since we installed at the telescope, and proceeds regardless of temperature, humidity, power, or move cycles. This tends to support theories like progressive crack propagation rather than wear mechanisms.

Of the positioners with failed gearmotors, we have found that a large number move a roughly constant fraction of the expected angular distance, i.e. with an output ratio like: $scale = \frac{angle~ moved}{angle~commanded} < 1.0$. We have tested operating these units (with individually calibrated $scale$ factors) on-sky during engineering time, to reasonably good effect. Such robots tend to position within $\sim$\,40\,\micron\,RMS. They take a longer time to get to their targets, by a factor $\sim 1/scale$. We also have found that some motors can be operated accurately by reducing the angular speed of the magnetic field that drives the rotor. As of this writing, we have chosen not to incorporate these back into the regular targeting yet. We currently use such units for sky background measurements instead. With an extra FVC image, faster FVC feedback, or tuned motor driver parameters, we suspect some may be re-introduced for targeting, as the survey progresses.

Another significant fraction of positioners developed communication errors. Unfortunately, a single bad actor on a CAN bus can sometimes impede communication across the entire bus. During the summer 2021 maintenance period, we replaced our electrical distribution boards with new ones that have individual relays per positioner. This allowed us to recover the use of several units, by switching their bad neighbors to a separate debug bus.

\subsection{Focal plane reconfiguration}
\label{sec:reconfig_performance}
In order to complete the 14,000 square degree DESI survey in 5 years, the average time to reconfigure the focal plane and to slew to a new field must be less than 2~minutes on average. All activities during this time between exposures are coordinated by the OCS sequencer of the Instrument Control System (ICS). During reconfiguration of the focal plane, many activities are done in parallel to minimize the total time between exposures. Central among these activities are two complete positioning loop cycles.

During the first iteration (the `blind' move), the robots are reconfigured to new positions while the telescope simultaneously slews to the new field. The reconfiguration process begins with processing a set of target requests. The anticollision calculation is performed, and move tables are generated for each device. The average duration of this step is 6.5 seconds. Next, we execute the move. This takes on average 28 seconds, which includes the time to transfer the move tables to the robots. The actual time the robots are moving is only a fraction of this number, typically 8\,--\,12 seconds, depending on details of move schedules. The blind move puts fibers $\sim$\,50\,\micron\,RMS from their target positions.

The fiber positions are imaged by the FVC and offsets from the target positions are derived by the PlateMaker application of the ICS. The FVC exposure time is 2 seconds and the readout time is $\sim$\,4 seconds. Processing of the FVC images takes on average 1.5 seconds. The extraction of positioning offsets as well as the conversion to focal plane coordinates typically requires another 15 seconds.  With this information in hand, a second positioning loop cycle is initiated for the `correction' move. A `handle fvc feedback' function analyzes positioner performance of the recent blind move, which takes on average 3 seconds. Position tracking data is updated and any highly deviant positioner is disabled for the night. Thereafter the correction move is performed. These motion paths are generally much simpler than the blind move, since corrections are small. Average times for the planning and the move execution steps are 3 and 20 seconds, respectively. Acquisition and processing of the FVC image after the correction move is done in parallel with other activities so that no further overhead is accrued. 

The combination of accuracy and speed of the robot system has thus allowed DESI to achieve the fast reconfiguration times ($<$~2~minutes) between exposures, needed to complete the survey on schedule.

\subsection{GFAs on-sky performance}
\label{sec:GFA_performance}

\begin{figure} 
	\includegraphics[width=\textwidth]{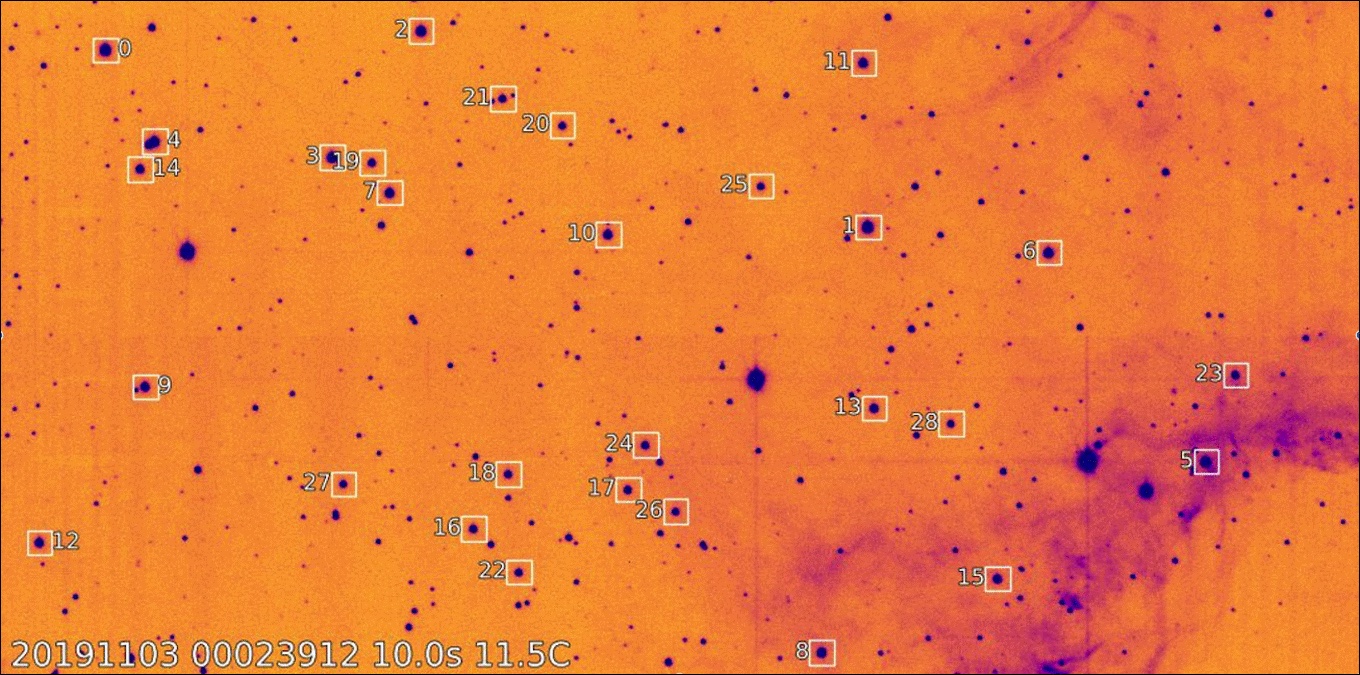}
	\caption{\label{fig:guide_image} A sample guide camera image of a region with high stellar density and dust.  White squares highlight the stars automatically selected for real-time monitoring of the atmospheric transparency and fiber acceptance fraction.  The data was captured with a 10s exposure on 3-Nov-2019.}
\end{figure}
\begin{figure} 
	\includegraphics[width=\textwidth]{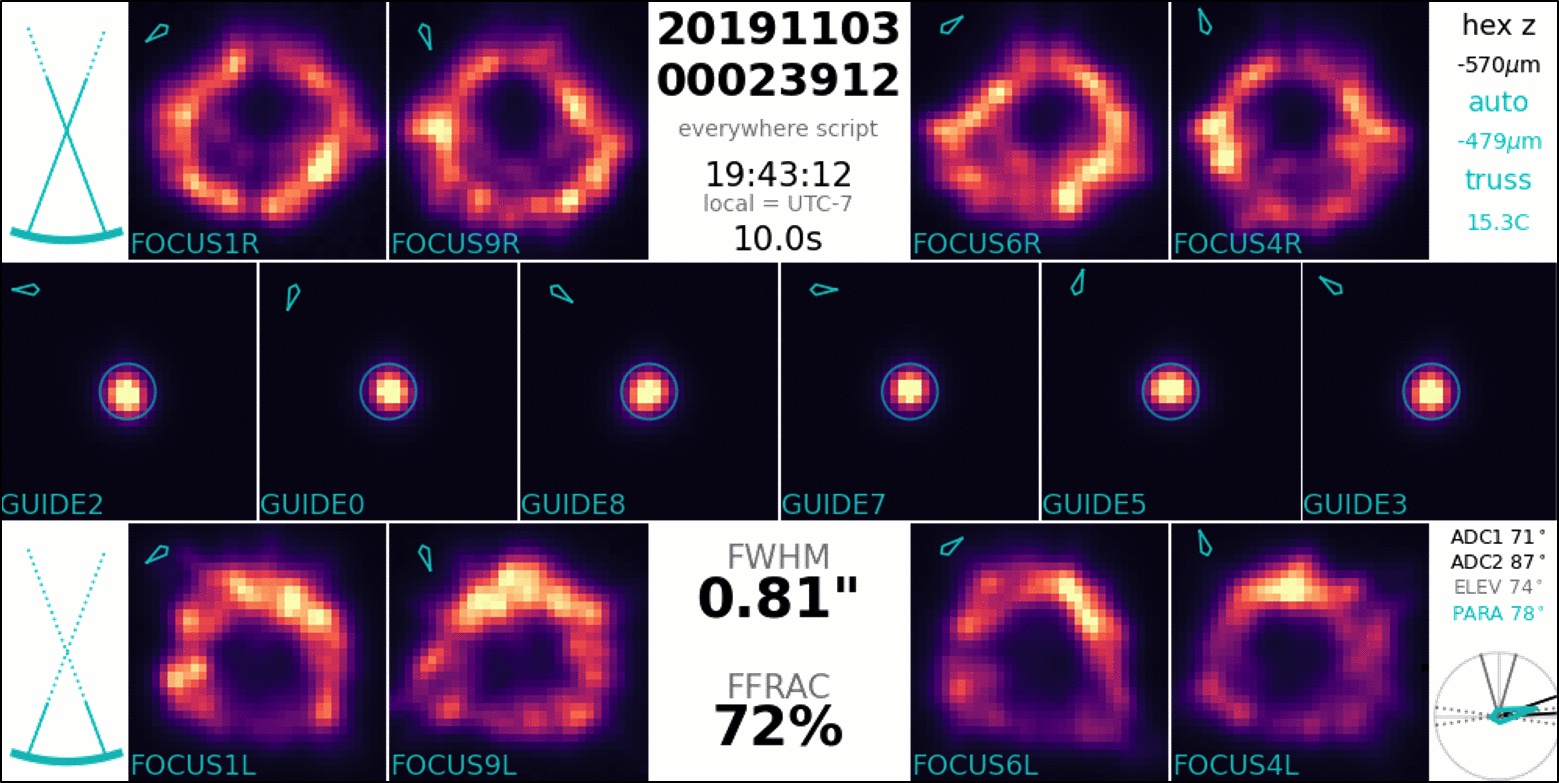}
	\caption{\label{fig:wavefront_image} Co-added images of stars observed in each GFA from a 10\,s exposure on 3-Nov-2019. The top and bottom rows show stars observed in the right and left  sides of the four focus sensors, where the image plane is above (below) the best focus relative to the primary mirror.  The similar sizes of the resulting donut-shaped images indicates that the focal plane is in focus.  The middle row shows stars observed in the six in-focus guide sensors, with a circle showing the aperture of a hypothetical \o\,100\,\micron~fiber (the DESI fibers have a \o\,107\,\micron~fiber). These images allow us to monitor the seeing ($0.81''$) and fraction (72\%) of light entering each fiber in real time during a spectrograph exposure.}
\end{figure}

The six guide GFAs (\S\ref{sec:GFA_hardware}) each acquire one 5~second integration at an $\sim$\,8~second cadence during standard on-sky spectroscopic exposures. The typical 10$\sigma$ point source depth achieved by the guider cameras during such a 5~second integration under ``normal'' observing conditions (transparency $>$~0.9, 1$'' < $ FWHM $ < 1.2''$, r-band AB sky mag per square arcsecond between 20.3 and 20.4) is r = 18.35 mag AB. This depth is sufficient for DESI guiding purposes, as it means that there are on average more than 3 Gaia stars per guider camera with S/N $>$~10, even at the highest Galactic latitudes. An example guide camera image is shown in figure \ref{fig:guide_image}. Example `donut' images from a wavefront camera are shown in figure \ref{fig:wavefront_image}.

During the design phase, we combined estimates of the KPNO atmospheric extinction, Mayall primary mirror reflectivity/area, DESI corrector throughput, GFA r-band filter transmission, and GFA detector quantum efficiency (QE), to predict the GFA r-band zeropoint magnitude at zenith under clear sky conditions. During DESI commissioning, we confirmed that the measured GFA r-band zeropoints agreed with those previously predicted at the $\sim$\,3\% level (with the measurements being $\sim$\,3\% deeper than the prediction on average); this validated both the GFA sensitivity/performance and the total Mayall/DESI (r-band) system throughput to the focal plane. The GFA gains are stable at the $\lesssim$\,1\% level from night to night, so that r-band transparency can be meaningfully measured from GFA photometry, with the dominant limitation being secular trends in the Mayall primary mirror reflectivity between cleanings.

The median r-band GFA FWHM, measured in 5~second guider integrations, is 1.11$''$. This FWHM includes any effects related to the GFA detector itself, such as the finite pixel size (15\,\micron). The median DESI guider FWHM of 1.11$''$ very closely matches the pre-DESI median Mayall/MOSAIC R-band delivered image quality (FWHM = 1.17$''$), and the Mayall/DESI Moffat $\beta$ parameter of $\beta = 3.5$ also matches that found with Mayall/MOSAIC \citep{meisner20, dey14}. The delivered image quality of the DESI GFAs therefore very closely matches the requirements of maintaining the performance of the foregoing Mayall/MOSAIC. 

\subsection{FVC Performance}
\label{sec:fvc_performance}

Key factors limiting the centroiding precision of backlit fibers and fiducials are their signal strength and the size of their PSF. Laboratory measurements with the FVC prior to installation \citep{baltay19} and at the Mayall telescope with  ProtoDESI \citep{fagrelius18a} established that sufficient precision is achieved with signal amplitudes of at least $\sim$\,10,000 electrons (signal-to-noise ratio of $\sim$\,100) and diffraction-limited PSF with FWHM of $\sim$\,2 pixels. With DESI, these requirements are met by suitably adjusting the strength of the illumination, the FVC exposure time, and choosing a lens aperture at least 20 times smaller than the focal length of the FVC lens. The resulting centroiding precision is $\sim$\,20 millipixels, which projects to $\sim$\,3\,\micron~on the DESI focal plane, which has a platescale of 73\,\micron/arcsec. With the FVC sensor cooled to -10\degree C, the  dark current and readout noise of the camera are not a significant factor.

\begin{figure} 
	\includegraphics[width=\textwidth]{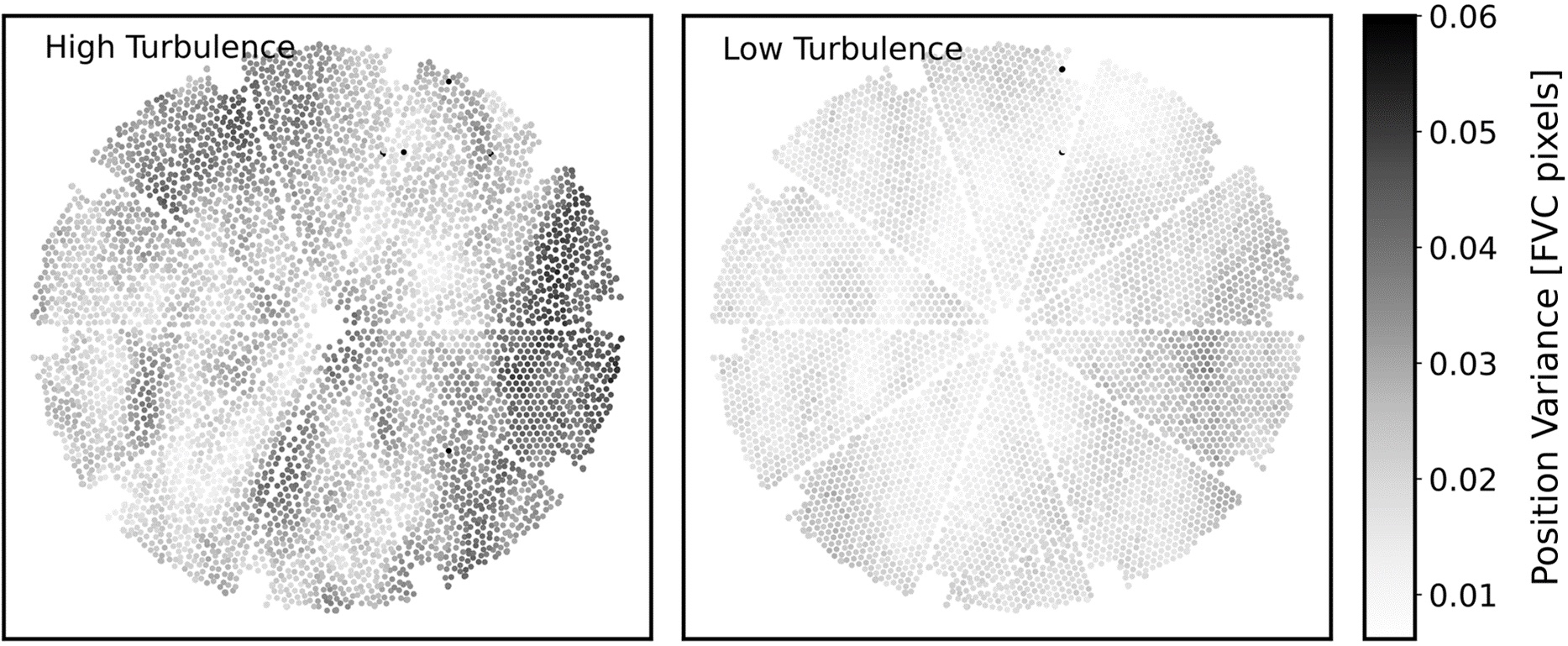}
	\caption{\label{fig:turbulence} Magnitude of the centroiding variance when the turbulence is high (left) and low (right). The highest turbulence occurs when the DESI focal plane is imaged with the dome closed, mirror cooling fans running, and the dome air temperature very different from the mirror temperature. The lowest turbulence occurs with the dome open, cooling fans off, and the mirror temperature matching the air temperature.  The small dots in these maps show the location of each measured source in the FVC image ($\sim$\,5500 sources). The brightness of each dot scales with the variance of the respective source position, determined from 10 repeated observations over a short time interval. Only the variance in the horizontal direction is represented, but the variance in the vertical direction is similar. The high turbulence map shows regions in the image where the variance is as high as 50 millipixels ($\sim$160\,\micron/pixel at the focal surface), depending on location in the field. The median is 30 millipixels. When the turbulence is low, the peak and median variance fall to 30 and 18 millipixels, respectively.}
\end{figure}

Given sufficient centroiding precision, the most significant factor affecting fiber positioning error is dome air turbulence between the FVC lens and the DESI corrector. Air density variations in this zone slightly shift the paths of rays emitted by the fibers and fiducials, randomly displacing their images on the FVC sensor. The variations occur on $\sim$\,1~sec time scales. With careful control of the dome and mirror temperature, the dome turbulence can be minimized, but it is usually the largest contributor to the  positioning error. Figure \ref{fig:turbulence} shows maps of the turbulence, as measured by centroiding variance.

Because the turbulence is correlated across the field, it can be measured and fit by comparing the measured fiducial positions to their known positions (from laboratory metrology or from  long-term average of their FVC measurements). Additionally, we can use previously disabled positioners with fibers intact as post-facto fiducials. The difference between the two reveals the effect of the turbulence at the time of the  FVC exposure. Fitting with a smooth function and extrapolating to the fiber locations, the turbulence can be subtracted. When the turbulence is high, this is an effective method to reduce the effect to an acceptable level.

\section{Lessons learned}\label{LESSONS_LEARNED}

The DESI Focal Plane has been highly successful; even at this early stage in the survey it has accurately measured more extragalactic redshifts  than all previous experiments combined. All projects have their pains, however, and we offer here some of the key things we would do differently in hindsight:

\begin{description}
    \item[Gearmotor internals] Specifications and QC for gearmotors should address not only externally observable performance characteristics (i.e. torque, coil resistance, mechanical dimensions, etc), but also internal assembly clearances and stresses, such as those that caused gear stack compression during our pre-production run, and latent, cracked pinion gears in final production.
    \item[Subcomponent testing] In mass-production, statistical assumptions about subcomponent quality control can fail to account for new process variations that arise unexpectedly. We found that every precision-manufactured part must be inspected upon arrival. Issues within a lot must be immediately fed back to the supplier.
    \item[Diagnostic resources] Planning should include personnel, time and budget for early, deep diagnostics of any failed robot units. In particular, high resolution 3D x-rays were found to be invaluable.
    \item[Individual disables] Bused electronics should include switches for disabling individual units on the bus (or moving them to a parallel debug bus), so that bad actors don't impair their neighbors.
    \item[Connectors / module size] Fusion splicing of fibers (as opposed to using connectors) led the modularity of spectrographs to drive the modularization of the focal plane (i.e. 10 spectrographs $\rightarrow$ 10 petals). For future instruments, from a total system throughput and instrument uptime perspective, the throughput loss incurred by fiber connectorization may be worth the gains of modularizing the focal plane in smaller rafts, with better economies of scale, easier serviceability, and faster iteration time in production.
    \item[Robot pre-production] Planning should include a large-scale pre-production run of robots, at the scale of 1,000 or more units. Smaller batches statistically do not uncover all issues, and furthermore lead to a scarcity of early, disposable units for integrated tests.
    \item[Robot electronics] Individual electronics boards for each positioner were a challenge to procure, test, and service. Integrating electronics over a somewhat larger number of robots (e.g. 20\,--\,80 units or so) with connectors to the motor wires, would ultimately have been preferable. In this spirit, broadcasting to 502 robots over 10 CAN buses from a single petalbox control computer proved more complicated than initially assumed. Reducing the number of robots per module by an order of magnitude would have thus saved significant time in integration and test.
    \item[Encoders] For ease of functional testing, calibration, and diagnostics, it would have been advantageous to have some position encoding of the motors built into the robots. (Even a single digital pulse per revolution would have sufficed, given our 337:1 gear ratio. High precision encoding is not necessary, since the FVC does a better job than any encoder of measuring final position.)
    \item[Test orientation] In hindsight, the cracked pinion gear issue might have been found earlier if large-scale testing of robots had been done at angles mimicking the telescope orientation. (Testing at elevated angles was performed, but not on a large enough sample to uncover the issue.)
    \item[Ferrule attachment] Some glass ferrules for fibers cracked under the stress of place-holding set screws when assembling into robots. This failure mode was difficult to completely avoid (due to variabilities in screw torque), and difficult to observe until well after the retaining epoxy had cured.
    \item[Calibration gauges] Tribology and potential for wear should be carefully considered for any repetitively-used mechanical alignment gauges (such as the rods used for fiber focusing). Gauges need regular, quantitative inspection and qualification.
    \item[GFA sensors] Early GFA CCD sensor selection was driven by dark noise criteria, but this turned out to be a relatively smaller contributor to final performance. A cheaper CCD with better area efficiency, simpler integration, or perhaps even a large CMOS, might have sufficed.
    \item[\# GFA fiducials] We designed GFAs with 2 fiducials, considering the 2D view of the FVC. In hindsight, a 3rd fiducial would have been helpful during metrology in the lab, to constrain the 3D rotation component of the sensor location with respect to the rest of the focal plane.
    \item[Simple FVC lens] We ultimately transitioned from a complex, multi-element zoom lens for the FVC to a single-element plano convex lens. This made analysis of distortions more straightforward.
    \item[Minimize high-friction furcation tubing] Thermoplastic polyester elastomer tubing (Hytrel), commonly-used for fiber routing, had a problematic combination of high friction against our polyimide-coated fibers, long stress-relaxation times, and large coefficient of thermal expansion. Lengths of this tubing greater than a few hundred mm had the ability, post-installation, to drive the fiber axially forward into the robotic mechanism by significant amounts, and to bow the fiber out into the adjacent robot’s envelope. The high friction then prevents the fiber from returning to its nominal position. Use of such tubing should be minimized or avoided. Polyimide tubing, while expensive, performed much better.
    \item[Service routing] Complete and accurate routing of services is entirely feasible in 3D CAD. Our computer models of the fibers, wires, and plumbing proved essential in planning and executing this dense system ($> 61,000$ free wires and fibers, $> 11,600$ electrical connectors, in a volume $< 1~m^3$).
    \item[Lower barriers to integrated tests] The modularization of the petal was useful, but it would have aided debugging and commissioning if we had been able to put together a multi-petal system in the lab. A less complex, smaller scale modularization would have made this more feasible.

\end{description}

\section{Summary}
\begin{figure} 
    	\includegraphics[width=\textwidth]{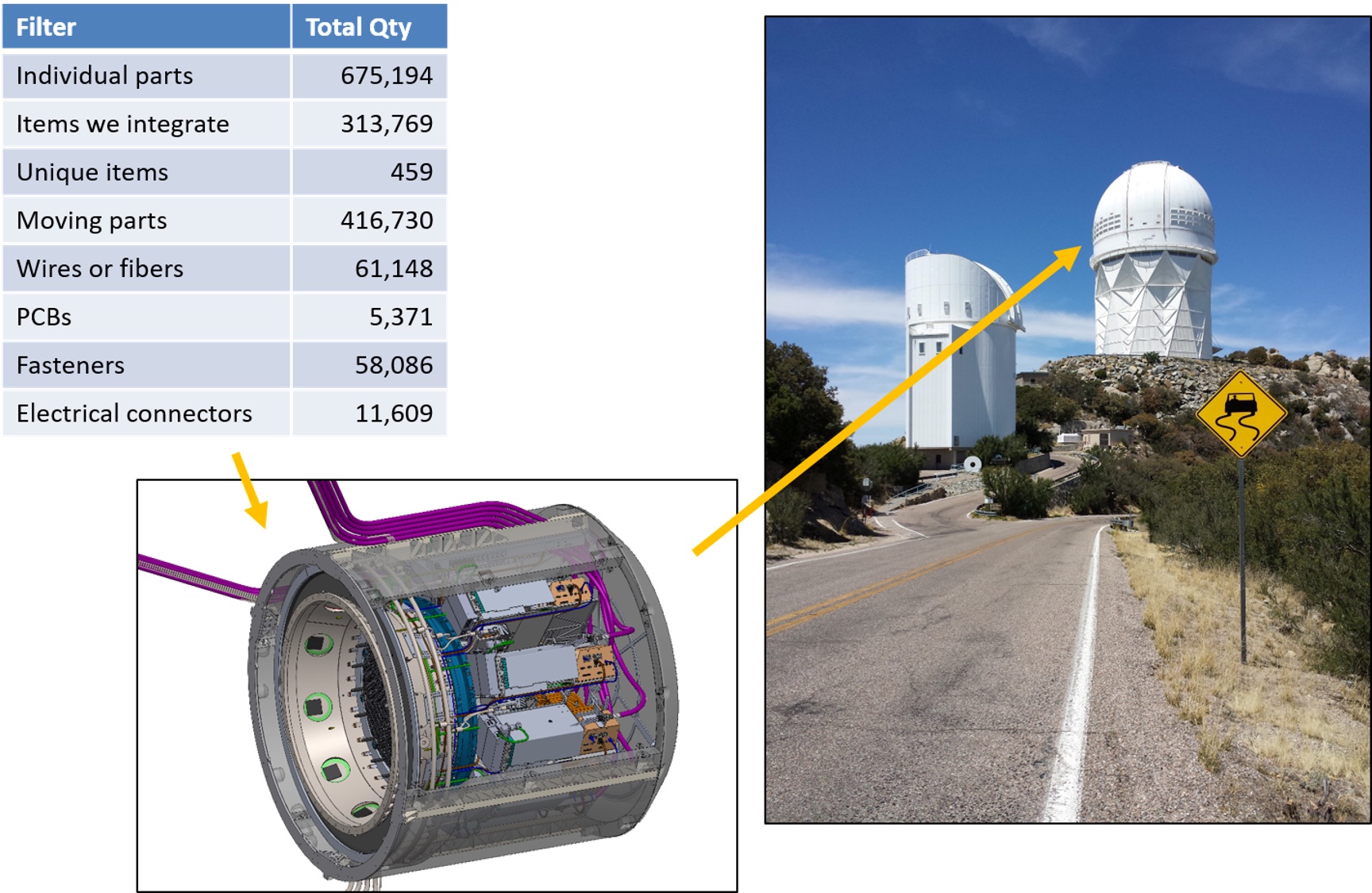}
    	\caption{\label{fig:parts} The DESI focal plane has over 675,000 individual parts. It was installed in 2019 on the Mayall Telescope, located atop Iolkam Du’ag (Kitt Peak) in the Tohono O’odham Nation (near Tucson, Arizona). Commissioning took place in 2020; survey operations began in 2021.}
    \end{figure}

    The DESI Focal Plane System, with 5,020 robotically mounted optical fibers, is enabling a survey of extragalactic redshifts which over the next several years will exceed all previous data sets by an order of magnitude. The focal plane incorporates over 675,000 individual parts \figref{fig:parts} and can rapidly position thousands of fibers in parallel to $<$~10\,\micron~accuracy. It is constructed of 10 identical petals, each supporting 502 robots.
    
    Each fiber robot has two rotational axes, driven by independent \o\,4\,mm DC brushless gearmotors. We operate these in two speed modes, a fast mode of $176.07~deg/s$ for general repositioning, always followed by a short, fine precision move at $2.67~deg/s$. The motors do not have encoders. The robots have precision at the \micron~scale, and when operated `blind' are accurate to $\sim$\,50\,\micron\,RMS, a number which reflects mostly our limits of calibration.
    
    After each blind move we measure the positions of all robots in parallel with a Fiber View Camera. It measures the centroids of the backlit fibers with a precision of $\sim$\,20 millipixels, which projects to $\sim$\,3\,\micron~at the focal surface. Based on this measurement, we send a single set of small correction moves to the robot array. Interspersed in the array are 123 fixed optical point sources (fiducials), which help us to remove lens, gravitational, and air turbulence distortions in the optical path between the camera and the focal plane. The robots ultimately achieve their commanded targets with an accuracy of 6\,\micron\,RMS, corresponding to 0.085\,arcsec on the sky.
    
    A guide signal to the telescope is provided by six custom GFA cameras. Each camera averages more than 3 Gaia stars per field with S/N $>$~10,  providing a tracking error signal better than $0.03''$. Four more GFAs are configured with split-thickness filters for wavefront measurements, providing a feedback signal for precise hexapod alignment of the DESI corrector and focal plane, with respect to the Mayall primary mirror.
    
    The focal plane instrument is housed within a thick, insulated enclosure. Interior air temperature is maintained within $\sim$\,+10 to +15\degree C, and a flow of clean, dry air suppresses interior dew point to $\sim$\,-50 to -20\degree C, preventing condensation. Heat is extracted by Novec 7100 coolant, delivered to the focal plane via 51\,m of hose.
    
    We assembled the 10 petals and fusion spliced them to 45\,m fiber cables in the laboratory. We shipped them as separate units from LBNL directly to Kitt Peak, and installed them one-by-one into the Mayall Telescope.
    
    We completed installation and commissioning of the focal plane in 2019 and 2020. Survey operations began thereafter. In the first 12 months of science operations, from November 2020 through October 2021, the DESI focal plane assumed 1840 field configurations, resulting in 9.2 million spectra. Excluding calibration targets and repeat observations, this yielded 4.2 million high-confidence, unique redshifts of galaxies and quasars, and 1.0 million unique stellar spectra. By January 2022, DESI had measured 7.5 million galaxies, delivering them at a rate of $\sim$\,1 million per month. We eagerly anticipate the scientific bounties that will be discovered in this unprecedented data set.
    
\acknowledgments
    The DESI collaboration is indebted to the late John Donaldson, Robin Lafever, Tammie Lavoie, and Glenn Roberts for their many contributions to the success of this project and mourns their passing.

    The authors in the Barcelona-Madrid Regional Participation Group (BCN-MAD RPG) wish to acknowledge support by MICIN under grants PID2019-111317GB, PGC2018-094773, PGC2018-102021, SEV-2016-0588, SEV-2016-0597, MDM-2015-0509, and CEX2020-001058-M, some of which include ERDF funds from the European Union. IFAE is partially funded by the CERCA program of the Generalitat de Catalunya.  

    MA, SB, NF, FP, GGR and JS acknowledge support from the Spanish MICINN grants MultiDark CSD2009-00064, IFT-UAM/CSIC Severo Ochoa Award SEV-2012-0249 and AYA2014-60641-C2-1-P, CSIC-AVS contract through MICINN grant AYA2010-21231-C02-01 and CDTI grant IDC-20101033, and support from the Campus of International Excellence UAM+CSIC.

    This research is supported by the Director, Office of Science, Office of High Energy Physics of the U.S. Department of Energy under Contract No. DE–AC02–05CH11231, and by the National Energy Research Scientific Computing Center, a DOE Office of Science User Facility under the same contract; additional support for DESI is provided by the U.S. National Science Foundation, Division of Astronomical Sciences under Contract No. AST-0950945 to the NSF’s National Optical-Infrared Astronomy Research Laboratory; the Science and Technologies Facilities Council of the United Kingdom; the Gordon and Betty Moore Foundation; the Heising-Simons Foundation; the French Alternative Energies and Atomic Energy Commission (CEA); the National Council of Science and Technology of Mexico (CONACYT); the Ministry of Science and Innovation of Spain (MICINN), and by the DESI Member Institutions: \url{https://www.desi.lbl.gov/collaborating-institutions}.

    The authors are honored to be permitted to conduct scientific research on Iolkam Du’ag (Kitt Peak), a mountain with particular significance to the Tohono O’odham Nation.

    For more information, visit \url{https://desi.lbl.gov/}.
    
    Source data files for all plots are archived at \url{https://zenodo.org/record/6286817}.
\vspace{5mm}
\facility{Mayall (DESI)}

\bibliography{focalplane}{}

\begin{thebibliography}{}
\expandafter\ifx\csname natexlab\endcsname\relax\def\natexlab#1{#1}\fi
\providecommand{\url}[1]{\href{#1}{#1}}
\providecommand{\dodoi}[1]{doi:~\href{http://doi.org/#1}{\nolinkurl{#1}}}
\providecommand{\doeprint}[1]{\href{http://ascl.net/#1}{\nolinkurl{http://ascl.net/#1}}}
\providecommand{\doarXiv}[1]{\href{https://arxiv.org/abs/#1}{\nolinkurl{https://arxiv.org/abs/#1}}}

\bibitem[{{Ahumada} {et~al.}(2020){Ahumada}, {Allende Prieto}, {Almeida},
  {Anders}, {Anderson}, {Andrews}, {Anguiano}, {Arcodia}, {Armengaud},
  {Aubert}, {Avila}, {Avila-Reese}, {Badenes}, {Balland}, {Barger},
  {Barrera-Ballesteros}, {Basu}, {Bautista}, {Beaton}, {Beers}, {Benavides},
  {Bender}, {Bernardi}, {Bershady}, {Beutler}, {Bidin}, {Bird}, {Bizyaev},
  {Blanc}, {Blanton}, {Boquien}, {Borissova}, {Bovy}, {Brandt}, {Brinkmann},
  {Brownstein}, {Bundy}, {Bureau}, {Burgasser}, {Burtin}, {Cano-D{\'\i}az},
  {Capasso}, {Cappellari}, {Carrera}, {Chabanier}, {Chaplin}, {Chapman},
  {Cherinka}, {Chiappini}, {Doohyun Choi}, {Chojnowski}, {Chung}, {Clerc},
  {Coffey}, {Comerford}, {Comparat}, {da Costa}, {Cousinou}, {Covey}, {Crane},
  {Cunha}, {da Silva Ilha}, {Dai}, {Damsted}, {Darling}, {Davidson}, {Davies},
  {Dawson}, {De}, {de la Macorra}, {De Lee}, {de Andrade Queiroz}, {Deconto
  Machado}, {de la Torre}, {Dell'Agli}, {du Mas des Bourboux},
  {Diamond-Stanic}, {Dillon}, {Donor}, {Drory}, {Duckworth}, {Dwelly},
  {Ebelke}, {Eftekharzadeh}, {Eigenbrot}, {Elsworth}, {Eracleous},
  {Erfanianfar}, {Escoffier}, {Fan}, {Farr}, {Fern{\'a}ndez-Trincado},
  {Feuillet}, {Finoguenov}, {Fofie}, {Fraser-McKelvie}, {Frinchaboy},
  {Fromenteau}, {Fu}, {Galbany}, {Garcia}, {Garc{\'\i}a-Hern{\'a}ndez}, {Garma
  Oehmichen}, {Ge}, {Geimba Maia}, {Geisler}, {Gelfand}, {Goddy},
  {Gonzalez-Perez}, {Grabowski}, {Green}, {Grier}, {Guo}, {Guy}, {Harding},
  {Hasselquist}, {Hawken}, {Hayes}, {Hearty}, {Hekker}, {Hogg}, {Holtzman},
  {Horta}, {Hou}, {Hsieh}, {Huber}, {Hunt}, {Ider Chitham}, {Imig}, {Jaber},
  {Jimenez Angel}, {Johnson}, {Jones}, {J{\"o}nsson}, {Jullo}, {Kim},
  {Kinemuchi}, {Kirkpatrick}, {Kite}, {Klaene}, {Kneib}, {Kollmeier}, {Kong},
  {Kounkel}, {Krishnarao}, {Lacerna}, {Lan}, {Lane}, {Law}, {Le Goff}, {Leung},
  {Lewis}, {Li}, {Lian}, {Lin}, {Long}, {Longa-Pe{\~n}a}, {Lundgren}, {Lyke},
  {Ted Mackereth}, {MacLeod}, {Majewski}, {Manchado}, {Maraston}, {Martini},
  {Masseron}, {Masters}, {Mathur}, {McDermid}, {Merloni}, {Merrifield},
  {M{\'e}sz{\'a}ros}, {Miglio}, {Minniti}, {Minsley}, {Miyaji}, {Mohammad},
  {Mosser}, {Mueller}, {Muna}, {Mu{\~n}oz-Guti{\'e}rrez}, {Myers}, {Nadathur},
  {Nair}, {Nandra}, {do Nascimento}, {Nevin}, {Newman}, {Nidever}, {Nitschelm},
  {Noterdaeme}, {O'Connell}, {Olmstead}, {Oravetz}, {Oravetz}, {Osorio},
  {Pace}, {Padilla}, {Palanque-Delabrouille}, {Palicio}, {Pan}, {Pan},
  {Parker}, {Paviot}, {Peirani}, {Pe{\~n}a Ram{\'r}ez}, {Penny}, {Percival},
  {Perez-Fournon}, {P{\'e}rez-R{\`a}fols}, {Petitjean}, {Pieri},
  {Pinsonneault}, {Poovelil}, {Povick}, {Prakash}, {Price-Whelan}, {Raddick},
  {Raichoor}, {Ray}, {Rembold}, {Rezaie}, {Riffel}, {Riffel}, {Rix}, {Robin},
  {Roman-Lopes}, {Rom{\'a}n-Z{\'u}{\~n}iga}, {Rose}, {Ross}, {Rossi},
  {Rowlands}, {Rubin}, {Salvato}, {S{\'a}nchez}, {S{\'a}nchez-Menguiano},
  {S{\'a}nchez-Gallego}, {Sayres}, {Schaefer}, {Schiavon}, {Schimoia},
  {Schlafly}, {Schlegel}, {Schneider}, {Schultheis}, {Schwope}, {Seo},
  {Serenelli}, {Shafieloo}, {Shamsi}, {Shao}, {Shen}, {Shetrone}, {Shirley},
  {Silva Aguirre}, {Simon}, {Skrutskie}, {Slosar}, {Smethurst}, {Sobeck},
  {Sodi}, {Souto}, {Stark}, {Stassun}, {Steinmetz}, {Stello}, {Stermer},
  {Storchi-Bergmann}, {Streblyanska}, {Stringfellow}, {Stutz}, {Su{\'a}rez},
  {Sun}, {Taghizadeh-Popp}, {Talbot}, {Tayar}, {Thakar}, {Theriault}, {Thomas},
  {Thomas}, {Tinker}, {Tojeiro}, {Toledo}, {Tremonti}, {Troup}, {Tuttle},
  {Unda-Sanzana}, {Valentini}, {Vargas-Gonz{\'a}lez}, {Vargas-Maga{\~n}a},
  {V{\'a}zquez-Mata}, {Vivek}, {Wake}, {Wang}, {Weaver}, {Weijmans}, {Wild},
  {Wilson}, {Wilson}, {Wolthuis}, {Wood-Vasey}, {Yan}, {Yang}, {Y{\`e}che},
  {Zamora}, {Zarrouk}, {Zasowski}, {Zhang}, {Zhao}, {Zhao}, {Zheng}, {Zheng},
  {Zhu}, \& {Zou}}]{ahumada20}
{Ahumada}, R., {Allende Prieto}, C., {Almeida}, A., {et~al.} 2020, \apjs, 249,
  3, \dodoi{10.3847/1538-4365/ab929e}

\bibitem[{{Baltay} {et~al.}(2019){Baltay}, {Rabinowitz}, {Besuner}, {Casetti},
  {Emmet}, {Fagrelius}, {Girard}, {Heetderks}, {Lampton}, {Lathem}, {Levi},
  {Padmanabhan}, \& {Silber}}]{baltay19}
{Baltay}, C., {Rabinowitz}, D., {Besuner}, R., {et~al.} 2019, \pasp, 131,
  065001, \dodoi{10.1088/1538-3873/ab15c2}

\bibitem[{{Besuner} {et~al.}(2020){Besuner}, {Allen}, {Baltay}, {Brooks},
  {Carton}, {Doel}, {Donaldson}, {Duan}, {Dunlop}, {Edelstein}, {Evatt},
  {Fagrelius}, {Gaztanaga}, {Guenther}, {Gutierrez}, {Hawes}, {Honscheid},
  {Jelinsky}, {Joyce}, {Karcher}, {Landriau}, {Levi}, {Magneville}, {Marshall},
  {Martini}, {Pappalardo}, {Poppett}, {Prada}, {Ross}, {Schubnell}, {Sharples},
  {Shourt}, {Silber}, {Sprayberry}, {Stupak}, {Tarle}, \& {Zhang}}]{besuner20}
{Besuner}, R., {Allen}, L., {Baltay}, C., {et~al.} 2020, in Society of
  Photo-Optical Instrumentation Engineers (SPIE) Conference Series, Vol. 11447,
  Society of Photo-Optical Instrumentation Engineers (SPIE) Conference Series,
  1144710, \dodoi{10.1117/12.2561507}

\bibitem[{Cui {et~al.}(2012)Cui, Zhao, Chu, Li, Li, Zhang, Su, Yao, Wang, Xing,
  Li, Zhu, Wang, Gu, Luo, Xu, Zhang, Liu, Zhang, Yang, Cao, Chen, Chen, Chen,
  Chen, Chu, Feng, Gong, Hou, Hu, Hu, Hu, Jia, Jiang, Jiang, Jiang, Jin, Li,
  Li, Li, Liu, Liu, Lu, Mao, Men, Qi, Qi, Shi, Tang, Tao, Wang, Wang, Wang,
  Wang, Wang, Wang, Wang, Wang, Wang, Wang, Wang, Wang, Xu, Xu, Yang, Yu, Yuan,
  Yuan, Zhai, Zhang, Zhang, Zhang, Zhao, Zhou, Zhou, Zhu, \& Zou}]{lamost}
Cui, X.-Q., Zhao, Y.-H., Chu, Y.-Q., {et~al.} 2012, Research in Astronomy and
  Astrophysics, 12, 1197, \dodoi{10.1088/1674-4527/12/9/003}

\bibitem[{{Dawson} {et~al.}(2013){Dawson}, {Schlegel}, {Ahn}, {Anderson},
  {Aubourg}, {Bailey}, {Barkhouser}, {Bautista}, {Beifiori}, {Berlind},
  {Bhardwaj}, {Bizyaev}, {Blake}, {Blanton}, {Blomqvist}, {Bolton}, {Borde},
  {Bovy}, {Brandt}, {Brewington}, {Brinkmann}, {Brown}, {Brownstein}, {Bundy},
  {Busca}, {Carithers}, {Carnero}, {Carr}, {Chen}, {Comparat}, {Connolly},
  {Cope}, {Croft}, {Cuesta}, {da Costa}, {Davenport}, {Delubac}, {de Putter},
  {Dhital}, {Ealet}, {Ebelke}, {Eisenstein}, {Escoffier}, {Fan}, {Filiz Ak},
  {Finley}, {Font-Ribera}, {G{\'e}nova-Santos}, {Gunn}, {Guo}, {Haggard},
  {Hall}, {Hamilton}, {Harris}, {Harris}, {Ho}, {Hogg}, {Holder}, {Honscheid},
  {Huehnerhoff}, {Jordan}, {Jordan}, {Kauffmann}, {Kazin}, {Kirkby}, {Klaene},
  {Kneib}, {Le Goff}, {Lee}, {Long}, {Loomis}, {Lundgren}, {Lupton}, {Maia},
  {Makler}, {Malanushenko}, {Malanushenko}, {Mandelbaum}, {Manera}, {Maraston},
  {Margala}, {Masters}, {McBride}, {McDonald}, {McGreer}, {McMahon}, {Mena},
  {Miralda-Escud{\'e}}, {Montero-Dorta}, {Montesano}, {Muna}, {Myers},
  {Naugle}, {Nichol}, {Noterdaeme}, {Nuza}, {Olmstead}, {Oravetz}, {Oravetz},
  {Owen}, {Padmanabhan}, {Palanque-Delabrouille}, {Pan}, {Parejko},
  {P{\^a}ris}, {Percival}, {P{\'e}rez-Fournon}, {P{\'e}rez-R{\`a}fols},
  {Petitjean}, {Pfaffenberger}, {Pforr}, {Pieri}, {Prada}, {Price-Whelan},
  {Raddick}, {Rebolo}, {Rich}, {Richards}, {Rockosi}, {Roe}, {Ross}, {Ross},
  {Rossi}, {Rubi{\~n}o-Martin}, {Samushia}, {S{\'a}nchez}, {Sayres}, {Schmidt},
  {Schneider}, {Sc{\'o}ccola}, {Seo}, {Shelden}, {Sheldon}, {Shen}, {Shu},
  {Slosar}, {Smee}, {Snedden}, {Stauffer}, {Steele}, {Strauss}, {Streblyanska},
  {Suzuki}, {Swanson}, {Tal}, {Tanaka}, {Thomas}, {Tinker}, {Tojeiro},
  {Tremonti}, {Vargas Maga{\~n}a}, {Verde}, {Viel}, {Wake}, {Watson}, {Weaver},
  {Weinberg}, {Weiner}, {West}, {White}, {Wood-Vasey}, {Yeche}, {Zehavi},
  {Zhao}, \& {Zheng}}]{dawson2013}
{Dawson}, K.~S., {Schlegel}, D.~J., {Ahn}, C.~P., {et~al.} 2013, \aj, 145, 10,
  \dodoi{10.1088/0004-6256/145/1/10}

\bibitem[{de~Jong {et~al.}(2012)de~Jong, Bellido-Tirado, Chiappini, Depagne,
  Haynes, Johl, Schnurr, Schwope, Walcher, Dionies, \& et~al.}]{4most2012}
de~Jong, R.~S., Bellido-Tirado, O., Chiappini, C., {et~al.} 2012, Ground-based
  and Airborne Instrumentation for Astronomy IV, \dodoi{10.1117/12.926239}

\bibitem[{{DESI Collaboration} {et~al.}(2016){DESI Collaboration}, {Aghamousa},
  {Aguilar}, {Ahlen}, {Alam}, {Allen}, {Allende Prieto}, {Annis}, {Bailey},
  {Balland}, {Ballester}, {Baltay}, {Beaufore}, {Bebek}, {Beers}, {Bell},
  {Bernal}, {Besuner}, {Beutler}, {Blake}, {Bleuler}, {Blomqvist}, {Blum},
  {Bolton}, {Briceno}, {Brooks}, {Brownstein}, {Buckley-Geer}, {Burden},
  {Burtin}, {Busca}, {Cahn}, {Cai}, {Cardiel-Sas}, {Carlberg}, {Carton},
  {Casas}, {Castand er}, {Cervantes-Cota}, {Claybaugh}, {Close}, {Coker},
  {Cole}, {Comparat}, {Cooper}, {Cousinou}, {Crocce}, {Cuby}, {Cunningham},
  {Davis}, {Dawson}, {de la Macorra}, {De Vicente}, {Delubac}, {Derwent},
  {Dey}, {Dhungana}, {Ding}, {Doel}, {Duan}, {Ealet}, {Edelstein},
  {Eftekharzadeh}, {Eisenstein}, {Elliott}, {Escoffier}, {Evatt}, {Fagrelius},
  {Fan}, {Fanning}, {Farahi}, {Farihi}, {Favole}, {Feng}, {Fernandez},
  {Findlay}, {Finkbeiner}, {Fitzpatrick}, {Flaugher}, {Flender}, {Font-Ribera},
  {Forero-Romero}, {Fosalba}, {Frenk}, {Fumagalli}, {Gaensicke}, {Gallo},
  {Garcia-Bellido}, {Gaztanaga}, {Pietro Gentile Fusillo}, {Gerard},
  {Gershkovich}, {Giannantonio}, {Gillet}, {Gonzalez-de-Rivera},
  {Gonzalez-Perez}, {Gott}, {Graur}, {Gutierrez}, {Guy}, {Habib}, {Heetderks},
  {Heetderks}, {Heitmann}, {Hellwing}, {Herrera}, {Ho}, {Holland}, {Honscheid},
  {Huff}, {Hutchinson}, {Huterer}, {Hwang}, {Illa Laguna}, {Ishikawa},
  {Jacobs}, {Jeffrey}, {Jelinsky}, {Jennings}, {Jiang}, {Jimenez}, {Johnson},
  {Joyce}, {Jullo}, {Juneau}, {Kama}, {Karcher}, {Karkar}, {Kehoe}, {Kennamer},
  {Kent}, {Kilbinger}, {Kim}, {Kirkby}, {Kisner}, {Kitanidis}, {Kneib},
  {Koposov}, {Kovacs}, {Koyama}, {Kremin}, {Kron}, {Kronig}, {Kueter-Young},
  {Lacey}, {Lafever}, {Lahav}, {Lambert}, {Lampton}, {Land riau}, {Lang},
  {Lauer}, {Le Goff}, {Le Guillou}, {Le Van Suu}, {Lee}, {Lee}, {Leitner},
  {Lesser}, {Levi}, {L'Huillier}, {Li}, {Liang}, {Lin}, {Linder}, {Loebman},
  {Luki{\'c}}, {Ma}, {MacCrann}, {Magneville}, {Makarem}, {Manera}, {Manser},
  {Marshall}, {Martini}, {Massey}, {Matheson}, {McCauley}, {McDonald},
  {McGreer}, {Meisner}, {Metcalfe}, {Miller}, {Miquel}, {Moustakas}, {Myers},
  {Naik}, {Newman}, {Nichol}, {Nicola}, {Nicolati da Costa}, {Nie}, {Niz},
  {Norberg}, {Nord}, {Norman}, {Nugent}, {O'Brien}, {Oh}, {Olsen}, {Padilla},
  {Padmanabhan}, {Padmanabhan}, {Palanque-Delabrouille}, {Palmese},
  {Pappalardo}, {P{\^a}ris}, {Park}, {Patej}, {Peacock}, {Peiris}, {Peng},
  {Percival}, {Perruchot}, {Pieri}, {Pogge}, {Pollack}, {Poppett}, {Prada},
  {Prakash}, {Probst}, {Rabinowitz}, {Raichoor}, {Ree}, {Refregier}, {Regal},
  {Reid}, {Reil}, {Rezaie}, {Rockosi}, {Roe}, {Ronayette}, {Roodman}, {Ross},
  {Ross}, {Rossi}, {Rozo}, {Ruhlmann-Kleider}, {Rykoff}, {Sabiu}, {Samushia},
  {Sanchez}, {Sanchez}, {Schlegel}, {Schneider}, {Schubnell}, {Secroun},
  {Seljak}, {Seo}, {Serrano}, {Shafieloo}, {Shan}, {Sharples}, {Sholl},
  {Shourt}, {Silber}, {Silva}, {Sirk}, {Slosar}, {Smith}, {Smoot}, {Som},
  {Song}, {Sprayberry}, {Staten}, {Stefanik}, {Tarle}, {Sien Tie}, {Tinker},
  {Tojeiro}, {Valdes}, {Valenzuela}, {Valluri}, {Vargas-Magana}, {Verde},
  {Walker}, {Wang}, {Wang}, {Weaver}, {Weaverdyck}, {Wechsler}, {Weinberg},
  {White}, {Yang}, {Yeche}, {Zhang}, {Zhao}, {Zheng}, {Zhou}, {Zhou}, {Zhu},
  {Zou}, \& {Zu}}]{desi16a}
{DESI Collaboration}, {Aghamousa}, A., {Aguilar}, J., {et~al.} 2016, arXiv
  e-prints, arXiv:1611.00036.
\newblock \doarXiv{1611.00036}

\bibitem[{{Dey} \& {Valdes}(2014)}]{dey14}
{Dey}, A., \& {Valdes}, F. 2014, \pasp, 126, 296, \dodoi{10.1086/675808}

\bibitem[{{Doel} {et~al.}(2014){Doel}, {Sholl}, {Liang}, {Brooks}, {Flaugher},
  {Gutierrez}, {Kent}, {Lampton}, {Miller}, \& {Sprayberry}}]{doel14}
{Doel}, P., {Sholl}, M.~J., {Liang}, M., {et~al.} 2014, in Society of
  Photo-Optical Instrumentation Engineers (SPIE) Conference Series, Vol. 9147,
  Ground-based and Airborne Instrumentation for Astronomy V, 91476R,
  \dodoi{10.1117/12.2057172}

\bibitem[{{Duan} {et~al.}(2018){Duan}, {Silber}, {Claybaugh}, {Ahlen},
  {Brooks}, \& {Tarl{\'e}}}]{duan18}
{Duan}, Y., {Silber}, J.~H., {Claybaugh}, T.~M., {et~al.} 2018, in Society of
  Photo-Optical Instrumentation Engineers (SPIE) Conference Series, Vol. 10706,
  Advances in Optical and Mechanical Technologies for Telescopes and
  Instrumentation III, ed. R.~{Navarro} \& R.~{Geyl}, 1070643,
  \dodoi{10.1117/12.2309875}

\bibitem[{{Fabricant} {et~al.}(2005){Fabricant}, {Fata}, {Roll}, {Hertz},
  {Caldwell}, {Gauron}, {Geary}, {McLeod}, {Szentgyorgyi}, {Zajac}, {Kurtz},
  {Barberis}, {Bergner}, {Brown}, {Conroy}, {Eng}, {Geller}, {Goddard},
  {Honsa}, {Mueller}, {Mink}, {Ordway}, {Tokarz}, {Woods}, {Wyatt}, {Epps}, \&
  {Dell'Antonio}}]{hectospec}
{Fabricant}, D., {Fata}, R., {Roll}, J., {et~al.} 2005, \pasp, 117, 1411,
  \dodoi{10.1086/497385}

\bibitem[{{Fagrelius} {et~al.}(2018){Fagrelius}, {Abareshi}, {Allen},
  {Ballester}, {Baltay}, {Besuner}, {Buckley-Geer}, {Butler}, {Cardiel}, {Dey},
  {Duan}, {Elliott}, {Emmet}, {Gershkovich}, {Honscheid}, {Illa}, {Jimenez},
  {Joyce}, {Karcher}, {Kent}, {Lambert}, {Lampton}, {Levi}, {Manser},
  {Marshall}, {Martini}, {Paat}, {Probst}, {Rabinowitz}, {Reil}, {Robertson},
  {Rockosi}, {Schlegel}, {Schubnell}, {Serrano}, {Silber}, {Soto},
  {Sprayberry}, {Summers}, {Tarl{\'e}}, \& {Weaver}}]{fagrelius18a}
{Fagrelius}, P., {Abareshi}, B., {Allen}, L., {et~al.} 2018, \pasp, 130,
  025005, \dodoi{10.1088/1538-3873/aaa225}

\bibitem[{{Fagrelius} {et~al.}(2020){Fagrelius}, {Duan}, {Fanning},
  {Honscheid}, {Kent}, {Martini}, {Poppett}, {Rabinowitz}, {Schubnell},
  {Silber}, {Brooks}, {Claybaugh}, {Doel}, {Gazta{\~n}aga}, {Levi},
  {Magneville}, {Prada}, \& {Tarl{\'e}}}]{fagrelius20}
{Fagrelius}, P., {Duan}, Y., {Fanning}, K., {et~al.} 2020, in Society of
  Photo-Optical Instrumentation Engineers (SPIE) Conference Series, Vol. 11447,
  Society of Photo-Optical Instrumentation Engineers (SPIE) Conference Series,
  114478K, \dodoi{10.1117/12.2561631}

\bibitem[{Fahim {et~al.}(2015)Fahim, Prada, Kneib, Sanchez, Hörler, Azzaro,
  Becerril, Bleuler, Bouri, Castano, Garrido, Gillet, Glez-de Rivera, Gómez,
  Gomez-lopez, Gonzalez-Arroyo, Jenni, Makarem, Yepes, \& Lachat}]{fahim2015}
Fahim, N., Prada, F., Kneib, J.-P., {et~al.} 2015, Monthly Notices of the Royal
  Astronomical Society, MNRAS, 794–806, \dodoi{10.1093/mnras/stv541}

\bibitem[{{Gaia Collaboration} {et~al.}(2018){Gaia Collaboration}, {Brown},
  {Vallenari}, {Prusti}, {de Bruijne}, {Babusiaux}, {Bailer-Jones}, {Biermann},
  {Evans}, {Eyer}, \& et~al.}]{gaia18}
{Gaia Collaboration}, {Brown}, A.~G.~A., {Vallenari}, A., {et~al.} 2018, \aap,
  616, A1, \dodoi{10.1051/0004-6361/201833051}

\bibitem[{{Gutierrez} {et~al.}(2018){Gutierrez}, {Besuner}, {Brooks}, {Doel},
  {Flaugher}, {Friend}, {Gallo}, {Stefanik}, \& {Tarle}}]{gutierrez18}
{Gutierrez}, G., {Besuner}, R.~W., {Brooks}, D., {et~al.} 2018, in Society of
  Photo-Optical Instrumentation Engineers (SPIE) Conference Series, Vol. 10702,
  \procspie, 107027Y, \dodoi{10.1117/12.2312728}

\bibitem[{{Honscheid} {et~al.}(2016){Honscheid}, {Elliott}, {Beaufore},
  {Buckley-Geer}, {Castander}, {daCosta}, {Fausti}, {Kent}, {Kirkby},
  {Neilsen}, {Reil}, {Serrano}, \& {Slozar}}]{honscheid16}
{Honscheid}, K., {Elliott}, A.~E., {Beaufore}, L., {et~al.} 2016, in Society of
  Photo-Optical Instrumentation Engineers (SPIE) Conference Series, Vol. 9913,
  Software and Cyberinfrastructure for Astronomy IV, ed. G.~{Chiozzi} \& J.~C.
  {Guzman}, 99130P, \dodoi{10.1117/12.2229835}

\bibitem[{{Honscheid} {et~al.}(2018){Honscheid}, {Elliott}, {Buckley-Geer},
  {Abreshi}, {Castander}, {da Costa}, {Kent}, {Kirkby}, {Marshall}, {Neilsen},
  {Ogando}, {Rabinowitz}, {Roodman}, {Serrano}, {Brooks}, {Levi}, \&
  {Tarle}}]{honscheid18}
{Honscheid}, K., {Elliott}, A.~E., {Buckley-Geer}, E., {et~al.} 2018, in
  Society of Photo-Optical Instrumentation Engineers (SPIE) Conference Series,
  Vol. 10707, \procspie, 107071D, \dodoi{10.1117/12.2311927}

\bibitem[{Kimura {et~al.}(2010)Kimura, Maihara, Iwamuro, Akiyama, Tamura,
  Dalton, Takato, Tait, Ohta, Eto, Mochida, Elms, Kawate, Kurakami, Moritani,
  Noumaru, Ohshima, Sumiyoshi, Yabe, Brzeski, Farrell, Frost, Gillingham,
  Haynes, Moore, Muller, Smedley, Smith, Bonfield, Brooks, Holmes, Lake, Lee,
  Lewis, Froud, Tosh, Woodhouse, Blackburn, Dipper, Murray, Sharples, \&
  Robertson}]{fmos}
Kimura, M., Maihara, T., Iwamuro, F., {et~al.} 2010, The Fibre Multi-Object
  Spectrograph (FMOS) for Subaru Telescope.
\newblock \doarXiv{1006.3102}

\bibitem[{{Leitner} {et~al.}(2018){Leitner}, {Aguilar}, {Ameel}, {Besuner},
  {Claybaugh}, {Heetderks}, {Schubnell}, {Kneib}, {Silber}, {Tarl{\'e}},
  {Weaverdyck}, \& {Zhang}}]{leitner18}
{Leitner}, D., {Aguilar}, J., {Ameel}, J., {et~al.} 2018, in Society of
  Photo-Optical Instrumentation Engineers (SPIE) Conference Series, Vol. 10706,
  Advances in Optical and Mechanical Technologies for Telescopes and
  Instrumentation III, 1070669, \dodoi{10.1117/12.2312228}

\bibitem[{Lynch {et~al.}(2016)Lynch, Marchuk, \& Elwin}]{CAN}
Lynch, K.~M., Marchuk, N., \& Elwin, M.~L. 2016, in Embedded Computing and
  Mechatronics with the PIC32, ed. K.~M. Lynch, N.~Marchuk, \& M.~L. Elwin
  (Oxford: Newnes), 249--265,
  \dodoi{https://doi.org/10.1016/B978-0-12-420165-1.00019-6}

\bibitem[{{Meisner} {et~al.}(2020){Meisner}, {Abareshi}, {Dey}, {Rockosi},
  {Joyce}, {Sprayberry}, {Besuner}, {Honscheid}, {Kirkby}, {Kong}, {Landriau},
  {Levi}, {Li}, {Marshall}, {Martini}, {Ross}, {Brooks}, {Doel}, {Duan},
  {Gazta{\~n}aga}, {Magneville}, {Prada}, {Schubnell}, \&
  {Tarl{\~N}{\'n}}}]{meisner20}
{Meisner}, A.~M., {Abareshi}, B., {Dey}, A., {et~al.} 2020, in Society of
  Photo-Optical Instrumentation Engineers (SPIE) Conference Series, Vol. 11447,
  Society of Photo-Optical Instrumentation Engineers (SPIE) Conference Series,
  1144794, \dodoi{10.1117/12.2574776}

\bibitem[{{Pogge} {et~al.}(2020){Pogge}, {Derwent}, {O'Brien}, {Jurgenson},
  {Pappalardo}, {Engelman}, {Brandon}, {Brady}, {Clawson}, {Shover}, {Mason},
  {Kneib}, {Araujo}, {Bouri}, {Kronig}, {Grossen}, {Gillet}, {Macktoobian},
  {Tuttle}, {Farr}, {S{\'a}nchez-Gallego}, \& {Sayres}}]{sdssVfps}
{Pogge}, R.~W., {Derwent}, M.~A., {O'Brien}, T.~P., {et~al.} 2020, in Society
  of Photo-Optical Instrumentation Engineers (SPIE) Conference Series, Vol.
  11447, Society of Photo-Optical Instrumentation Engineers (SPIE) Conference
  Series, 1144781, \dodoi{10.1117/12.2561113}

\bibitem[{{Poppett} {et~al.}(2018){Poppett}, {Sharples}, {Edelstein},
  {Schmoll}, \& {Bramall}}]{poppett18}
{Poppett}, C., {Sharples}, R., {Edelstein}, J., {Schmoll}, J., \& {Bramall}, D.
  2018, in Society of Photo-Optical Instrumentation Engineers (SPIE) Conference
  Series, Vol. 10702, \procspie, 107027O, \dodoi{10.1117/12.2312176}

\bibitem[{{Poppett} {et~al.}(2020){Poppett}, {Jelinsky}, {Guy}, {Edelstein},
  {Jelinsky}, {Aguilar}, {Sharples}, {Schmoll}, {Bramall}, {Tyas}, {Martini},
  {Fanning}, {Levi}, {Brooks}, {Doel}, {Duan}, {Tarle}, {Gazta{\~n}aga}, \&
  {Prada}}]{poppett20}
{Poppett}, C., {Jelinsky}, P., {Guy}, J., {et~al.} 2020, in Society of
  Photo-Optical Instrumentation Engineers (SPIE) Conference Series, Vol. 11447,
  Society of Photo-Optical Instrumentation Engineers (SPIE) Conference Series,
  1144711, \dodoi{10.1117/12.2562565}

\bibitem[{Poppett {et~al.}(2014)Poppett, Edelstein, Besuner, \&
  Silber}]{poppett14}
Poppett, C.~L., Edelstein, J., Besuner, R., \& Silber, J.~H. 2014, in
  Ground-based and Airborne Instrumentation for Astronomy V, ed. S.~K. Ramsay,
  I.~S. McLean, \& H.~Takami, Vol. 9147, International Society for Optics and
  Photonics (SPIE), 1937 -- 1946, \dodoi{10.1117/12.2054454}

\bibitem[{{Ross} {et~al.}(2018){Ross}, {Martini}, {Coles}, {Derwent},
  {Honscheid}, {O'Brien}, {Pappalardo}, {Tie}, {Brooks}, {Schubnell}, \&
  {Tarle}}]{ross18}
{Ross}, A.~J., {Martini}, P., {Coles}, R., {et~al.} 2018, in Society of
  Photo-Optical Instrumentation Engineers (SPIE) Conference Series, Vol. 10702,
  \procspie, 1070280, \dodoi{10.1117/12.2312885}

\bibitem[{{Schmoll} {et~al.}(2018){Schmoll}, {Besuner}, {Bramall}, {Brooks},
  {Edelstein}, {Jelinsky}, {Levi}, {Murray}, {Poppett}, {Sharples}, {Tyas}, \&
  {Schlegel}}]{schmoll18}
{Schmoll}, J., {Besuner}, R., {Bramall}, D., {et~al.} 2018, in Society of
  Photo-Optical Instrumentation Engineers (SPIE) Conference Series, Vol. 10702,
  Ground-based and Airborne Instrumentation for Astronomy VII, ed. C.~J.
  {Evans}, L.~{Simard}, \& H.~{Takami}, 107027N, \dodoi{10.1117/12.2312140}

\bibitem[{{Schubnell} {et~al.}(2016){Schubnell}, {Ameel}, {Besuner},
  {Gershkovich}, {Heetderks}, {Hoerler}, {Kneib}, {Heetderks}, {Silber},
  {Tarl{\'e}}, \& {Weaverdyck}}]{schubnell16}
{Schubnell}, M., {Ameel}, J., {Besuner}, R.~W., {et~al.} 2016, in Society of
  Photo-Optical Instrumentation Engineers (SPIE) Conference Series, Vol. 9908,
  Ground-based and Airborne Instrumentation for Astronomy VI, 990892,
  \dodoi{10.1117/12.2233370}

\bibitem[{Schubnell {et~al.}(2018)Schubnell, Aguilar, Ameel, Caseiro, Fanning,
  Freudenstein, Gershkovich, Heetderks, Hörler, Leitner, Levi, Silber, Sun,
  Tarlé, Weaverdyck, Zhang, \& Brooks}]{schubnell18}
Schubnell, M., Aguilar, J., Ameel, J., {et~al.} 2018, in Advances in Optical
  and Mechanical Technologies for Telescopes and Instrumentation III, ed.
  R.~Navarro \& R.~Geyl, Vol. 10706, International Society for Optics and
  Photonics (SPIE), \dodoi{10.1117/12.2311573}

\bibitem[{Sharp {et~al.}(2006)Sharp, Saunders, Smith, Churilov, Correll,
  Dawson, Farrel, Frost, Haynes, Heald, \& et~al.}]{aaomega}
Sharp, R., Saunders, W., Smith, G., {et~al.} 2006, Ground-based and Airborne
  Instrumentation for Astronomy, \dodoi{10.1117/12.671022}

\bibitem[{{Shectman} {et~al.}(1996){Shectman}, {Landy}, {Oemler}, {Tucker},
  {Lin}, {Kirshner}, \& {Schechter}}]{lascampanas}
{Shectman}, S.~A., {Landy}, S.~D., {Oemler}, A., {et~al.} 1996, \apj, 470, 172,
  \dodoi{10.1086/177858}

\bibitem[{{Shourt} {et~al.}(2020){Shourt}, {Besuner}, {Silber}, {Dunlop},
  {Evatt}, {Brooks}, {Doel}, {Duan}, {Fanning}, {Gazta{\~n}aga}, {Martini},
  {Prada}, {Schubnell}, \& {Tarle}}]{shourt20}
{Shourt}, W., {Besuner}, R., {Silber}, J., {et~al.} 2020, in Society of
  Photo-Optical Instrumentation Engineers (SPIE) Conference Series, Vol. 11445,
  Society of Photo-Optical Instrumentation Engineers (SPIE) Conference Series,
  114456J, \dodoi{10.1117/12.2562687}

\bibitem[{Silber {et~al.}(2012)Silber, Schenk, Anderssen, Bebek, Becker,
  Besuner, Cepeda, Edelstein, Heetderks, Jelinsky, Johnson, Karcher, Perry,
  Post, Sholl, Wilson, \& Zhou}]{silber12}
Silber, J.~H., Schenk, C., Anderssen, E., {et~al.} 2012, in Modern Technologies
  in Space- and Ground-based Telescopes and Instrumentation II, ed. R.~Navarro,
  C.~R. Cunningham, \& E.~Prieto, Vol. 8450, International Society for Optics
  and Photonics (SPIE), 1064 -- 1076, \dodoi{10.1117/12.926457}

\bibitem[{Smee {et~al.}(2013)Smee, Gunn, Uomoto, Roe, Schlegel, Rockosi, Carr,
  Leger, Dawson, Olmstead, \& et~al.}]{smee2013}
Smee, S.~A., Gunn, J.~E., Uomoto, A., {et~al.} 2013, The Astronomical Journal,
  146, 32, \dodoi{10.1088/0004-6256/146/2/32}

\bibitem[{{Wang} {et~al.}(2020){Wang}, {Huang}, {Chen}, {Kimura}, {Wen}, {Yan},
  {Karr}, {Chou}, {Chang}, {Hsu}, {Hu}, {Ling}, {Reiley}, {Roberts}, {Gunn},
  {Loomis}, {Lupton}, {Siddiqui}, {Murray}, {Ferreira}, {dos Santos}, {Souza
  Oliveira}, {de Oliveira}, {Marrara}, {Tamura}, {Moritani}, \&
  {Takato}}]{pfs2020}
{Wang}, S.-Y., {Huang}, P.-J., {Chen}, H.-Y., {et~al.} 2020, in Society of
  Photo-Optical Instrumentation Engineers (SPIE) Conference Series, Vol. 11447,
  Society of Photo-Optical Instrumentation Engineers (SPIE) Conference Series,
  114477V, \dodoi{10.1117/12.2561194}

\bibitem[{Zhang {et~al.}(2018)Zhang, Silber, Heetderks, Leitner, Schubnell,
  Levi, Wang, Fanning, Fagrelius, Dobson, \& Aguilar}]{zhang18}
Zhang, K., Silber, J.~H., Heetderks, H.~D., {et~al.} 2018, in Advances in
  Optical and Mechanical Technologies for Telescopes and Instrumentation III,
  ed. R.~Navarro \& R.~Geyl, Vol. 10706, International Society for Optics and
  Photonics (SPIE), 1358 -- 1369, \dodoi{10.1117/12.2314666}

\end{thebibliography}
\bibliographystyle{aasjournal}

\end{document}